\documentclass[prb,twocolumn,superscriptaddress,showpacs]{revtex4}

\usepackage{times}
\usepackage{amsmath}
\usepackage{amssymb}
\usepackage{graphicx}
\usepackage[noxcolor]{pstricks}

\newcommand{\ZZ}{\mathbb{Z}}

\newcommand{\MA}{\mathcal{A}}
\newcommand{\MC}{\mathcal{C}}
\newcommand{\MT}{\mathcal{T}}
\newcommand{\MU}{\mathcal{U}}
\newcommand{\MD}{\mathcal{D}}

\begin{document}

%\begin{flushright}
%ITFA-2002-13\\
%EMPG-02-08\\
%HWM-02-16
%\end{flushright}
%\vspace{0.0cm}

{}\vspace*{2cm}

\title{Condensate induced transitions between topologically ordered phases}

\author{F. A. Bais}
\affiliation{Institute for Theoretical Physics, University of Amsterdam, Valckenierstraat 65, 1018 XE Amsterdam, The Netherlands}
\affiliation{Santa Fe Institute, Santa Fe, NM 87501, USA}
\email{F.A.Bais@uva.nl}
\author{J.K.Slingerland}
\affiliation{Dublin Institute for Advanced Studies, School of Theoretical Physics, 10 Burlington Rd, Dublin, Ireland.}
\email{slingerland@stp.dias.ie}

%\thanks{}

\begin{abstract}
We investigate transitions between  topologically ordered phases in two spatial dimensions induced by the condensation of a bosonic quasiparticle. To this end, we formulate an extension of the theory of symmetry breaking phase transitions which applies to phases with topological excitations described by quantum groups or modular tensor categories. This enables us to deal with phases whose quasiparticles have non-integer quantum dimensions and obey braid statistics. Many examples of such phases can be constructed from two-dimensional rational conformal field theories and we find that there is a beautiful connection between quantum group symmetry breaking and certain well-known constructions in conformal field theory, notably the coset construction, the construction of orbifold models and more general conformal extensions. Besides the general framework, many representative examples are worked out in detail. 
\end{abstract}

\date{\today}

%\pacs{%PACS numbers:
%71.10.Pm, %Fermions in reduced dimensions (anyons, composite fermions, Luttinger liquid, etc.
%73.43.-f, % Quantum Hall effects
%05.30.Pr,	% Fractional statistics systems (anyons, etc.)
%11.25.Hf.	% Conformal field theory, algebraic structures
%}
%\maketitle must follow title, authors, abstract and \pacs

\maketitle

\section{Introduction}
In both high energy and condensed matter physics, there is a long tradition of studying systems which exhibit topological excitations. Recently, this field has received a new impetus, since it has been realized that such topological excitations may permit fault tolerant storage and manipulation of quantum information~\cite{Kitaev03,Freedman98,Nayak2007}. In connection with this, there are current experimental efforts to prove the existence of nontrivial topological phases in the fractional quantum Hall effect\cite{Dolev2008,Radu2008,Willett2008,Godfrey2007} and to construct such phases in Josephson junction networks \cite{Gladchenko2008}.

Topological excitations are usually introduced at the classical level as solutions to the equations of motion and the observables that distinguish them are directly linked to topologically invariant properties of these solutions. This places topological particles in marked contrast to the more usual perturbative (quasi)particles. The latter are described as low energy quantum fluctuations over a given vacuum state. The group of symmetries of the system that fixes the vacuum state will act on the fluctuations and cause them to organize into multiplets. As a result these perturbative particle states form irreducible representations of the symmetry group. For topological particles no such labeling is obviously present. 

A similar dichotomy exists when considering the ground states of different phases, or the order parameters that distinguish between phases. Traditionally, when a phase exhibits ground state degeneracy, the different ground states would be related by the action of symmetry operators, but in topological phases, ground state degeneracies appear for models on topologically nontrivial spatial manifolds without the obvious intervention of any symmetry, and in fact the different ground states can often not be mixed by any local operator. Analogously, traditional phases can be distinguished by the expectation values of local order parameters, while different topological phases may exist which are not distinguished by any local order parameter. As a result a number of indicators for \emph{topological order} which are not based on symmetry have emerged, notably the dimensions of the spaces of ground states on spatial surfaces of non-trivial topology \cite{Einarsson90,Wen90} and the topological entanglement entropy \cite{Levin06,Kitaev06}. 

Despite the fact that topological phases may not be fully characterized by their symmetries (or at least not by symmetries represented by local operators), one may often still organize the excitation spectra of such phases by using `symmetries' that are not obvious from the Hamiltonian or Lagrangian of the system and which may in fact not be realized locally. Thus one may hope for a generalization or analogue of the theory of symmetry breaking phase transitions which applies topological phases, by allowing for such `topological symmetries'. One of the main goals of this paper is to set up such a formalism for the particular case of phase transitions which occur due to the formation of a condensate of bosonic quasiparticles. Before we go into a further description of this formalism, let us make some remarks which we hope may prevent confusion in reading the rest of the paper. 

First of all, we do not want to limit ourselves to `strictly topological phases', which have no nontrivial symmetries represented by local operators. In fact we will include theories which have a discrete symmetry represented by local operators but no nontrivial topology as a special case. We would like to point out that in gauge theories, electric charges, which are supposedly `non-topological particles', coming from the locally represented gauge symmetry of the system, can have nontrivial topological interactions with magnetic fluxes through the Aharonov-Bohm effect, so in order to describe the full topological order of gauge theories it is necessary to take the usual gauge symmetry into account. 

Secondly, we will often speak of `symmetry breaking' when some of the symmetries involved may be gauge symmetries. Gauge theories can be interpreted to a certain extent as constrained systems; some of the gauge degrees of freedom are auxiliary and could in principle be eliminated at the price of introducing very complicated interactions among the true physical degrees of freedom. In a theory with a gauge symmetry the physical states are the gauge invariant states, so the spectrum does not manifestly exhibit the degeneracies of nontrivial representations in the spectrum and one may wonder at the idea of a symmetry breaking phase transition. However, despite the absence of gauge-variant states, the physics of gauge theories certainly depends on the invariants characterizing the representations that are present in the model and one speaks of gauge symmetries as `hidden symmetries'. A similar situation occurs in topological field theory, where the particles often do not have internal degrees of freedom on which a symmetry could act, but nevertheless, their fusion rules can be described by the representation theory of a quantum group. 

Hidden symmetry breaking is to a large extent analogous to the usual breaking of global, non-gauge symmetries.  In the global case, there is typically a local order parameter that breaks the symmetry and as a result, in the broken phase the degeneracies due to the original symmetry are (partially) lifted, and the spectrum is now organized in representations of the smaller residual symmetry group.  A gauge symmetry cannot be broken by a local order parameter (by Elitzur's theorem\cite{Elitzur1975}). Yet, condensates with invariant order parameters are allowed and the hidden symmetry can effectively be reduced due to such a condensate. This phenomenon is usually referred to as the Higgs effect or the `breaking' of a hidden (or local)  symmetry. Bearing this warning in mind our philosophy is to use the term `breaking' in this cavalier way. 

A well known example of topological symmetry occurs in gauge theories with gauge group $\ZZ_N$ defined on a lattice in \mbox{2+1} dimensions\cite{thooft:1977} 
In such gauge theories, the spectrum consists of charges, whose internal state transforms under the $\ZZ_N$ gauge group, magnetic fluxes (where flux is the topological quantum number), which are gauge invariant and composites of charge and flux, called dyons. One sees immediately that the $\ZZ_N$ representation label on the particles is not enough to completely fix the sector, since it is blind to the flux. However, one may introduce a second \lq\lq dual'' $\ZZ_N$ symmetry -- which is not a gauge symmetry, and which acts on fluxes in the same way that the original gauge group acts on charges and which leaves states without flux invariant. The topological sectors of the theory are then completely distinguished by their behavior under the full $\ZZ_N\!\times\!\ZZ_N$ symmetry and the addition of flux and charge quantum numbers is also captured by the tensor product of $\ZZ_N\!\times\!\ZZ_N$ representations. One may even include the Aharonov-Bohm braid interactions between charges and fluxes by introducing a new structure on $\ZZ_n\!\times\!\ZZ_n$ called the universal $R$-matrix.

In general, one cannot expect to capture the full particle spectrum and topological interactions of a physical system using only group theory. Still, it is believed that every type of topological order in 2+1 dimensional systems can be described using the representation theory of a modular tensor category or, dually, a quantum group.  The different types of quasiparticle correspond to the irreducible representations of the quantum group and the fusion and braiding interactions are described by the tensor product of representations and the $R$-matrix respectively. This asks for generalization of the theory of symmetry breaking phase transitions to include topological symmetries described by quantum groups.
%Since the situation is so analogous to the usual classification of particles and couplings using group theory, it makes sense to ask for a generalization of the theory of symmetry breaking phase transitions which applies to this more general situation. We are interested in situations where this \lq\lq quantum group symmetry breaking'' is caused by the condensation of one of the quasiparticles (necessarily a boson).  
In earlier work\cite{BSS02,BSS03}, we have developed such a theory of `spontaneous' quantum group symmetry breaking and  applied it to discrete gauge theories. This theory was later refined and applied to phase transitions in quantum nematics and other systems\cite{baismathy06a,baismathy06b,baismathy06c}. Recent work of Bombin and Martin-Delgado \cite{Bombin2007,Bombin2008} also provides interesting realizations of such transitions in models based on Kitaev's toric code model\cite{Kitaev03}, which exhibits the same topological order as the discrete gauge theories.

\begin{figure}[hbt]
\begin{center}
\includegraphics[width=3.2in]{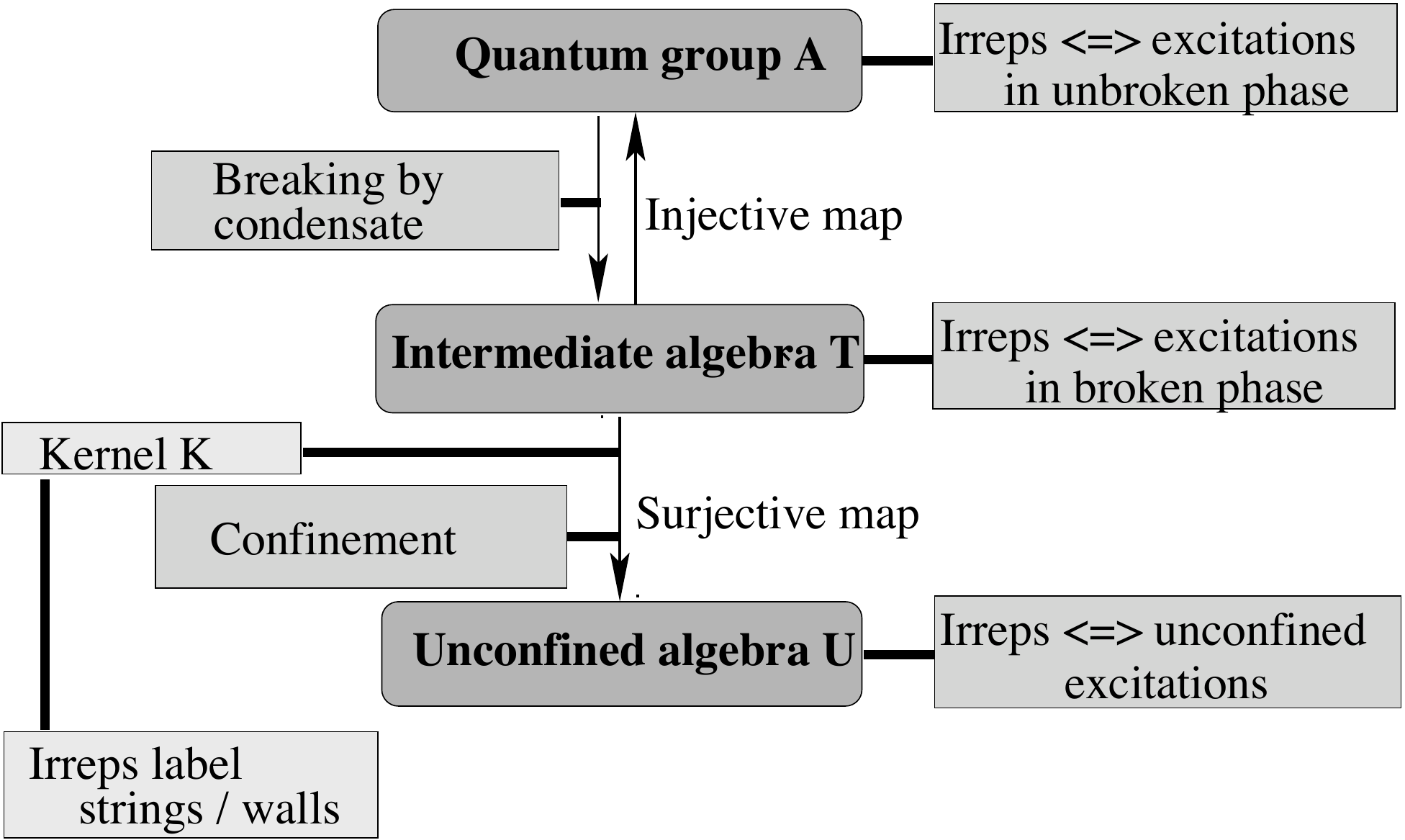}
\end{center}
\caption{\footnotesize A schematic of quantum group symmetry breaking. After breaking we arrive at an intermediate algebra $\MT$ that may have irreps which are in fact confined. The low energy effective theory of the condensed phase is based on the unconfined algebra $\MU$.}
\label{fig:symscheme}
\end{figure}
The general features of our symmetry breaking scheme are as follows (see figure \ref{fig:symscheme}). Before condensation, the system is described by a quantum group $\MA$ and each particle carries an irreducible representation (irrep) of $\MA$.  When particles carrying the representation $\pi_c$ condense, the condensate will have an order parameter which is a state in the module of the representation $\pi_c$. The symmetry of the condensed phase should leave this order parameter invariant and hence the quantum group $\MA$ is broken down to a Hopf subalgebra $\MT\subset \MA$ whose representations characterize the excitations of the condensed phase. Depending on their braiding interaction with the condensed particles, these excitations may or may not be confined. In particular, if an excitation over the condensate has nontrivial braiding with the condensed particle then the order parameter of the condensate will not be single valued near this excitation and the excitation will pull a string or wall in the condensate and be confined (the energy required for the creation of the string will be linear in the string's length, since the condensate is destroyed near the string). The non-confined particles are particles in the true sense of the word, that is, point-like excitations, and their interactions are described by the representation theory of a `Hopf quotient' $\MU$ of $\MT$ ($\MU$ is the image of $\MT$ under a surjective map that preserves the Hopf algebra structure). The strings pulled by the confined particles can also be studied and they are classified by the representations of a subalgebra of $\MT$ which is determined by the Hopf map from $\MT$ onto $\MU$ and which is analogous to the kernel of a homomorphism between groups. 
\begin{figure}[hbt]
\begin{center}
\includegraphics[width=2.5in]{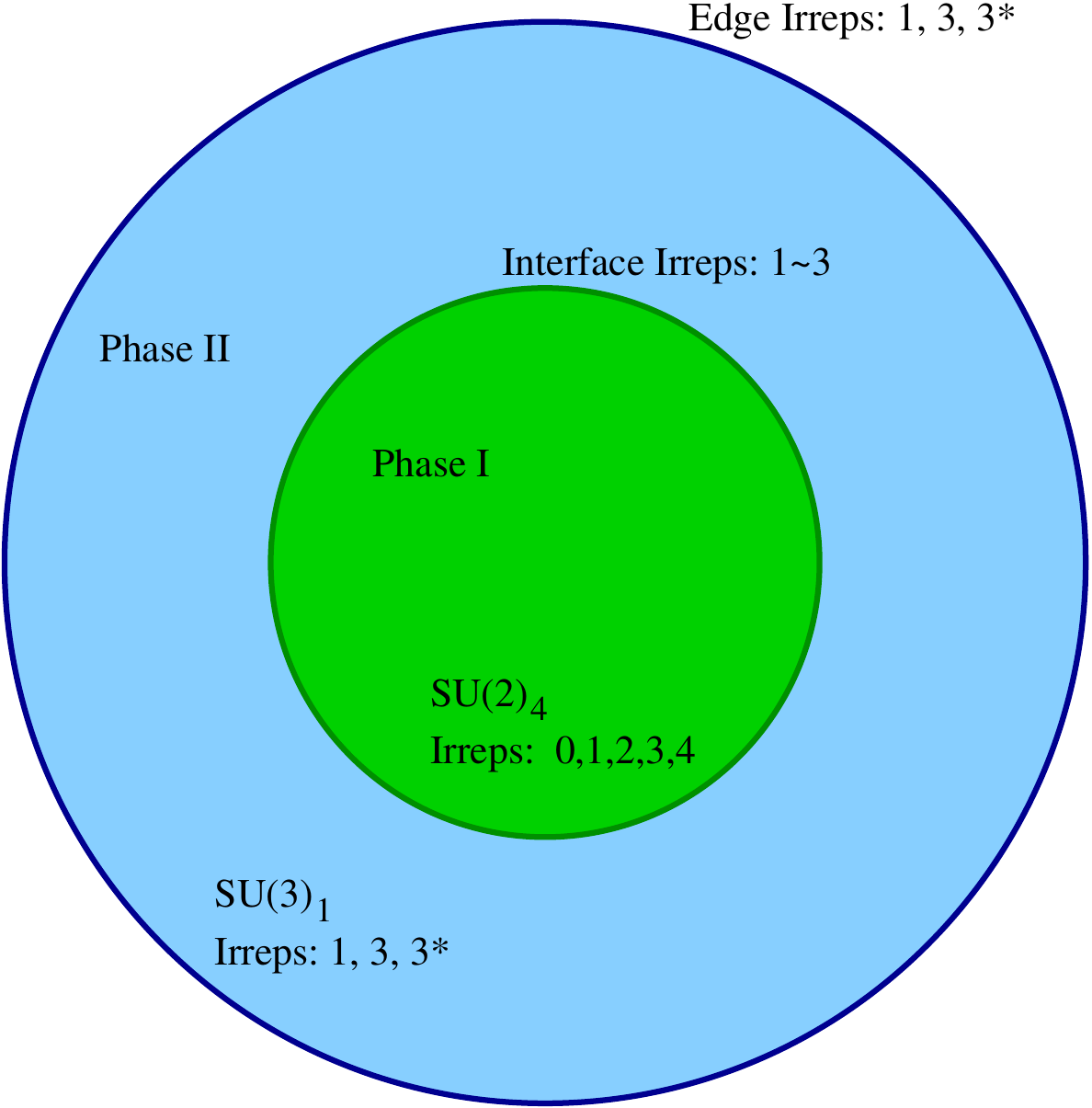}
\end{center}
\caption{\footnotesize A geometry with an interface between two topological phases related by the proposed breaking mechanism. In the inner region there is an unbroken phase, while the outer region is in a broken phase. The interface states correspond to the representations that are confined in the outer region (i.e.~the $1$ and $3$ representations of $SU(2)_4$). The states on the outer edge belong to representations that are not confined in the outer region (the $3$ and the $3^{*}$ of $SU(3)_1$).}
\label{fig:edgeinterface}
\end{figure}
In the sequel, we will devise a treatment of Hopf symmetry breaking which generalizes the treatment given in our earlier papers, while keeping the amount of quantum group or Hopf algebra theory that is needed to a minimum. Therefore, it will not be necessary to flesh out the details of the algebras contained in diagram~\ref{fig:symscheme} (detailed definitions can be found in the original papers). However, the structure of the diagram will be preserved, in that there will still be two levels to our treatment of condensation-induced phase transitions. First, the symmetry is broken, leading to a spectrum of excitations that occur in the broken phase, or on an interface between domains that support the broken and unbroken phases. Then these excitations are separated into confined and non-confined excitations, where the word `confined' means that the excitations are either bound to a boundary between a domain in the broken phase and a domain in the unbroken phase, or bound together like quarks in hadrons. This should lead us finally to a description of the fusion and braiding properties of the non-confined particles and ultimately also to a description of the strings pulled by the confined particles and a classification of \lq\lq hadronic'' composites of confined particles.

The breaking schemes we discuss may also have interesting applications in describing the physics of spatial geometries where interfaces between different topological phases occur, that may be enforced by external means (by applying different magnetic fields for example). We give an example of such a two phase geometry in figure \ref{fig:edgeinterface}, where the interface carries only those  edge states of the interior disc (with phase I), which are confined in phase II of the outer region, the outer edge carries (non-confined) states that are also allowed in the outer region. We will return to the specific phases indicated in the figure later on. In line with this application one may also  draw conclusions on the boundary theory of certain two layer systems as will be explained  in sections \ref{sec:embeddings2} and \ref{sec:doubleC-S}.

\section{Setting the stage}
\label{sec:tqftsec}

\subsection{Fusion rules, spin and monodromy}

Let us quickly review the minimal knowledge of $(2+1)$-dimensional topological field theory that we will need for the rest of the paper. For much more detail, the reader may consult for instance Refs.~\onlinecite{Preskill-lectures,Kitaev06a,Turaev94,Kassel95}. First of all, we assume that the theory has a finite number of topological sectors, labeled by some finite set of labels. We will call these \emph{anyonic charges}, or in some cases \emph{topological charges}, especially when it is not obvious if we are dealing with charges characterizing point-like excitations.  We can think of these charges as topological quantum numbers, but also as charges related to some (group) symmetry and sometimes, like in the case of $\ZZ_N$ gauge theory described before, we can think of them either way. In many physical situations, it is necessary to introduce superselection sectors which correspond to the same topological charges but which have different non-topological quantum numbers (these may for instance characterize short-range interactions). For the purposes of this paper, we will ignore these and consider such sectors to be the same. As a result we can describe theories which include electric charge and other gauge charges as if they have only finitely many sectors.

% However, we will not introduce labels to distinguish particles which have identical topological properties, for example in the fractional quantum Hall effect, we will not distinguish a (hypothetical) bound state of two electron holes from the vacuum, because such a state would not have topological interactions with any other state. Because of this focus on topology, the Hall systems will have finitely many topological sectors, even though there are of course infinitely many superselection sectors, given by the electric charges.

The basic interactions between topological sectors in 2+1 dimensions are {\em fusion} and {\em braiding}. Fusion may be summarized using {\em fusion rules} of the form
\begin{equation}
a\times b=\sum_{c}N^{ab}_{c} c.
\end{equation}
Here $a$ and $b$ are the two topological charges which are to be fused, the labels $c$ are the possible overall topological charges of the result of the fusion and the integer coefficients $N^{ab}_{c}$ indicate the number of independent couplings between $a$ and $b$ that give $c$. Concretely, a zero fusion coefficient means that the charges $a$ and $b$ cannot fuse to $c$ and a nonzero coefficient means they can. One may think of the fusion coefficient $N^{ab}_{c}$ as the dimension of the space of low energy states of a piece of topological medium which has overall topological charge $c$ and which contains two topological excitations with charges $a$ and $b$.  A physical requirement on fusion rules is that they must be associative, that is 
\begin{equation}
(a\times b)\times c= a\times(b\times c)
\end{equation}
%. 
We also require that there is a unique vacuum sector, labeled $0$ or $1$, depending on the context, which has the property that 
\begin{equation}
1\times a=a\times 1=a
\end{equation}
for any other sector $a$. Finally, we require that any sector $a$ has a charge conjugate sector, denoted $\bar{a}$, with which it can fuse to the vacuum in a unique way, i.e. 
\begin{eqnarray}
a\times \bar{a}=1+\sum_{c\neq 1}N^{a\bar{a}}_{c} c \\
\bar{a}\times a=1+\sum_{c\neq 1}N^{\bar{a}a}_{c} c.
\end{eqnarray}
For systems with well defined braiding interactions (that is, all two-dimensional systems with pointlike excitations), we will also have `symmetry of the fusion interaction', that is $a\times b=b\times a$.\footnote{This is not necessarily true for systems with non-pointlike excitations (for example systems with excitations that are attached to a boundary by a string) and we will see an example of this in paragraph~\ref{D3sec}.} 

Any topological charge $a$ has a {\em spin factor} $\theta_a$ associated with it. This is a phase factor that the wave function of the anyonic system picks up when the anyon is rotated (twisted) over a $2\pi$ angle. We can think of the particle as being in an eigenstate of two dimensional spin and the spin factor is the effect of a $2\pi$ rotation on this eigenstate. We will also use the spin $h_a$ of the particle, which is related to the spin factor by $\theta_a=e^{2\pi h_a}$. For systems with finitely many topological sectors, the spins are always rational\cite{Vafa1988}. 

\begin{figure}[hbt]
\begin{center}
\includegraphics[width=3.0in]{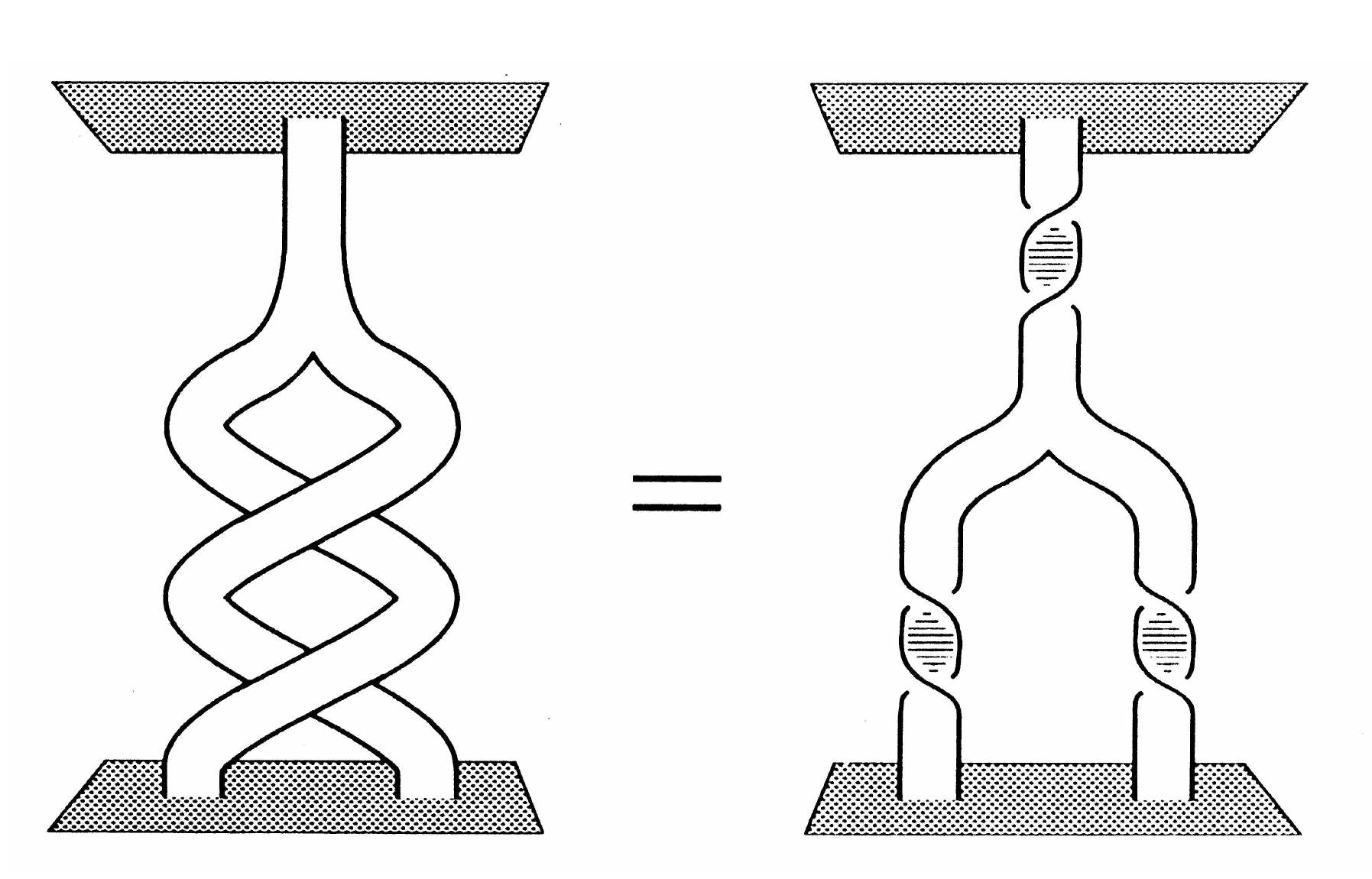}
\end{center}
\caption{\footnotesize The ribbon equation, relating fusion to topological spin and monodromy.}
\label{fig:ribbon}
\end{figure}

Adiabatic exchanges of the particles (without twisting) also have an effect on the wave function of the system. This is the analogue of the statistical interaction of fermions or bosons (for bosons the fact that there is no effect of the exchanges actually tells us a lot about collective behaviour). In two dimensions these exchanges are governed by the braid group, rather than the permutation group. In particular this means that left over right exchanges are not the same as their inverses, the right over left exchanges. The product of two right over left exchanges is often called the {\em monodromy}. It returns the excitations to their original positions, but may nevertheless have a nontrivial effect on the state of the system.  This effect may be described in terms of fusion and twisting, using the so called `ribbon equation' whose pictorial representation is shown in figure~\ref{fig:ribbon}. The braiding process is topologically the same as a full twist of the region containing both charges (i.e.~a full twist of their fusion product), combined with full twists of the charges themselves in the opposite direction.  Hence, given two anyonic charges $a$ and $b$, the effect of the monodromy on the fusion channels that yield overall charge $c$ is to introduce a phase factor $\theta_c (\theta_a \theta_b)^{-1}$, or $e^{2\pi i (h_c-h_a-h_b)}$.  
% This is an analogue of the spin--statistics theorem for two-dimensional systems. Except that it doesn't give anything non-trivial in higher dimensions, because monodromy is completly trivial there - we would need to have something about exchange here for a real generalization. See for instance Pachos and Brennen for that. 
%

\subsection{Quantum dimensions and modular group
% from fusion and spin
}

Like the monodromy, many of the other properties of the topological phase may be obtained directly from the fusion rules and the spin factors. Examples are the quantum dimension $d_a$ of a particle labeled $a$, the modular matrices $S$ and $T$ and the topological central charge $c$. We give formulas for all these quantities here. 

The quantum dimension $d_a$ of the sector $a$ gives the asymptotic number of fusion channels available when many particles of type $a$ are fused together. If there are $N$ such particles, then the total number of fusion channels scales asymptotically as $(d_a)^N$ for large $N$. We may find $d_a$ as the Perron-Frobenius eigenvalue of the fusion matrix $N_a$ whose elements are defined by $(N_a)_{b,c}=N^{a,c}_{b}$. 
% Another way to find $d_a$ is the following. First of all, define $c_{a}^{b}(N)$ by 
% %
% \begin{equation}
% (a\times{\bar{a}})^{\times N}=\sum_{b}c_{a}^{b}(N) b, 
% \end{equation}
% %
% then 
% %
% \begin{equation}
% \log(d_a)=\lim_{N\rightarrow \infty} \frac{1}{2N}\log(c_{a}^{1}).
% \end{equation}
% %
The quantum dimensions are real and positive and they have the important property that they are conserved under fusion, that is
\begin{equation}
\label{dimtens}
d_a d_b= \sum_{c}N_{ab}^{c}d_c.
\end{equation}
The total quantum dimension $\MD$ is defined by 
\begin{equation}
\MD=\sqrt{\sum_{i}d_i^2}
\end{equation}
The quantity $\MD^2$ can be considered the quantum dimension of the quantum group that underlies the system. The topological entanglement entropy of the ground state of a system in a topological phase is proportional to the logarithm of $\MD$. 

The topological central charge $c$ is defined by the following formula. It is only determined up to a multiple of $8$.   
% %
% \begin{eqnarray}
% P^{\pm}&=&\sum_{i}d_i^2\theta_i^{\pm 1}\nonumber\\
% P^{+}P^{-}&=&\MD^{2}\nonumber\\
% \frac{P^{+}}{\MD}&=&e^{i\frac{\pi}{4}c} \Longleftrightarrow (c \mod 8) = \arg(P^{+}).
% \end{eqnarray}
% %
\begin{equation}
e^{i\frac{\pi}{4}c}=\frac{1}{\MD}\sum_{i}d_i^2\theta_i
\end{equation}
Any conformal field theory whose topological sectors have these quantum dimensions and spin factors must have conformal central charge $\tilde{c}$ equal to $c$ modulo $8$.
It is in fact impossible to find out more about the conformal central charge of a CFT from the TQFT corresponding to the CFT than its value up to multiples of $8$. To see this one may consider the $E_{8}$ WZW model at level $1$. This is a CFT at $c=8$ which has only one primary field, or topological sector. Tensoring any CFT with copies of this $E_8$ model 
%or its mirror model at $c=-8$ 
allows one to change the central charge of the CFT by multiples of $8$ without changing the corresponding TQFT.

On a surface of nontrivial topology, a TQFT will have a set of degenerate ground states. On the torus there is one state for every topological sector (for anyon models described by a \emph{modular} tensor category). The mapping class group of the surface acts projectively on this ground state space. In the case of the torus this means we can work with ordinary representations of the double cover of the mapping class group, the famous modular group generated by two elements $S$ and $T$ subject to the relations
\begin{equation}
\label{eq:modulargroup}
S^2=\MC,~~~~~(ST)^3=\MC.
\end{equation}
Here $\MC$ is an element of order $2$. On the basis of states labeled by the particle sectors $\MC$ is represented by the charge conjugation matrix, that is $\MC_{ab}=\delta_{a\bar{b}}$.
 
In a CFT that realizes this TQFT, the modular $S$-matrix and $T$-matrix are given by\cite{dsm}
\begin{eqnarray}
\label{modularmatrices}
S_{ab}&=&\frac{1}{\MD}\sum_{c} N^{c}_{a\bar{b}}\frac{\theta_c}{\theta_a\theta_{b}}d_c\nonumber\\
T_{ab}&=&e^{-2\pi i \frac{\tilde{c}}{24}}\theta_{a}\delta_{a,b}
\end{eqnarray}
The numbers $S_{ab}$ can also be defined within TQFT, as the trace of the monodromy acting on particles with labels $a,b$. This, together with the ribbon equation, leads to the formula above. The formula for $T$ is interesting, because it depends on the value of the conformal central charge $\tilde{c}$ modulo $24$. We just noted that the value of the central charge $c$ in a TQFT is only determined modulo $8$, so it seems that there might be a problem defining the above modular group representation for TQFTs. However, it turns out that any choice of $\tilde{c}$ modulo $24$, given that $\tilde{c}=c$ modulo $8$, gives a good representation of the double cover of the modular group. The different choices just change the action of $T$ by a third root of unity and this factors through the relations~(\ref{eq:modulargroup}) that $S$ and $T$ must satisfy.

Still, it is remarkable that CFTs are able to see topological information that TQFTs are blind to, namely a factor of a third root in the action of the modular matrix $T$. Of course one might include this third root into the definition of a TQFT. If this is done, then there will be three TQFTs with only one sector, realized in CFT for example by the trivial, $E_8$ level $1$ and $(E_8)_1\otimes (E_8)_1$ theories. Taking this third root into account in any definition of {\em topological order} would also necessarily mean that \lq topologically ordered' phases arise whose ground states on the torus are not degenerate, but do transform nontrivially under the action of $T$. Such phases can be realized by $E_8$ Chern--Simons theories at levels $1$ and $-1$ respectively.

\subsection{Contact with experiment}

Fusion rules, spin factors and the quantities that can be expressed in terms of these are the only elements of topological field theory that will be essential for an understanding of the main thread of the rest of this paper. An important question one may ask is thus whether fusion rules and spin factors are information that can be accessed through experiment. Experimental probes of topological systems (notably quantum Hall systems) that have been proposed in recent years include measurements of the tunneling current at point contacts and especially interferometric measurements of tunneling currents through double point contacts \cite{Chamon97,Fradkin98,DasSarma05,Stern06a,Bonderson06a,Bonderson06b,Bonderson07a,Bonderson07b,Chung06,Ilan2008}. In the case of fractional quantum Hall states with Abelian topological order, experiments with double point contacts have in fact recently been performed\cite{Camino05a,Camino05b,Godfrey2007,Willett2008}. 

The tunneling current at a single point contact in a Hall system is dominated at weak tunneling by the quasiparticle with the lowest scaling dimension. By observing the behavior of the tunneling current as a function of temperature, voltage or the size of the system\cite{Kane1992,Fendley1994,Fendley1995}, it should be possible, in principle, to extract this scaling dimension and from that the topological spin of the particle. Extracting the spins of the other particles will probably be considerably more difficult.

Interference experiments with double point contacts are governed by the \emph{monodromy matrix}\cite{Bonderson06a}. This is just a normalized version $M$ of the $S$-matrix, given by
\begin{equation}
M_{ab}=\frac{S_{ab}S_{11}}{S_{a1}S_{b1}}=\frac{1}{d_a\theta_a d_b\theta_{b}}\sum_{c} N^{c}_{a\bar{b}}d_c\theta_c
\end{equation}
It seems likely that at least some elements of the monodromy matrix $M$ will become available through interferometric measurements in any topological system that allows these. On the other hand, it should be noted that for example in the Hall systems, it will be difficult to get elements that do not involve the quasiparticle of lowest scaling dimension, which naturally dominates the tunneling. If enough of the matrix elements of $M$ are known, it might be possible to reconstruct the whole $M$ matrix and its close relative, the S-matrix. The $M$ and $S$-matrices are strongly constrained by various TQFT-identities, so a relatively small number of measured elements may be enough to determine them completely. From the $S$-matrix one may reconstruct the fusion rules using Verlinde's formula,
\begin{equation}
N_{ab}^{c}=\sum_{x}\frac{S_{ax}S_{bx}S_{\bar{c}x}}{S_{1x}}
\end{equation}
The $S$-matrix also gives information on the spins of the particles. For example, given the $S$-matrix and the fusion rules, the second line of equation (\ref{modularmatrices}) becomes a system of equations for the spin factors $\theta_i$. Another set of equations for the spin factors comes from the fact that $S$ and $T$ satisfy the modular group relation $(ST)^3=S^2$. These equations do not always allow for a full determination of the $\theta_i$: two theories which have the same $S$-matrix, but different spin factors are the Ising model and the $SU(2)$ Chern Simons theory at level $2$. These theories are distinguishable by their central charges, which are $\frac{1}{2}$ and $\frac{3}{2}$, respectively. However, there also exist examples of distinct theories with the same $S$-matrix as well as equal central charge, but with different spin factors, for example the two $c=0$ theories based on the quantum doubles of the finite groups $D_4$ (the symmetry group of a square) and $\bar{D}_2$ (the group of unit quaternions). Nevertheless, it is clear from Ocneanu rigidity (see section~\ref{sec:reconstructionphantasies}) that for given fusion rules (and hence for any given $S$-matrix), there can only be finitely many possible solutions for the spinfactors.

\subsection{From fusion and spin to a full TQFT}
\label{sec:reconstructionphantasies}

The mathematics of topological phases obviously involves more than just fusion rules and spin factors. The requirements on fusion we have given may be fleshed out with more mathematical structure to give the definition of a tensor category. Similarly, including braiding and spin, we may get to the definition of a ribbon tensor category.  The Hilbert spaces and transition amplitudes of the topological systems we are interested in may then be viewed as coming from  representations of such categories. The categories themselves in turn may be viewed as the representation categories of (appropriately generalized) quantum groups. While such structures will certainly be of importance for a more mathematically rigorous treatment of transitions between topological phases, that is not the aim of this paper and we intend instead to go into these details in a separate publication\cite{BaisWIP}. 

Still, one may ask at this point wether knowledge of the fusion rules and spin factors would allow one to reconstruct the full TQFT describing the system. An important step towards answering this question is a mathematical theorem which is usually referred to as Ocneanu rigidity \cite{Etingof2005}. This theorem states that, given a set of fusion rules, there can only be a finite set of inequivalent TQFTs, ribbon tensor categories, or just plain tensor categories corresponding to it. 
% {\em Note: the proof in Kitaev 's paper only shows that the set is discrete, but I'm pretty sure it's actually finite and he doesn't care to mention that. Finiteness is mentioned in one the other reference}. 
Since we are given not only the fusion rules but also the spin factors, it seems likely that a TQFT will be uniquely determined by this information in most cases. In fact, we are not aware of any example of a pair of inequivalent TQFTs with the same fusion rules and spin factors and even if such pairs do exist in nature, it will be difficult to separate them by experimental observation, for instance because they have the same $M$-matrix.

% {\em Add Frobenius-Schur indicators:}
% %
% \begin{equation}
% \kappa_{c}=\frac{1}{\MD^2}\sum_{a,b}N^{ab}_{c}\frac{\theta_{a}}{\theta_b}d_a d_b
% \end{equation}
% %

\section{On Bosons}
\label{sec:bosons}
In 3+1 or more dimensions, bosons can be characterized either as particles with integer spin or as 
particles which have trivial exchange interactions, that is, wave functions for many identical bosons are invariant under exchanges of the bosons. These two properties are equivalent by the spin statistics theorem \cite{Pauli40,StreaterandWightman}. In 2+1 dimensions, the requirements of integer spin and trivial statistics are no longer equivalent. There may be particles which have the property that when two of them are fused together, multiple fusion products may arise and the braiding of the original particles is trivial or nontrivial depending on the fusion channel they are in. 
% When the particles are in a superposition of fusion channels, the total state of the system is not an eigenstate of braiding and this is precisely what allows braiding to be used to do interesting quantum computations. 
Therefore it is not completely obvious what constitutes a boson in dimension 2+1. Two necessary conditions for a particle $a$ to be a boson are the following. 
\begin{itemize}
\item $a$ should have trivial spin, that is $\theta_a=1$, or $h_a\in \ZZ$.
\item $a$ should have partially (or completely) trivial self-monodromy. 
\end{itemize}
By partially trivial self-monodromy, we mean that there should be \textit{at least} one fusion channel in $a\times a$ for which the monodromy factor equals $1$. In other words, if $a\times a=\sum_{c}N^{aa}_{c} c$, then there should be at least one charge $c$ in the sum such that $\theta_c (\theta_a)^{-2}=1$. Since we already required that $\theta_a=1$, this comes down to the the requirement that there is a fusion channel $c$ with trivial twist, $\theta_c=1$. 

Both of these conditions are special cases of a more general condition,
\begin{itemize}
\item For every number $N$ of identical particles of type $a$, there should be at least one state in the Hilbert space for $a^{\times N}$ which is completely invariant under monodromy.
\end{itemize}
This condition is a reasonable requirement for particles which should be able to condense, because for any particle number, it provides at least one state which will not notice any `stirring'. This is analogous to the requirement that any `order parameter' for the condensate should be single valued. The general condition is much more difficult to check than the two special cases mentioned earlier. A thorough treatment would also require that we introduce much more of the formalism of topological field theory. However, we can make two useful remarks. First of all, for any particle $a$ with quantum dimension $d_a=1$, one may show that the requirement of trivial spin actually implies the general condition above (and hence it also implies trivial self-monodromy). Secondly, in a number of cases with $d_a\neq 1$, we have been able to show explicitly that there are in fact states with trivial monodromy for any number of particles of type $a$.

For the rest of this paper, we will ignore the general condition and work with condensates of particles with trivial spin and partially or fully trivial self-monodromy. These conditions seem to be sufficient for the condensate transitions we have studied. 

Finally let us note that we have asked only for trivial monodromy and not for completely trivial braiding. This means that in principle, the exchanges in multi-particle states may be represented  non-trivially (with eigenvalues $\pm 1$), so that the particles would behave in some ways like fermions, though spinless ones. We could of course exclude this behavior, but this restriction is unnecessary for our purposes and so we will allow for the more general situation.
% Such issues may possibly be resolved by using the concept of a Frobenius algebra in a category, as developed by \cite{runkelenco}, but this would lead us too far afield here. 

\subsection{Examples}

It turns out that bosons, that is, particles with trivial spin and partially trivial self-monodromy, are a very common occurrence in topological field theories and particularly in the theories that arise from proposed models for topological quantum computation. Let us give a few examples. 

\subsubsection{Non-Abelian Hall states}

The quantum Hall effect is currently the only physical system where theory predicts the existence of anyonic excitations in a parameter regime that is accessible in the laboratory. Recently, direct observation of anyonic statistics has been claimed \cite{Camino05a,Camino05b} and much indirect experimental evidence is also available. Anyons with non-Abelian braiding are also expected to exist at filling fraction $\nu=\frac{5}{2}$ and predicted at a number of other filling fractions, including $\nu=\frac{12}{5}$. These non-Abelian Hall states are currently the most experimentally advanced avenue toward topological quantum computation.  

Hall states are described by Chern-Simons theory in the bulk of the sample and by a corresponding chiral conformal field theory on the edge. In the case of non-Abelian theories, the CFT on the edge is often better understood than the bulk theory. The Moore-Read (MR) state\cite{mooreread} for $\nu=\frac{5}{2}$ and its generalizations, the Read-Rezayi (RR) series of states \cite{readrez}, which includes a candidate wave function for $\nu=\frac{12}{5}$, are described by an $SU(2)_k$ parafermionic CFT coupled to a $U(1)$ theory describing an electrically charged chiral boson. Electrically neutral excitations of these models may be described more simply using the even spin subsectors of an $SU(2)_k$ Wess-Zumino-Witten model. The $SU(2)$ WZW-model at level $k$ has $k+1$ primary fields which we will label $0,\ldots,k+1$ by their $SU(2)$ weights. The field labeled $\Lambda$ corresponds to a topological sector with spin factor given by $e^{2\pi i h_{\Lambda}}$, with $h_{\Lambda}=\frac{\Lambda(\Lambda+2)}{4(k+2)}$. In particular, the field labeled by $\Lambda=k$ has spin factor $e^{\frac{i\pi k}{2}}$, which equals $1$ whenever $4$ divides $k$. It also has quantum dimension equal to $1$, so trivial spin is enough to guarantee that this field is a boson. We may also see directly that this field has trivial self-monodromy, since it fuses to the vacuum sector with itself. Thus the RR-states based on the $SU(2)$ WZW-theories at levels which are a multiple of $4$ all have bosons. If we want to consider excitations that have nonzero electric charge we have to introduce labels corresponding to the $U(1)$ part of the CFT describing the RR-states, in addition to the $SU(2)$ labels. Taking the full spectrum into account, we then find even more bosons. However, the condensation of such charged bosons should lead to superconductivity as well as a change in the filling fraction and so it may be physically more difficult to imagine. 
% In particular, we find such a charged boson in the $k=2$ theory, which is the Moore-Read or Pfaffian state~\cite{mooreread} which is the most popular proposed description of the quantum Hall plateau at $\nu=\frac{5}{2}$. 
It is perhaps interesting to note that the $k=3$ RR-state, which could describe the Hall plateau at $\nu=\frac{12}{5}$, is one of the few low-lying states in the Read-Rezayi series that does not admit any bosons. We will present an inventory of bosons in the proposed non-Abelian Hall states and the expected condensed phases that result from them in a separate publication\cite{BaisWIP}.  

\subsubsection{Non-chiral theories}

Most of the known local models with anyons are in fact non-chiral; they have vanishing central charge. Important examples of this class of model are Kitaev's toric code models for discrete groups\cite{Kitaev03} and Levin and Wen's string net condensates\cite{Levin05a} (the loop and dimer models of Refs.~\onlinecite{Freedman04a,Freedman05a,Freedman05b} can also be viewed as a special case of these\cite{fendley08}). 

The toric code models exhibit the same topological order as the discrete gauge theories\cite{dgt1,dgt2,dgt3,dgt4,dgt5}, described by a quantum group called the quantum double $D(H)$ of the finite gauge group $H$. We have treated quantum group symmetry breaking in these models in our earlier works\cite{BSS02,BSS03}, so we will discuss them only briefly here. In the model with gauge group $H$, topological sectors are labeled by a conjugacy class $A\subset H$ and by an irreducible representation $\alpha$ of the centralizer group $N_A$ of an element $g_A$ of $A$ (the choice of $g_A$ does not matter). The spin factor of the sector labeled $(A,\alpha)$ is $\frac{1}{d_{\alpha}}\mathrm{Tr}(\alpha(g_{A}^{-1}))$, where $d_{\alpha}$ is the dimension of the irrep $\alpha$. Depending on the group $H$ there may be many bosonic sectors, but in general there are two classes of bosons that are always present: the electric sectors which have $A=\{e\}$ (where $e\in H$ is the unit element) and the magnetic sectors which have $\alpha=1$, the trivial representation of $N_{A}$. It is clear that both electric and magnetic sectors have trivial spin. Under fusion, the electric sectors only produce new electric sectors, so they also have trivial monodromy (in fact not just for two particle states, but for arbitrary numbers of particles). The purely magnetic sectors may fuse to give sectors which have nontrivial electric charges (that is, nontrivial centraliser labels), called \lq Cheshire charges'. The sectors with Cheshire charge will usually have nontrivial spin and so one may wonder if the magnetic particles are always true bosons. However, it is not difficult to show that the fusion of identical magnetic particles always contains at least one channel without Cheshire charge, so that the requirement of partially trivial self-monodromy is satisfied. In fact, one may go further and show that the topological Hilbert space for arbitrarily many identical magnetic particles always contains at least one state which has completely trivial monodromy. This state is basically the gauge invariant magnetic condensate state proposed earlier\cite{BSS02,BSS03}, but to make contact with the present formalism one must project this onto a subspace of the Hilbert space with fixed total topological charge (for example the space of topological singlets).

String net condensates are also described by quantum doubles, but more typically by quantum doubles of quantum groups like $U_{q}(sl(2))$ and its generalizations. These models are directly related to doubled Chern-Simons or WZW-models, with gauge groups $G\times G$, where $G$ is now a Lie group and the left and right hand copies of $G$ occur with the opposite levels, i.e.~these are $G_{k}\otimes G_{-k}$ theories. The topological sectors of such models are labeled by pairs of representations of $G$ that are admissible at level $k$. The spin of the sector labeled $(\lambda_1,\lambda_2)$ is given by $\theta_{\lambda_1}\theta_{\lambda_2}^{-1}$ and we see that all \lq diagonal' fields $(\lambda_1,\lambda_2)$ with $\lambda_1=\lambda_2$ have trivial spin. The fusion of two diagonal fields always yields at least one diagonal field, so the requirement of partially trivial self-monodromy is also satisfied. However, as with the case of the toric code, one may show that there are in fact states with totally trivial self-monodromy for any number of identical diagonal fields. The reason is that the monodromy of any state with $n$ diagonal fields $(\lambda,\lambda)$ may be described using the tensor product of the braid group representation associated with $n$ copies of $\lambda$ and its dual braid group representation. This tensor product contains a canonical singlet representation of the braid group and the corresponding state has totally trivial monodromy.    

\section{Condensation, Symmetry Breaking and Confinement}
\label{sec:qbreak}

\subsection{Symmetry Breaking:\\Branching Rules and Physical Requirements}
Suppose we can change the parameters of the microscopic Hamiltonian underlying our anyonic system in such a way that the particle with charge $a$ condenses ($a$ should be a boson in order for this to happen). Then we can ask what the overall effect of this condensation on the topological excitation spectrum of the system will be. The basic idea of this paper is that the condensate breaks down a symmetry underlying the spectrum. That is, before condensation the charge sectors correspond to irreducible representations of some quantum group $\MA$, the fusion rules are described by decomposition of the tensor products of these irreps etcetera. Then condensation breaks $\MA$ down to a subalgebra $\MT$ and afterward the excitations carry irreps of $\MT$.  When a quantum group is broken down to a subalgebra, two things generically happen to its irreducible representations. First of all, some irreps of the original quantum group $\MA$ will not be irreducible as representations of the subalgebra $\MT$. These representations will split into multiple irreps of $\MT$. Secondly, some representations which are inequivalent under the full $\MA$-action will have equivalent $\MT$-actions and hence become identified. More concretely, all this can be described by {\em branching rules} of the form 
\begin{equation}
a\rightarrow \sum_{b}n^{a}_{t} t.
\end{equation}
where $a$ is an irreducible representation of the original quantum group $\MA$, the $t$'s are irreps of the algebra $\MT$ that is left after symmetry breaking and the $n^{a}_{t}$ are multiplicities. We will now make an important conceptual step and put these representation labels and branching rules center stage, forgetting for the moment about quantum groups and their subalgebras. So assuming that we have a set of labels $\{a,b,c,\ldots\}$ which characterise the charge sectors of the unbroken theory, as well as fusion rules and spin factors for these labels, then we will say that symmetry breaking means that to each of these labels we associate a branching rule
\begin{equation}
a\rightarrow \sum_{i}n^{a}_{i} a_i.
\end{equation}
We will call the right hand side of this equation the {\em restriction} of $a$ (we are still thinking of it as the restriction of a representation $a$ of $\MA$ to $\MT$). The $n^{a}_{i}$ are again multiplicities and we have introduced a new notation where instead of labeling the components of the restriction of $a$ directly by sectors of the broken theory (which we think of as labels of irreps of $\MT$), we simply label them $a_1$, $a_2$ etc. Of course the new labels $a_i$ that occur will most likely not all correspond to distinct sectors of the new theory, for different choices of $a$. However, the notation introduced here is quite useful in the process of finding out exactly what the new set of sectors actually is. 

We want the new labels to be the labels for the excitations of the broken phase, so we will make the physical assumption that they have their own set of fusion rules satisfying the requirements of section~\ref{sec:tqftsec}, namely, associativity, existence of a vacuum label and conjugate representations, and a unique way for each conjugate pair of labels to annihilate to the vacuum. We will not require the new fusion rules to be symmetric, that is, we may allow that $a_i\times b_j\neq b_j\times a_i$ for some pairs $a_i,b_j$. We also do not require a well defined spin or monodromy of the $a_i$ at this point. The reason that we do not impose these requirements is that  we want the new set of labels to capture not only pointlike excitations of the condensate vacuum,  but also topological excitations which pull strings or alternatively, excitations which are confined to a boundary between the broken and unbroken phase. The string pulling excitations are expected to be the same as the excitations which occur only on the boundary, since a confined boundary excitation may be visualized as a thread or string 
extending from the boundary into the broken phase, ending at a string-pulling excitation of that phase. 

Apart from the requirement that the new theory has sectors with associative fusion and unique duals, there are two more important assumptions that go into the determination of the new set of labels and and their fusion rules. First of all, the sector that contains the condensed excitation should be indistinguishable from the vacuum sector in the condensed phase. Hence we require that the restriction of the condensed sector $c$ of the original theory contains the vacuum label $1$ of the new theory. In other words
\begin{equation}
c\rightarrow (c_1\equiv 1)+\sum_{i>1}n^{c}_{i} c_i.
\end{equation}
Secondly, we require that the fusion of the old and new labels is compatible with the branching, that is, restriction and fusion commute and we have 
\begin{equation}
\label{tensreq}
a\times b  =  \sum_{c} N^{ab}_{c} c \Rightarrow 
(\sum_{i}n^{a}_{i} a_i)\times (\sum_{j}n^{b}_{i} b_i) = \sum_{c,k}N^{ab}_{c}n^{c}_{k} c_k
\end{equation}
and
\begin{equation}
a  \rightarrow \sum_{i}n^{a}_{i} a_i \Rightarrow \bar{a}\rightarrow \sum_{i}n^{a}_{i} \overline{a_i}.
\end{equation}
%
% \begin{eqnarray}
% \label{tensreq}
% \left( a\times b  =  \sum_{c} N^{ab}_{c} c\right) \;\;\;\;\;\;\;\;\;\;& &\nonumber\\
% \Rightarrow
% \left( (\sum_{i}n^{a}_{i} a_i)\times (\sum_{j}n^{b}_{i} b_i) \right. &=& \left.\sum_{c,k}N^{ab}_{c}n^{c}_{k} c_k\right) \nonumber\\
% \left( a  \rightarrow \sum_{i}n^{a}_{i} a_i \right) &\Rightarrow& \left(\bar{a}\rightarrow \sum_{i}n^{a}_{i} \overline{a_i} \right) \hfill
% \end{eqnarray}
%
The equations above, together with the uniqueness of the unit of the new theory also imply that 
\begin{equation}
1\rightarrow 1_1\equiv 1.
\end{equation}
Here, we introduce a slight abuse of notation that we will utilize throughout, namely, if a sector branches to a unique new sector, we will denote the old and new sectors by the same label, as long as the meaning is clear from the context. 

The compatibility of fusion and restriction has another important consequence: it implies that the quantum dimensions are preserved under branching, that is, for every label $a$ of the unbroken phase, we have
\begin{equation}
\left(a\rightarrow \sum_{b}n^{a}_{b} b \right) \Rightarrow \left(d_a= \sum_{b}n^{a}_{b} d_b\right).
\end{equation}

\subsubsection{Example: breaking $\mathbf{SU(2)_4}$.}

We will now give a simple example of how one can determine the set of labels for the broken phase and their fusion rules directly, given the assumptions above. Consider the representation theory of $SU(2)_q$ at $q=e^{2\pi i/6}$. This is the quantum group for the $SU(2)_4$ WZW model of conformal field theory and for the $SU(2)$ Chern-Simons theory at level $4$.
%  and it plays an important role in one of the applications to the Hall effect that we will discuss later. 
There are five different topological sectors in this theory which are simply denoted by Dynkin labels $0,\ldots,4$, with $0$ denoting the vacuum. The quantum dimensions, spins and fusion rules for these sectors are given in table~\ref{tab:su24} and from this table, we may read off that the sector labeled $4$ is bosonic (we have $h_4\in \ZZ$, $4\times 4=0$ and $h_0-2h_4\in\ZZ$).   
\begin{table}[hbt]
\begin{center}
$\begin{array}{|l|l}
\hline SU(2)_4 {\rm~unbroken}\\
\hline
\hline
\begin{array}{l|ll}
0&d_0=1&h_0=0  \\
1&d_1=\sqrt{3}&h_1=\frac{1}{8}\\
2&d_2=2&h_2=\frac{1}{3}\\
3&d_3=\sqrt{3}&h_3=\frac{5}{8}\\
4&d_4=1&h_4=1
\end{array}\\
\hline
\hline
\begin{array}{lll}
1\times 1=0+2~&& \\
1\times 2=1+3&2\times 2=0+2+4~& \\
1\times 3=2+4&2\times 3=1+3&3\times 3=0+2 \\
1\times 4=3&2\times 4=2&3\times 4=1\;\;\;\;\;\; 4\times 4=0\\ 
\end{array}\\
\hline
\end{array}$
\end{center}
\caption{\footnotesize Spins, quantum dimensions and nontrivial fusion rules for $SU(2)_4$ (the fusion rules are symmetric and fusion rules for the vacuum have been omitted).}
\label{tab:su24}
\end{table}
If an excitation in the $4$-sector condenses, then $4$ will have to branch to the new vacuum and possibly other new labels. However, since $d_4=1$ and the quantum dimension of the new vacuum is also necessarily equal to $1$ and quantum dimensions are preserved under branching, we find that 
\begin{equation}
4\rightarrow 0.
\end{equation}
From here, we may conclude immediately that the restrictions of $3$ and $1$ must equal each other, since $4\times 1=3$ and $4\times 3=1$. Also, the restriction of $1$ (or $3$) can only have one part, because each part would contribute at least a numerical value of $1$ to the quantum dimension of the label $1$ and the value of this quantum dimension is less than $2$. Now let us look at the fusion of the restriction of $2$ with itself.  We have 
\begin{equation}
\label{eq:2x2}
2\times 2=0+2+4 \rightarrow 0 + \sum_{i}n^{2}_{i}2_i + 0.
\end{equation}

Since the vacuum appears twice on the right hand side, $2$ must branch into at least two parts (if there was only one part, it would be able to annihilate with itself in two different ways). 
Since the quantum dimension of $2$ equals $2$, this is possible, and in fact there must be exactly two parts $2_1$ and $2_2$, each with quantum dimension $1$. Note that neither $2_1$ nor $2_2$ can be identified with the vacuum sector $1$, since this would imply the splitting of $1$ through the fusion rule $1\times 1=0+2=0+2_1+2_2$ and this is impossible, since $d_1<2$. Looking back at the fusion $2\times 2$ we then conclude also that $2_1\neq 2_2$. We have now completely identified the branching rules for this transition and we turn to the fusion rules. These are straightforward for the new labels $0$ and $1$(using~(\ref{tensreq})), but for $2_1$ and $2_2$, we have two options, in principle. Either these sectors are both self dual, giving $2_1\times 2_1=2_2\times 2_2=0$, or they are dual to each other, giving $2_1\times 2_2=2_2\times 2_1=0$. Now rewriting equation (\ref{eq:2x2}) with our current knowledge, we see that
%
%{\small 
\begin{equation}
\begin{array}{rcl}
2\times 2&=&(2_1+2_2)\times \!(2_1 + 2_2)\\
&=&2_1\!\times \! 2_1+2_1\!\times \! 2_2+2_2\! \times \! 2_1+2_2\! \times \! 2_2\\ 
&=& 0+2_1+2_2+0.  
\end{array}
\end{equation}
%}
%
Hence if we assume that $2_1$ and $2_2$ are self-dual, it follows that either $2_1\times 2_2=2_1$ and $2_2\times 2_1=2_2$ or $2_1\times 2_2=2_2$ and $2_2\times 2_1=2_1$. In either case, one quickly checks that associativity of the fusion rules is violated, by evaluating $(2_1\times 2_2)\times 2_1$ and $2_1\times(2_2\times 2_1)$. Hence $2_1$ and $2_2$ must be dual to each other. Now we just have to decide whether $2_1\times 2_1$ equals $2_1$ or $2_2$ (and similarly for $2_2\times 2_2$). A similar associativity argument as before quickly yields that we must have $2_1\times 2_1=2_2$ and $2_2\times 2_2=2_1$. Hence we can straightforwardly obtain the full new set of sectors, as well their fusion rules. We summarize these results in table~\ref{tab:su24broken}
\begin{table}[hbt]
\begin{center}
$\begin{array}{|l|}
\hline SU(2)_4 {\rm ~broken}\\
\hline
\hline
\begin{array}{l|l}
0\rightarrow 0&d_0=1\\
1\rightarrow 1&d_1=\sqrt{3}\\
2\rightarrow 2_1+2_2~~&d_{2_1}=d_{2_2}=1\\
3\rightarrow 1&\\
4\rightarrow 0&
\end{array}\\
\hline
\hline
\begin{array}{llll}
1\times 1=0+2_1+2_2~&&& \\
1\times 2_1=1&2_1\times 2_1=2_2~&& \\
1\times 2_2=1&2_1\times 2_2=0&2_2\times 2_2=2_1~& \\
\end{array}\\
\hline
\end{array}$
\end{center}
\caption{\footnotesize Branching rules, quantum dimensions and nontrivial fusion rules for $SU(2)_4$ after condensation in the $4$-sector (the fusion rules are symmetric).}
\label{tab:su24broken}
\end{table}
Note that while the fusion rules of the broken theory  turn out to be symmetric, we did not put this in by hand and it is in fact just a particular feature of this theory that is not reproduced in general. 

\subsection{Confinement}

Not all of the excitations of the broken phase will be pointlike; some will pull strings in the condensate. These excitations will be confined, since a string is just a part of the medium where the original symmetry is restored and will cost an amount of energy proportional to its length. The excitations which do not pull a string will be the particlelike excitations of the new phase and they should have well-defined fusion and braiding interactions, in particular well defined monodromies and spin factors. Intuitively, an excitation should pull a string when it has nontrivial monodromy with the condensed excitation, since such nontrivial braiding would lead to a branch cut singularity in the condensate order parameter. Though not very rigorous, it is probably best to say that it is not possible to have a smooth  single valued order parameter field enclosing a particle which has a nontrivial monodromy, and that therefore that particle has to pull a string upon entering such a phase. One expects at least that the presence of the condensate does not interfere with the monodromy of the non-confined particles. In particular, we expect to be able to assign spin factors to the non-confined sectors by \lq lifting' them into the unbroken phase. The lifts of a sector $b$ of the broken theory are just all labels $b^i$ of the original theory that have $b$ in their restriction. A necessary condition for a sector to not be confined is the following: 
\begin{itemize}
\label{confreq}
\item If a sector $b$ is not confined then all its lifts $b^{i}$ must have equal spin factors.
\end{itemize}
In the other direction, sectors which do not satisfy this condition will be confined. Note that it is only natural that we should not be able assign spin factors to string-like excitations, since twisting such an excitation leads to a physically observable twist in the string connected to it and we should not expect that such a change can be absorbed by a change of the phase of the wave function.

There are a number of other physical criteria on the set of non-confined particles which we could impose separately, but which in practice turn out to be implied by the simple requirement above in all cases we have investigated. First of all, the non-confined sectors must form a closed set under fusion, since pointlike excitations must fuse to pointlike excitations. Also, this set must contain the vacuum. In particular, this means that all lifts of the vacuum must have trivial spin. This is of course a criterion that is intimately related to the nature of the condensate; we can impose it already at the symmetry breaking stage, or even view it as part of the definition of a \lq bosonic' condensate. Finally, we can go so far as to require that there is a unitary braided tensor category describing the fusion and spins of the set of unconfined excitations. Proving such a thing is beyond the scope of this paper, but again it does turn out to be true in all our examples. Also, we would like to stress once more that it is likely that if such a braided tensor category exists, it will actually be fixed uniquely by the fusion and spins of the unconfined sectors.

{}From our assignment of spin factors, we may derive the monodromy of the non-confined particles using the ribbon equation (assuming that the set of non-confined particles closes under fusion). The resulting monodromy is just the same as the monodromy of the lifts of the particles. More specifically, let $a$, $b$ and $c$ be sectors of the broken theory which are not confined and let $c\in a\times b$. Also, let $a^{i}$, $b^{j}$ and $c^{k}$ be arbitrary lifts of $a$, $b$ and $c$ with the property that $c^{k}\in a^{i}\times b^{j}$. Then the monodromy of $a^{i}$ and $b^{j}$ in the fusion channel $c^{k}$ is given by the combination of spin factors  $\theta_{c^{k}}/(\theta_{a^{i}}\theta_{b^{j}})$, but since the spin factors of the lifts of $a$, $b$ and $c$ are all equal, this factor does not actually depend on the choice of lifts $a^{i}$, $b^{j}$ and $c^{k}$ and we may as well write $\theta_{c}/(\theta_{a}\theta_{b})$, which is the monodromy of $a$ and $b$ in the fusion channel $c$. An important special case of this argument is the case $b=1$. In this case we are looking at the monodromy of the lifts of $a$ with the lifts of the vacuum, which are of course the condensed sectors. The argument we just gave now says precisely that the lifts of the non-confined particle $a$ have trivial braiding with the condensed particles, so we have managed to give a more precise meaning to the intuition about confinement that we mentioned at the start of this section.

We could in fact turn the whole argument above around and start by requiring that all lifts of  a non-confined sector should have trivial monodromy with lifts of the vacuum sector (i.e.~with condensed sectors). From that assumption we can get back to the confinement criterion given above if two conditions are satisfied. First of all, all lifts of the vacuum of the broken theory should have trivial spin (we also required this before) and secondly, it must be possible to obtain all lifts of a sector of the broken theory by fusion with lifts of the vacuum.
As we remarked, the first requirement is intimately connected with the nature of the condensate and with the question what exactly constitutes a boson in $2+1$ dimensions. If the second requirement does not hold, then it would seem that we have identified sectors which should be distinguishable. It is not clear to us at this point whether these two requirements follow from the conditions on the condensate and on the symmetry breaking scheme that we had imposed already, though they do in all our examples. In any case, assuming that these two requirements do hold, we can regain our previous confinement criterion. Any lift of the fusion channel $b\times 1=b$ is of the form $b^{i}\times 1^{j}=b^{k}$ and the monodromy factor in this lift is $\theta_{b^{i}}/(\theta_{1^{j}}\theta_{b^{k}})$. Now using the two conditions above, we see that $\theta_{1^j}=1$ for all $j$ and also, for any $j$, the possible $b^k$ run through all lifts of $b$. Thus for all these monodromy factors to be equal to $1$ is equivalent to $\theta_{b^i}=\theta_{b^k}$ for all $i$ and $k$. In other words, excitations in the $b$ factor are not confined precisely when all lifts of $b$ have the same spin factor and we are back at our original criterion.

\subsubsection{Back to the $\mathbf{SU(2)_4}$ example}

Applying the confinement criterion to our $SU(2)_4$ example, we see that the sectors labeled $0$, $2_1$ and $2_2$ are not confined. For $2_1$ and $2_2$ this is immediate, since they have a unique lift, and $0$ lifts to either $0$ or $4$, both of which have spin factor $1$. The sector with label $1$ is confined, since $1$ lifts to $1$ and $3$, which have different spin factors ($e^{i\pi/4}$ and $-e^{i\pi/4}$ respectively). This gives the results in table~\ref{tab:su24nonconf}.
\begin{table}[hbt]
\begin{center}
$\begin{array}{|l|}
\hline SU(2)_4 {\rm ~after~confinement}\\
\hline
\hline
\begin{array}{l|l}
0&\theta_0=1\\
1&{\rm confined}\\
2_1&\theta_{2_1}=e^{2\pi i /3}\\
2_2&\theta_{2_2}=e^{2\pi i /3}\\
\end{array}\\
\hline
\hline
\begin{array}{lll}
2_1\times 2_1=2_2~&& \\
2_1\times 2_2=0&2_2\times 2_2=2_1~& \\
\end{array}\\
\hline
\end{array}$
\end{center}
\caption{\footnotesize Spin factors, and nontrivial fusion rules for the non-confined sector of $SU(2)_4$ with a condensate in the $4$-sector.}
\label{tab:su24nonconf}
\end{table}
This result actually fixes the topological order of the non-confined sector of the broken theory uniquely, since there is only one solution to the consistency conditions for topological field theories (notably the pentagon and hexagon equations\cite{MS}) which has these fusion rules and spin factors\cite{BondersonWIP,Rowell2007}.

\subsection{Classification of strings}

We have given a description of the spectrum of topological excitations in a theory which has undergone a condensation transition. We have seen that the broken theory has excitations which are pointlike as well as confined excitations which pull strings. These confined excitations can exist as boundary excitations, when their strings are attached to a phase boundary, or as `hadronic' composites, when two or more confined excitations form a cluster whose overall topological charge is not confined. In such clusters, the confined particles are joined together by their strings. It is interesting to try and characterize the different types of string themselves in some non-redundant way. A redundant labeling is given by the set of labels of confined particles. Many confined particles will likely pull the same type of string in the condensate vacuum, since each confined particle can be fused with any non-confined particle to give some other confined particle, and this should not change the type of string that occurs. Therefore, we propose to label the different types of string by equivalence classes of confined sectors modulo fusion with excitations from non-confined sectors. To be more precise, let us introduce an equivalence relation on the sectors of the broken theory as follows,
\begin{equation}
\begin{array}{rcl}
a\sim b &\Leftrightarrow & \exists \; c, c' {\rm ~not ~confined,}\\
~&~&{\rm ~~~~such~that~}(b\in a\times c)\wedge(a\in b\times c')
\end{array} 
\end{equation}
We clearly have $a\sim a$, just take $c$ and $c'$ trivial. Also $a\sim b\Leftrightarrow b\sim a$, since the definition of the relation is symmetric. Finally if $a\sim b$ and $b\sim c$ then $a\sim c$. To see this, note that from $a\sim b$, we have unconfined sectors $d,d'$ with $a\in b\times d$ and $b\in a\times d'$. Similarly, from $b\sim c$, we have unconfined sectors $e,e'$ with $b\in c\times e$ and $c\in b\times e'$. Hence, we have $a\in c\times e\times d$ and $c\in a\times d'\times e'$ and since $d$, $e$, $d'$ and $e'$ are not confined, neither are $e\times d$ or $d'\times e'$, so $a\sim c$ and we have a good equivalence relation. 

The different types of string should be uniquely labeled by the equivalence classes, which are some sort of `orbits' under fusion with non-confined excitations. As a check, we note that all non-confined representations are equivalent to each other, which is what we want, since they all correspond to the situation with no string.  To see this note that if $a$ and $b$ are not confined, then neither are $\bar{a}\times b$ and $\bar{b}\times a$. But $b\in a\times (\bar{a}\times b)$ and $a \in b \times (\bar{b}\times a)$, so indeed $a\sim b$.

For our $SU(2)_4$ example the classification of strings is rather trivial, since there is only one type of confined particle. Hence there are just two classes under the equivalence above, the class consisting of the confined particle $1$, which pulls a string and the class consisting of the unconfined particles $0$, $2_1$ and $2_2$, which pull no string. 
%We will see more interesting examples later on.

\subsection{Summary and comparison to our earlier approach}

%{\em Begin with some general statements that can be compared with the corresponding CFT statements
%Explain briefly what the advantages of this approach are (non-integer quantum dimensions included etc., that the old approach is included as a special case and that the old approach gives some more info that we are not going into here.} 

In the previous sections we have given general principles for the treatment of the phenomenon of \lq breaking' a quantum symmetry $\mathcal{A}$ through the formation of a boson condensate,  as well as a detailed example. The analysis proceeds in three stages:\\
1. \textit{Criteria for a condensate}. We formulated some criteria that have to be satisfied for a \lq field' $c$ to be a \lq boson', and to serve as a possible candidate to form a condensate.  Two necessary conditions are that it should have trivial spin factor ($\theta_c=1$) and partially trivial self monodromy, i.e.~there is at least one fusion channel $f\in c\times c$ with $\theta_f=1$.  \\
2. \textit{Consistent branching}. Our analysis is based on the construction of a set of branching rules giving the decomposition of the topological sectors of the unbroken theory into sectors of the broken theory. We can think of this as branching representations of the quantum group $\mathcal{A}$ describing the unbroken phase into representations of an intermediate algebra $\mathcal{T}$ (whose structure is not discussed a priori). There are a number of consistency conditions on these branching rules that have to be met and these in fact appear to determine the possible branchings uniquely. A crucial condition is that branching commutes with fusion. This implies in particular that the total quantum dimension is preserved under the branching rule and that the old vacuum branches into the new. We furthermore require that the condensate has the new vacuum in its branching.\\
3. \textit{Confinement}. We observe that we can determine which representations in the broken phase have a nontrivial braiding with the condensate, and it is clear that these will pull a string in the new vacuum and hence are confined. The effective topological low energy theory is then described by the fusion and braiding rules of the non-confined representations (these must form a closed fusion ring) and these can then presumably be identified as the irreducible representations of some quantum group $\mathcal{U}$.

To our knowledge applying these conditions and performing these steps determines the breaking pattern uniquely. In previous papers on this subject we restricted our attention mostly to theories described by finite dimensional quasitriangular Hopf algebras, especially the so-called discrete gauge theories, whose hidden symmetry corresponds to the quantum double of the discrete gauge group. In those cases the analysis of the breaking phenomena was done by explicitly considering the algebraic structure of  $\mathcal{T}$ and $\mathcal{U}$. However following that route directly in case one is dealing with representations that carry non-integer quantum dimensions is problematic, and that is why in this paper our analysis is based on the \lq dual' route, directly studying the breaking pattern on the level of topological sectors (\lq representations') and their branching rules. 

Indeed, in the explicit example we treated in the previous section of this paper we showed that the present approach, using the branching rules directly, can also be applied to topological systems which have sectors carrying non integer quantum dimensions, for example the systems described by the quantum groups based on quantum deformations of semisimple Lie algebras which  show up in relation to conformal field theories of the Wess-Zumino-Witten (WZW) type. This is an important extension of the possible applications of the breaking mechanism which will allow applications in physical contexts like the fractional quantum Hall effect.

\section{General features of the condensation transition}
\label{generalmethodsec}
\subsection{Simple current condensates}
\label{sec:simplecurrents}
Let us uncover some features of the condensate transition that are mostly independent of the details of the topological phase we start with. One can get surprisingly far with this if the condensed sector is a {\em simple current}. A simple current in CFT is a primary field $J$ whose fusion rules are such that the fusion of $J$ with any other field contains only one channel, i.e.~for any primary field $\phi$, the fusion rule $J\times \phi$ has only one primary field on the right hand side. We will use the analogous definition in the context of TQFT. It is easy to see that a topological sector labeled $J$ is a simple current precisely if $d_J=1$. 
First of all, since there are only finitely many sectors, we must have $J^{\times p}=1$ for some $p$. We call this integer $p$ the order of $J$ and denote it $|J|$. Now using formula (\ref{dimtens}) repeatedly and noting that $d_1=1$ (which also follows from formula (\ref{dimtens})), we see that $(d_J)^{p}=1$. But since $d_J$ is real and positive, it follows that $d_J=1$. For the converse, let us assume that $d_J=1$. Then it is immediate that all fusion powers of $J$ are dimension $1$ sectors and there will be some $J$ for which $J^{\times p}=1$. Now if there would be some sector $b$ for which the fusion $J\times b$ has multiple channels, then this would imply that $J^{p}\times b$ also has multiple channels, or multiplicities greater than $1$. However since $J^{p}=1$, this is a contradiction and hence $J$ is a simple current.

If a bosonic simple current $J$ condenses, we can immediately see that the restrictions of a number of fields of the original theory will be identified. First of all, the fusion powers of $J$ all branch to the vacuum. More generally, for any sector $a$, there is an orbit of $a$ under the action of fusion with powers of $J$ and the restrictions of the fields $J^{\times l}\times a$ in this orbit are all identified. 

If the orbits are all of the maximal size, $|J|$, then these identifications lead directly to a new fusion theory, without any further identifications or splittings. The $J$-orbits of the old theory correspond to the excitations of the condensed theory. The lifts of the new vacuum sector are precisely the sectors $1,J,\ldots, J^{|J|-1}$ of the old theory and using that $\theta_J=1$, we find that $\theta_{J^{l}}=1$ for all $l$, so all lifts of the vacuum have trivial spin. more generally, the non-confined excitations of the new medium are precisely those $J$-orbits for which the spin factor is the same for all particles in the orbit. 

If there are $J$-orbits of less than maximal size, the sectors in these orbits will split. To see this, let $a$ be sector in a non-maximal $J$-orbit and let $p$ be the smallest integer for which $J^{p}\times a=a$ (note that $p$ must divide $|J|$). Then we have 
\begin{equation}
a\times \bar{a}=1+\ldots=(J^{p}\times a)\times \bar{a}= J^{p}\times (a\times \bar{a})=J^{p}+\ldots
\end{equation}
and so $a\times\bar{a}=1+J^{p}+\ldots$. An analogous argument shows that we must have $a\times\bar{a}=1+J^{p}+\ldots+J^{|J|-p}+\ldots$ and so the restriction of $a\times \bar{a}$ contains at least $\frac{|J|}{p}$ copies of the new vacuum sector, which implies that $a$ (and $\bar{a}$) must split. If there are no multiplicities $n_{a}^{i}$ greater than $1$ in the restriction of $a$, then it must in fact split into at least $\frac{|J|}{p}$ parts, but there may be extreme cases where $a$ restricts to $\sqrt{|J|/p}$ copies of the same sector of the new theory.
% This footnote is almost certainly wrong (need to find a counterexample).
% \footnote{The above considerations actually lead to a small conjecture on the structure of TQFTs, or ribbon tensor categories. Let $a$ be an object in a ribbon tensor category, let $J$ be a bosonic simple current in this category and let $s$ be the smallest integer such that $J^{\times s}\times a=a$, then we conjecture that the quantum dimension of $a$ must be greater than or equal to $s$. Note that the argument in the main text already shows that $d_s\ge\sqrt{s}$, but the more stringent bound has to be true if $a$ is able to split into $s$ parts, each of which must have quantum dimension greater than or equal to $1$.}

%This may well be true:
%It is natural to conjecture that no further splitting is in fact ever necessary. 
To obtain the fusion rules for the parts of the split sectors, new input about the theory is necessary; we have examples where two parts obtained in this way are dual to each other (like $2_1$ and $2_2$ in the broken $SU(2)_4$ theory) as well as examples where they are self dual (see for instance the discussion of $SU(2)_8$ in section~\ref{su28sec}). 

\subsection{More general condensates}
When the condensed sector is not a simple current it becomes much more difficult to say anything general about the symmetry broken and confined theories.  In this case the sector $c$ that condenses branches to a number of copies of the vacuum and possibly to other sectors, i.e.~$c\rightarrow n^{c}_{0}1+\sum_{i\neq 0}n^{c}_{i}c_{i}$, where we have chosen $c_0=1$. Here $n^{c}_{0}\ge 1$ and some of the $n^{c}_{i}$ with $i\neq 0$ may be greater than zero. If $d_{c}$ is not an integer, then there have to be such non-vacuum components of the restriction of $c$, in order to preserve the quantum dimension. In fact, some of the restrictions of $c$ may be confined. One may heuristically interpret this \lq partial condensation' by thinking of particles in the topological sector labeled by $c$ as having a hidden internal Hilbert space of dimension $d_c$ and condensing in a particular state in this internal space. The condensed state (and possibly some other states in the internal space) will then branch to the vacuum, but other internal states will not and may even be confined. In our earlier work, where we restricted ourselves to a class of theories with integer quantum dimensions, this interpretation could be made completely rigorous. However, in the present context, this seems more difficult. It is likely better to think of the number $n_{c}^{0}$ as a measure for the number of states in a system of $N$ identical particles of type $c$ which would be indistinguishable from the vacuum in the condensed phase (one would expect this number to grow as $(n^{0}_{c})^{N}$). 
%Possibly the issue might be resolved by going to a framework where the topological sectors are labeled by irreducible representation of some weak quantum group, like the $q$-deformed universal enveloping algebra $U_{q}(g)$ of a Lie algebra $g$, with $q$ being a root of unity and truncated tensor product. However, this does not seem straightforward and is beyond the scope of this paper.

With the condensate $c$ not a simple current, there will still be identifications, but they are more difficult to obtain. In general, the fusion rules $c\times a=\sum_{b}N^{b}_{ca}b$ just tell us that the components of the restriction of $a$ are identified with some of the components of the restrictions of the sectors $b$ appearing on the right. Similarly, if a fusion $a\times b$ contains the condensed sector $c$, then this tells us that some component of the restriction of $a$ must be identified with a component of the restriction of $\bar{b}$. To get more information, we need to use the requirement that the broken theory is once again a good fusion theory. 

It is possible to make some general predictions on splitting, though not as strong as the ones for simple currents. If, for some sector $a$, the fusion $a\times\bar{a}$ contains $n^{c}_{a\bar{a}}\ge 1$ copies of $c$, then the restriction of this sector must split in order to produce the at least $n^{c}_{a\bar{a}}+1$ copies of the new vacuum in the restriction of $a\times\bar{a}$.  Again, if the components of the restriction are all distinct then there must be at least $n^{c}_{a\bar{a}}+1$ of them, but in extreme cases, we may have just one component with multiplicity $\sqrt{n^{c}_{a\bar{a}}+1}$.  Note that if $c$ is a simple current, then $n^{c}_{a\bar{a}}\le 1$, since if $n^{c}_{a\bar{a}}\ge 2$, then we would have $n^{1}_{c^{|c|-1}\times a,\bar{a}}\ge 2$, contradicting the axiom of fusion that says that sectors can only fuse the the vacuum in a unique way. Also, in the simple current case, one may see easily that the splittings deduced here are a special case of the splittings of sectors in non-minimal orbits discussed before. Similar arguments to the above show that for any pair of sectors $a$, $b$ for which $N^{c}_{ab}\ge 2$, at least one of $a$ and $b$ must have a restriction which splits.

Another case in which splitting of a sectors $a$ occurs for arbitrary condensates $c$ is if the fusion $c\times a$ contains $a$ and no other sectors whose quantum dimension is greater than or equal to that of $a$. Of course if $c$ is a simple current this just says that $a$ is a fixed point. In the general case the restriction of the fusion $c\times a\times \bar{a}$ must contain at least $n^{a}_{ca}+1$ copies of the new vacuum sector, one from the $c$ in $c\times(a\times\bar{a})$ and $n^{a}_{ca}$ from the copies of $a\times\bar{a}$ in $(c\times a)\times \bar{a}$. Now if $a$ (and $\bar{a}$) do not split, we see that there must be at least $n^{a}_{ca}+1$ copies of the restriction of $a$ in the restriction of the fusion of $c\times a$. However, none of the components of the restrictions of other fields in $c\times a$ can be identified with the restriction of $a$, since the quantum dimensions of these other fields are smaller than $d_a$ by assumption. Hence, it follows that $a$ must split.

\subsection{Observations on $c$ and $\MD$}
\label{sec:observations}
From the examples we have calculated, we observe that the central charges and total quantum dimensions of anyon models seem to follow certain general rules under condensation, if the anyon model that one starts with is \emph{modular}. Modularity is equivalent to the requirement that the monodromy is nondegenerate, that is, for every topological sector $a$ there is at least one topological sector $b$ such that $a$ and $b$ have nontrivial monodromy (see Ref.~\onlinecite{Kitaev06a}, section E.5). Another useful characterization of modularity is that the S-matrix of the theory must be unitary. This requirement is satisfied for many models that have been studied in physics, for example for all models coming from conformal field theories with bosonic chiral algebras. However, there are examples where modularity does not apply, notably in systems with excitations which behave like the vacuum under monodromy but which have nontrivial, necessarily fermionic, exchange behavior. Typical examples of such excitations are the actual electrons in quantum Hall systems. 

Given modularity, we observe that
\begin{itemize}
\label{cDobservations}
\item The topological central charges of the unbroken theory and the unconfined theory are equal
\item Denoting the total quantum dimensions of the original, broken and unconfined theories by $\MD_{\MA}$, $\MD_{\MT}$ and $\MD_{\MU}$, we have 
$\frac{\MD_{\MA}}{\MD_{\MT}}=\frac{\MD_{\MT}}{\MD_{\MU}}$. 
\end{itemize}
We also note that generally (independently of modularity) we have $\MD_{\MA}\ge\MD_{\MT}\ge\MD_{\MU}$.
In the remainder of this paper we will study connections between the quantum group symmetry breaking scheme we have described so far and constructions in conformal field theory. We will argue that quantum group symmetry breaking in CFT is dual to conformal extension of the chiral algebra. This should also clarify the observation that central charge is conserved.  It appears more difficult to get an intuition for the identity between total quantum dimensions from the CFT side.

\section{Quantum Group breaking vs.~Conformal Extensions}

In an idealized system with topological order, topological quantum numbers cannot be changed by the application of any local operator. In other words, topological observables are conserved quantities which commute with the full algebra of local observables. Of course in realistic systems, the situation is often more complicated than this. First of all, the topological features are often emergent only at low energies and different \lq topological sectors' of the Hilbert space may be mixed by high energy (virtual) excitations. Secondly, any real system has a finite size, which implies that a product of finitely many local operators can actually become a \lq topologically nontrivial' operator and relate states with different topological quantum numbers. In other words, there is no clean separation between \lq local' and \lq topological' observables. However, in the setting of conformal field theory, such a separation does exist. Here, the role of the local algebra is played by the chiral algebra, which can be the Virasoro algebra, or some more complicated algebra like a Kac-Moody algebra or W-algebra. The Hilbert space of the theory splits into sectors on which this chiral algebra acts, but which are not mixed with each other by this action. These chiral sectors correspond to the topological sectors of the CFT. Hence it is natural to expect that there should be a TQFT, or modular tensor category, which describes the fusion and exchange interactions of states from the different sectors. It is also natural to introduce operators for topological charges which commute with the full chiral algebra and one may in fact hope to find a quantum group whose representation category is precisely the modular tensor category that describes the system's topological interactions and which has an action on the Hilbert space of the CFT that commutes with the action of the chiral algebra. Operators for topological charges can then be the analogs of Casimir operators for this quantum group. While this picture of a chiral algebra and a quantum group with commuting actions seems to be part of the lore of CFT, there does not appear to be a detailed mathematical understanding of this picture for general CFTs. We will nevertheless attempt to flesh it out  for Wess-Zumino-Witten models in the remainder of this section, and note in advance that the picture just sketched does provide useful intuition about the connection between quantum group symmetry breaking and some well known constructions in CFT. 

Since the quantum group and the chiral algebra are morally each other's commutants, we expect that breaking down quantum group symmetry from a large quantum group to a smaller one should be accompanied by an extension of the chiral algebra on the \lq local' side of things. In fact, we find that there is a beautiful connection between quantum group symmetry breaking and conformal extension of the chiral algebra. In conformal extensions, we start with a chiral algebra that has a representation which is bosonic, but topologically nontrivial. Then we enlarge the algebra by adding an intertwining operator between the vacuum representation and this nontrivial bosonic representation (i.e.~a creation operator for a topologically nontrivial particle). Before the conformal extension, the theory would have a topological (or more precisely, chiral) sector corresponding to this bosonic field, but afterwards, this sector has become part of the vacuum sector of the new chiral algebra. On the quantum group side of the story we can interpret this merger of a bosonic sector with the vacuum sector as condensation of the bosonic particle and describe its effects using the formalism proposed here.  In retrospect, this intuitive argument explains the similarity of some of our constructions and criteria to those mentioned for conformal extensions in Moore and Seiberg's famous work on the classification of rational CFT's\cite{mszoo}. Since many of the common constructions of CFTs, most notably the coset construction\cite{Goddard1984}, can be described in terms of conformal extension of the chiral algebra, these constructions now obtain a physical interpretation as being due to condensation of bosonic quasiparticles. In fact, this suggests that some CFT constructions will have a direct physical realization in phase transitions which occur in systems described by CFTs.

After this somewhat abstract discussion let us turn to WZW models.
In these models\cite{Witten1984,Gepner1986} the physical states  are organized into integrable representations of an extended conformal symmetry, a Kac-Moody algebra based on a finite dimensional Lie algebra $G$, at a certain level $k$, which we will denoted as $G_k$. These representations  correspond to chiral primary fields and there is a finite number of them. This theory has a central charge equal to 
\begin{equation}
c(G,k) =\frac{k \,{\rm dim}\,G}{k+h} \;\; ,
\end{equation}
with $h$ is the dual Coxeter number of G.
The chiral primary fields are operators which create the lowest energy states of the different topological sectors of the theory from the vacuum and one may obtain the fusion rules and braiding of the topological sectors directly from the CFT by calculating the correlators, or more precisely the conformal blocks, of these chiral primary fields. There is a one to one correspondence between chiral primary fields in the WZW model and irreducible representations of the quantum group $U_q(G)$, where $q=e^{\frac{2\pi i}{k+h}}$ and in fact, it is known\cite{tsuka87,tsuka88} that the fusion and braid relation obtained in this way are exactly the fusion and braiding of these quantum group representations. 
%This relation can also be understood from the point of view of Topological Quantum Field theory in particular Chern Simons theory and conformal field theory \cite{Witten1989}. ??
Explicit representations of the quantum group in terms of operators acting on the Hilbert space of the theory can also be obtained, within the Coulomb gas formalism\cite{bmccp91,gomsier90,gomsier91}. 
%As sketched earlier in this section, the Affine algebra moves you around in the chiral representations, while the quantum group tells us how the these representations combine under fusion. 
All of this goes a long way toward establishing the picture that we sketched earlier in this section, of a chiral algebra and a quantum group normalizing each other. The relation between WZW theory and quantum groups is quite useful even as a calculational tool, because
many properties of multi(quasi)particle states in conformal field theory, such as their braiding properties, can be determined by just using the properties of the quantum group (see e.g. Ref.~\onlinecite{sliba}). 
%The reason is that the action of the braid group on such states commutes with the quantum group action, which implies that n-particle states can be decomposed into product representations of $B_n\otimes G_k$.   
%One may think of the quantum group operations as the normalizer of the chiral algebra in the operator algebra of the theory. And this perception naturally suggests that  by breaking the quantum symmetry to some smaller symmetry, the chiral symmetry of the broken topological phase may be enlarged!   

In the remainder of the paper we will pursue the relation between the breaking mechanism and CFT in some detail, exhibiting the connections between the two formalisms in explicit examples, most of them based on WZW theories. We will find that the breaking of quantum symmetries is indeed closely related to conformal extensions and we will show how well known constructions like the coset construction\cite{Goddard1984}, conformal embeddings\cite{Bais1987,Schellekens1986} and orbifolding\cite{Dixon1985,Dixon1986a,Dixon1986b}  
%and the construction based on simple currents \cite{schellekens} 
acquire a direct physical relevance and interpretation in the present context.

\section{Conformal embeddings}
\label{sec:confemb}
%Explain embeddings very briefly, with $SU(2)_4 \subset SU(3)_1$ example and comparison back. Also indicate that the \lq\lq identification currents'' don't have to be simple currents and give an example (maybe use $SU(5)_1$ here?)

Conformal embeddings are embeddings of affine Lie algebras $H_{k'}\subset G_{k}$ with the property that the central charges are equal,
\begin{equation}
c(G,k)=c(H,k').
\end{equation}
As a result, the  corresponding cosets $G_{k}/H_{k'}$ have central charge equal to zero and are therefore trivial. We will have more to say about this in section~\ref{sec:cosets}. Here we will focus on the embeddings themselves. General (non-conformal) embeddings of affine Lie algebras do not conserve the central charge, but for all embeddings, the levels $k$ and $k'$ are related by the Dynkin index $l$ of the corresponding embedding of $H$ into $G$, one has $k'=lk$. The conformal embeddings of affine Lie algebras have been classified in Refs.~\onlinecite{Bais1987,Schellekens1986}; they form a number of infinite series and a finite list of special cases. In these papers it is proved that for a conformal embedding the level of ${G_k}$ is always unity: $k=1$. Conformal embeddings also have the remarkable property that the (infinite dimensional) highest weight representations of ${G_k}$ branch to finitely many highest weight representations of $H_{k'}$ \cite{Bais1986}.  

Let us return to the breaking of $SU(2)_4$ that we studied in section~\ref{sec:qbreak} and let us show how it is connected to the well known conformal embedding of  ${SU}(2)_4$ in ${SU}(3)_1$; indeed conformal because  both have central charge $c=2$. In table \ref{tab:su3_1_and_so5_1} we give the representations and fusion algebra of $SU(3)_1$.
\begin{table}[hbt]
\begin{center}
$
\begin{array}{cc}
\begin{array}[t]{|l|}
\hline SU(3)_1 \\
\hline
\hline
\begin{array}{l|ll}
1&d_1=1&h_0=0\\
3&d_3=1&h_3=\frac{1}{3}\\
\bar{3}& d_{\bar{3}} =1&h_3=\frac{1}{3}\\
\end{array}\\
\hline
\hline
\begin{array}{llll}
3\times 3=\bar{3}~&&& \\
3\times \bar{3}=1&\bar{3}\times \bar{3}=3~&& \\
\end{array}\\
\hline
\end{array}
&
\begin{array}[t]{|l|}
\hline SO(5)_1 \\
\hline
\hline
\begin{array}{l|ll}
1&d_1=1&h_0=0\\
4&d_4=\sqrt{2}&h_4=\frac{5}{16}\\
5& d_{5} =1&h_5=\frac{1}{2}\\
\end{array}\\
\hline
\hline
\begin{array}{llll}
4\times 4= 1+5~&&& \\
4\times 5=4& 5\times 5=1~&& \\
\end{array}\\
\hline
\end{array}
\end{array}$
\end{center}
\caption{\footnotesize  Spins, quantum dimensions and nontrivial fusion rules for $SU(3)_1$ and $SO(5)_1$.}
\label{tab:su3_1_and_so5_1}
\end{table}
From the embedding we obtain the branching of the corresponding Kac-Moody representations:
\begin{equation}
\begin{array}{rcl}
 1 &\rightarrow &0+4   \\
 3 &\rightarrow &2   \\
  \bar{3}& \rightarrow  &2   
\end{array}
\label{eq:branchingsu31}
\end{equation}
Indeed, this finite branching is possible because the $0$ and $4$ representations of
${SU}(2)_4$ are degenerate, in the sense that their conformal weights differ by an integer. So one way to understand the conformal embeddings is to say that because the $H_{k'}$ representations are degenerate there is a larger symmetry realized in the spectrum, i.e.~${G_k}$. Interestingly, the (bosonic) singlet module of ${G_k}$ decomposes into bosonic representations under $H_{k'}$, as we see in the first line of (\ref{eq:branchingsu31}) and if we now return to our analysis of section III we see that it is exactly the nontrivial bosonic component in the branching (i.e.~the 4 of ${SU}(2)_4$) that is assumed to form the condensate. On the other hand we make the remarkable observation that after breaking and subsequent confinement the residual symmetry $\mathcal{U}$, i.e.~the  representations and their fusion rules as given in table \ref{tab:su24nonconf} are precisely those of ${SU}(3)_1$! The conclusion is that there is a unique correspondence between the  conformal embedding  
${H}_{k'} \subseteq {G}_k$ and the breaking of the quantum group for $H_{k'} \rightarrow G_k$, where it should be noted that on the side of the chiral algebras, the embedded algebra is \lq smaller', while on the quantum group side the residual quantum group corresponding to $G_{k}$ is \lq smaller' than the one for the embedded alegebra $H_{k'}$. All this is in good agreement with our intuition that the fusion algebra is somehow the normalizer of the chiral algebra in the operator product algebra of the CFT. In fact, we can think of the chiral algebra of $SU(3)_1$ as an extension of the chiral algebra of $SU(2)_4$ by the intertwining operator between the vacuum sector of the $SU(2)_4$ theory and the sector labeled by $\Lambda=4$. The breaking of quantum symmetries is thus related to enlarging the conformal symmetry, and the construction of new conformal models with larger chiral symmetries, starting with models related to Kac--Moody algebras, has in the present context acquired a very direct physical meaning and relevance, namely the formation of a bosonic condensate in the phase with the smaller chiral symmetry.  

It is instructive to discuss  another example of a conformal embedding, where applying the breaking formalism is less straightforward. Let us consider the conformal embedding ${SO}(5)_1 \supseteq {SU}(2)_{10}$ both with $c=5/2$. We have listed the quantum dimensions, spins and fusion of the $SO(5)_1$ theory in table~\ref{tab:su3_1_and_so5_1}. The spins and quantum dimensions for $SU(2)_{10}$ are given in table~\ref{tab:su2_10}. 
\begin{table}[hbt]
$
\begin{array}{|l|}
\hline SU(2)_{10}\\
\hline
\hline
\begin{array}{l|ll}
0&d_0=1&h_0=0\\
1&d_1=\sqrt{2+\sqrt{3}}&h_1=\frac{1}{16}\\
2&d_2=1+\sqrt{3}&h_2=\frac{1}{6}\\
3&d_3=\sqrt{2}+\sqrt{2+\sqrt{3}}&h_3=\frac{5}{16}\\
4&d_4=2+\sqrt{3}&h_4=\frac{1}{2} \\
5&d_5=2\sqrt{2+\sqrt{3}}& h_5=\frac{35}{48}\\
6&d_6=2+\sqrt{3}&h_6=1\\
7&d_7=\sqrt{2}+\sqrt{2+\sqrt{3}}&h_7=\frac{21}{16}\\
8&d_8=1+\sqrt{3}&h_8=\frac{5}{3}\\
9&d_9 =\sqrt{2+\sqrt{3}}&h_9=\frac{33}{16}\\
10&d_{10}=1&h_{10}=\frac{5}{2}\\
\end{array}\\
\hline
\end{array}
$
\caption{\footnotesize  Spins, quantum dimensions for $SU(2)_{10}$.}
\label{tab:su2_10}
\end{table}

\noindent The fusion rules for $SU(2)_k$, and in particular for $SU(2)_{10}$, are given by 
\begin{equation}
\label{su2_k_fusion}
\Lambda_{1}\times\Lambda_{2}=\sum_{\Lambda=|\Lambda_1-\Lambda_2|}^{\min\,\{\Lambda_{1}+\Lambda_2, \,2k-\Lambda_1-\Lambda_2 \}}\Lambda,
\end{equation}
where the sum runs over those $\Lambda$ in the indicated range for which $\Lambda_1+\Lambda_2-\Lambda$ is even (i.e.~$\Lambda$ is incremented by $2$).

Let us now consider the breaking mechanism. The $6$ representation is the only nontrivial bosonic representation, and it  has a trivial self-braiding channel because the fusion product with itself contains the identity representation. We see that it has a quantum dimension
$d_6=2+\sqrt{3}$ which tells us that we have to split the representation $6\rightarrow 6_1+6_2$ where  we assume $6_1$ to have unit quantum dimension and to be the component that condenses (indeed: $6_1 \times 6_1 = 0$) while we consider the $6_2$ component with quantum dimension $1+\sqrt{3}$ for the moment as an independent  field in the broken phase. 

Starting with this splitting of the $6$ and using the fusion rules in a similar fashion as we did in section III we see that also other representations have to split and furthermore other identifications have to be made. The net result of this straightforward analysis is given in table \ref{tab:su210identifications}.
\begin{table}[hbt]
\begin{center}
$\begin{array}{|l|}
\hline SU(2)_{10} {\rm ~broken}\\
\hline
\hline
\begin{array}{l|l}
{\rm ~splittings} &{\rm ~ identifications} \\
3:= 3_1 + 3_2&  0 \leftrightarrow  6_1 \\
4:= 4_1 + 4_2 & 1 \leftrightarrow 5_1 \leftrightarrow 7_1 \\
5:= 5_1 + 5_2 & 2 \leftrightarrow 4_2 \leftrightarrow 6_2 \leftrightarrow 8 \\
6:= 6_1 + 6_2 & 3_1 \leftrightarrow 5_2 \leftrightarrow 9 \\
7:= 7_1 + 7_2 & 3_2 \leftrightarrow 7_2 \\
                       & 4_1 \leftrightarrow 10 \\
\end{array}\\
\hline
\end{array}$
\end{center}
\caption{\footnotesize  Splitting and identifications of representations after breaking by the $6_1$ condensate.}
\label{tab:su210identifications}
\end{table}
It is easy to see that the new representations have the following quantum dimensions:
$d_{3_1}=\sqrt{2+\sqrt{3}}, ~d_{3_2}=\sqrt{2}$ and $d_{4_1}=1$. 
%Splitting obviously does not affect the conformal weight.  
At this intermediate (broken) level we are left with five representations which have the fusion rules given in table \ref{tab:su210brokenfusions}:
\begin{table}[hbt]
\begin{center}
$\begin{array}{|l|}
\hline {\rm ~Fusion~rules~ in~ the~ broken~ phase~of~SU(2)_{10} } \\
\hline
\hline
\begin{array}{llll}
1 \times 1= 0 + 2 & &\\
1 \times 2= 1+ 3_1 + 3_2 & 2 \times 2= 0+2+2+4_1 & \\
1 \times 3_1= 2+4_1 & 2 \times 3_1= 1+ 3_1 + 3_2 &\\
1 \times 3_2= 2  & 2 \times 3_2= 1+ 3_1 &\\
1 \times 4_1= 3_1 & 2 \times 4_1= 2 & \\
&&\\
3_1 \times 3_1= 0 + 2 &  & \\
3_1 \times 3_2= 2 & 3_2 \times 3_2= 0+ 4_1 & \\
3_1 \times 4_1= 1 & 3_2 \times 4_1= 3_2 & 4_1 \times 4_1= 0 \\ 
\end{array}\\
\hline
\end{array}$
\end{center}
\caption{\footnotesize  Fusion rules of the broken phase  with the $6_1$ condensate.}
\label{tab:su210brokenfusions}
\end{table}
These fusion rules together with the conformal weights of the parent representations in the unbroken phase now allow us to determine which representation will be confined in the broken phase. As we said before, representations will not be confined if all their lifts have equal spin factors, or equal spins up to integers. For example if we want to know whether $3_1$ will be confined, we have to check whether $h_a-h_b \in Z$ for all combinations  $a,b$ with $a$ and $b$ taken from the list of fields that restrict to $3_1$ according to table \ref{tab:su210identifications}, i.e.~$\{3,5,9\}$. Since these have conformal weights $5/16,~35/48$ and $33/16$ respectively, we conclude that the $3_1$ representation will be confined. For the $3_2$ which is identified with the $7_2$ we have parent conformal weights $5/16$ and $21/16$  so that that representation will {\em not } be confined. 
The upshot of this analysis is that only the $0$, the $3_2$ and the $4_1$ survive after confinement, of course with the fusion rules given in table \ref{tab:su210brokenfusions}.
We see that indeed our residual set of fields and their fusion and spin factors are isomorphic to the $SO(5)_1$ algebra under the map $0 \leftrightarrow 1,~ 3_2 \leftrightarrow 4$ and $4_1 \leftrightarrow 5$. Clearly the fusion algebra is also isomorphic to the Ising model or the $SU(2)_2$ model, but these have to be rejected because the conformal weights do not match. 

If we furthermore look at the branching rules  for the conformal embedding:
\begin{equation}
\begin{array}{rcl}
 1 &\rightarrow & 0+6  \\
 4 & \rightarrow & 3+7   \\
 5 &\rightarrow & 4+10  
\end{array}
\label{eq:so51branching}
\end{equation}
we confirm  that they are fully consistent with this correspondence. Note that in these rules we clearly have matching (modulo $\ZZ$) of the conformal weights. Representations can only branch into representations with the same conformal weights up to integers and hence the spin factors of the representations are preserved under the branching. The conclusion is that also in this more complicated situation we find that the quantum group $\mathcal{U}$, which appears after breaking ${H}_k$ by a bose condensate and subsequent confinement, is the expected quantum group $G_1$ appearing in the conformal embedding.
 %\begin{widetext}

\subsection{Finding new embeddings}
\label{su28sec}
 
Using the quantum group symmetry breaking formalism, we can now conjecture new conformal embeddings which are not contained in the classification of conformal embeddings of Refs.~\onlinecite{Bais1987,Schellekens1986}, for example because one or both of the theories involved in the embedding is not a WZW theory. We can start with an arbitrary TQFT or CFT which has a boson, condense the boson, find the theory describing the nonconfined excitations of the broken phase and then conjecture that the original theory can be conformally embedded in a CFT with the same topological order as the unconfined broken theory. 

As an example let us consider $SU(2)_8$. The quantum dimensions and spin factors for this theory are given in table~\ref{su2_8_spintab}. The fusion rules for $SU(2)_8$ follow from formula~(\ref{su2_k_fusion}).
\begin{table}[htb]
\begin{center}
$
\begin{array}[t]{|l|}
\hline SU(2)_{8}\\
\hline
\hline
\begin{array}{l|ll}
0&d_0=1&h_0=0\\
1&d_1=  \sqrt{\frac{5+\sqrt{5}}{2}}&h_1=\frac{3}{40}\\
2&d_2= \frac{3+\sqrt{5}}{2}&h_2=  \frac{1}{5}\\
3&d_3= \sqrt{5+2\sqrt{5}}&h_3=\frac{3}{8}\\
4&d_4= 1+\sqrt{5}&h_4=  \frac{3}{5}\\
5&d_5= \sqrt{5+2\sqrt{5}}& h_5= \frac{7}{8}\\
6&d_6= \frac{3+\sqrt{5}}{2}&h_6=\frac{6}{5} \\
7&d_7= \sqrt{\frac{5+\sqrt{5}}{2}}&h_7= \frac{63}{40}\\
8&d_8= 1&h_8= 2 \\
\end{array}\\
\hline
\end{array}
$
\end{center}
\caption{\footnotesize  Spins and quantum dimensions for $SU(3)_2$.}
\label{su2_8_spintab}
\end{table}

\noindent The $\Lambda=8$ representation  is the only bosonic field that meets the requirements  for a condensate. Analysis of the fusion rules after condensation of this field, using the methods of section~\ref{generalmethodsec} leads to identifications of the $\Lambda=p$ with the $\Lambda=8-p$ sectors for $p\in\{0,1,2,3\}$, while the $4$ has to split: $4:= 4_1 + 4_2$, in two parts which have equal quantum dimension. Without going through the details we summarize the branching and the fusion rules of the symmetry broken theory, in table~\ref{tab:SU(2)_8_broken}. 
\begin{table}[hbt]
\begin{center}
%{\scriptsize 
$\begin{array}{|c|}
\hline SU(2)_8 {\rm ~broken}\\
\hline
\hline
\begin{array}{l|l}
0,8\rightarrow 0&d_0=1\\
1,7\rightarrow 1&d_1=\sqrt{\frac{5+\sqrt{5}}{2}}\\
2,6\rightarrow 2&d_{2}=\frac{3+\sqrt{5}}{2}\\
3,5\rightarrow 3&d_{3}=\sqrt{5+2\sqrt{5}}\\
4\rightarrow 4_{1}+4_{2}&d_{4_1}=d_{4_2}=\frac{1+\sqrt{5}}{2}\\
\end{array}\\
\hline
\hline
\begin{array}{lll}
1\times 1=0+2 &\\
1\times 2=1+3&2\times 2=2+4_1+4_2\\
1\times 3=2+4_1+4_2& 2\times 3=1+1+3 \\
1\times 4_1=3& 2\times 4_1=2+4_2 \\
1\times 4_2=3& 2\times 4_2=2+4_1 \\
& \\
3\times 3=0+2+4_1+4_2& \\
3\times 4_1=1+3& 4_1\times 4_1=0+4_1\\
3\times 4_2=1+3& 4_1\times 4_2=2 \;\;\;\;\;\;\; 4_2\times 4_2=0+4_2 \\
\end{array}\\
\hline
\end{array}$
\end{center}
\caption{\footnotesize Branching rules, quantum dimensions and nontrivial fusion rules for $SU(2)_8$ after condensation in the $8$-sector (the fusion rules are symmetric).}
%}
\label{tab:SU(2)_8_broken}
\end{table}
%\end{widetext}

\noindent The $ 1$ and $3$ representations will become confined, so that we are left with four  fields: $0, 2, 4_1$ and $4_2$. 
We see that the fusion rules of these non-confined fields are just those of the direct product of two Fibonacci theories. The proper identification of the algebra $\mathcal{U}$ is in fact the quantum group $SU(3)_2/Z_3 \otimes SU(3)_2/Z_3$ with the identifications $0\Leftrightarrow (1,1)$, $4_1\Leftrightarrow (8,1)$, $4_2\Leftrightarrow (1,8)$, $2 \Leftrightarrow (8,8)$.  This quantum group has identical fusion rules, quantum dimensions  and conformal weights as the ones given in the tables. Alternatively, one might use $\overline{(G_2)_1\otimes (G_2)_1}$. All of this strongly suggests that there should be a conformal field theory with the same topological order as $\overline{(G_2)_1\otimes (G_2)_1}$ which has the property that the $SU(2)_8$ theory can be conformally embedded into it.

\subsection{Modular invariants}
%If we can do it quickly: try to understand some non-diagonal modular invariants as coming from our new embeddings, for example $SU(2)_{4n}$. We have $SU(2)_8$ breaks to $Fib\times Fib$ and $SU(2)_{12}$ breaks to $SU(3)_4/\ZZ_3$ (see Gepner).  

In the previous section, we have shown that the quantum group breaking allows us to conjecture many new conformal embeddings. Now we will show how these conformal embeddings can be used to generate non-diagonal modular invariants for certain conformal field theories. In fact for what follows, it will not be necessary to know the exact CFT describing the theory into which the embedding takes place (i.e.~the symmetry broken theory): it is enough to know the corresponding modular group representation, which is precisely what we get from the construction in the previous section.

In standard cases of conformal embeddings the branchings of representations can be used to construct the non-diagonal invariants for $H_{k'}$ using the standard diagonal modular invariant for ${G_k}$\cite{Bais1986,Bouwknegt1986}. For example for the simple  conformal embedding  ${SU}(2)_4$ in ${SU}(3)_1$ discussed in section \ref{sec:confemb} we had the branching rules (\ref{eq:branchingsu31}), and substituting these branchings into the modular invariant partition function for 
${SU}(3)_1$ :
\[ Z= |\chi_1|^2 + |\chi_3|^2 + |\chi_{\bar{3}}|^2, \]
yields the exceptional  ${SU}(2)_4$ invariant:
\[ Z=|\chi_0 + \chi_4|^2 + 2|\chi_3|^2. \] This is the lowest member of the socalled $A$ series series of non diagonal ${SU}(2)_k$ invariants with $k=4p\;\; (p\ge 1)$ in the classification of invariants by Cappelli, Itzykson and Zuber\cite{Cappelli1986,Cappelli1987}:
\[ Z = \sum_{n=0}^{p-1} |\chi_{2n} +\chi_{4p-2n}|^2 + 2 |\chi_{2p}|^2.    \]
These invariants follow from the breaking scheme of the quantum group $SU(2)_{4p}$ with a condensate in the highest i.e.~$\Lambda= 4p$ representation. This representation corresponds to a simple current under which the representations $\Lambda=q$ and $\Lambda=4p-q$ get identified, while the $\Lambda=2p$ representation has to split as $ 2p \rightarrow (2p)_1 + (2p)_2$, furthermore the odd representations with $q=2n-1$ get confined. This leaves us with a fusion algebra $\mathcal{U}$ of some CFT with the modular invariant partition function given above.
\begin{table}[hbt]
\begin{center}
$\begin{array}{cc}
\begin{array}[t]{|c|}\hline
{SU}(3)_2 \otimes {SU}(3)_2 \supseteq {SU(2)}_{8} \\
\hline
\hline
\begin{array}{ll}
(1,1) \rightarrow 0+8 & h_{1,1}=0\\
(8,1) \rightarrow 4_1 & h_{8,1}=\frac{3}{5}\\
(1,8) \rightarrow 4_2 & h_{1,8}=\frac{3}{5}\\
(8,8) \rightarrow 2+6 & h_{8,8}=\frac{6}{5}\\
\end{array} \\
\hline
\end{array} 
&
\begin{array}[t]{|c|}
\hline
{SU(3)}_4/\ZZ_3 \supseteq {SU(2)}_{12} \\
\hline
\hline
\begin{array}{ll}
1 \rightarrow 0+12 & h_0=0\\
8 \rightarrow 2+10 & h_8=\frac{3}{7}\\
10 \rightarrow 6_1& h_{10}=\frac{6}{7}\\
\bar{10} \rightarrow  6_2 & h_{\bar{10}}=\frac{6}{7}\\
27 \rightarrow 4+8 & h_{27} = \frac{8}{7}\\
\end{array}\\
\hline
\end{array}\\
\end{array}
$
\end{center}
\caption{\footnotesize  Branching of conformal representations and their spins, used to construct non-standard modular invariants.}
\label{tab:confbranchings}
\end{table}

\noindent Let us give some details for the cases $p=2$ and $3$.
The first case is the formation of a condensate in the $\Lambda=8$ representation in $SU(2)_8$. Here we have the identifications $0 \leftrightarrow 8$ and  $2 \leftrightarrow 6$ while the $\Lambda=4$ splits $4=4_1+4_2$. We have discussed this case already in detail in section \ref{sec:confemb} on conformal embeddings, in particular tables \ref{su2_8_spintab} and \ref{tab:SU(2)_8_broken}. This leaves us after confinement of the odd representations with four fields described by a $Fibonaci \otimes Fibonaci$ theory, or equivalently a theory with the same topological order as ${SU}(3)_2 \otimes {SU}(3)_2$. The branchings are given in the left part of table \ref{tab:confbranchings}.

Finally the case $SU(2)_{12}$ .  One is left with a theory with  five primary fields, which is easily identified  as the  $SU(3)_4/\ZZ_3$ (for identification one may consult for example Ref.~\onlinecite{Gepner94}). This theory has the fields corresponding to the  $1, 8, 10, \overline{10}$ and $27$ dimensional representations with the branchings and conformal weights given in the right hand part of table \ref{tab:confbranchings}.
 
We see that the breaking mechanism allows us to systematically construct many new conformal embeddings and thereby it will also generate a large number of non-diagonal  modular invariants for non-chiral CFTs.

\section{The Coset Construction}
\label{sec:cosets}
%Explain about branching rules, identification currents etc. Do a simple example with simple currents (maybe an example relevant to the Hall effect)?

The coset construction\cite{Goddard1984} is a way to construct a new conformal field theory, starting from $G_{k}$ and $H_{k'}$ WZW models based on Lie groups $G$ and $H$ with $H\subset G$. Given an embedding of $H$ into $G$ with Dynkin index $l$, this embedding will fix the relation between the levels $k'$ and $k$ as $k'=lk$. This also implies that  $c(G,k)\geqslant c(H,k')$. {}The canonical generators of the conformal algebra for the coset are just the differences of the conformal generators of the WZW theories (which are Sugawara bilinears in the currents of the chiral algebras). Equivalently, the energy momentum tensor of the coset is defined as the difference of the energy momentum tensors for the $G$ and $H$ theories,
\begin{equation}
T_{G/H} = T_G -T_H.
\end{equation}
This gives the coset central charge as $c(G/H,k',k)= c(G,k')-c(H,k)$. 
%In general one may decompose the integrable representations of $G_k$ into irreducible representations of $H_{k'}$. This decomposition the coefficients are usually called  branching functions because the branchings are infinite in general.  
One physical interpretation of the coset models is that they correspond to gauging the $H$ subgroup of $G$ in the WZW model based on $G$\cite{Gawedzki88,Karabali88,Bardakci88}. 
%In this paper we will encounter a different interpretation. 

Coset CFTs play an important role in for example the description of fractional quantum Hall states with non-Abelian anyonic excitations. For example the Moore-Read state and the series of Read--Rezayi states involve the cosets:
\begin{equation}
\label{eq:suncosets}
SU(n)_1\otimes SU(n)_1 / SU(n)_{2}
\end{equation}
with central charges $c(n)=\frac{2n-2}{n+2} $

It is in general nontrivial to the determine the full chiral algebra and the set of primary fields of a coset theory  and to determine their fusion and braiding properties. One way to approach this problem is through the character theory of affine Lie algebra representations (see for instance Ref.~\onlinecite{dsm}). 
The $G_k$ highest weight representations $r_\Lambda$ branch into $H_{k'}$ representations $r_{\Lambda'}$. Both the $r_{\Lambda}$ and the $r_{\Lambda'}$ are infinite dimensional and in most cases the branching of $r_{\Lambda}$ yields either infinitely copies of $r_{\Lambda'}$ or no copies at all. However, the subspaces of the $r_{\Lambda}$ and the $r_{\Lambda'}$ at any fixed eigenvalue of $L_0$, the chiral Hamiltonian, are finite dimensional. The character of an affine Lie algebra representation is just a generating function for the dimensions of the eigenspaces of $L_0$ in that representation. Hence there is an identity between the characters of the integrable $G_k$ representations and the integrable $H_{k'}$ representations into which they decompose. We have 
\begin{equation}
\chi_{\Lambda} = \sum_{\Lambda'}  \chi_{\Lambda;\Lambda'} \chi_{\Lambda'}, 
\end{equation}
where $\chi_{\Lambda}$ and $\chi_{\Lambda'}$ are the characters of the representations $r_{\Lambda}$ and $r_{\Lambda'}$ and the $\chi_{\Lambda;\Lambda'}$ are so called \emph{branching functions}. One approach to coset models is to consider the branching functions directly as characters of the representations of the coset theory. In other words, one does not explicitly construct the coset chiral algebra, but instead one says that there is a non-zero chiral primary field of the coset theory for any nonzero branching function $\chi_{\Lambda;\Lambda'}$. The requirement that the branching function should be nonzero means that there will not be a coset primary field for any combination $(\Lambda;\Lambda')$ but only for those combinations allowed by the branching rules. On top of the branching rules, there are so called \emph{field identifications} which say that some of the coset primary fields may be labelled by various different combinations of weights $(\Lambda;\Lambda')$, or in other words, some of the pairs $(\Lambda;\Lambda')$ are identified if they are used as labels for coset primaries. Basically the pairs $(\Lambda_1;\Lambda'_1)$ and $(\Lambda_2;\Lambda'_2)$ are identified when the corresponding branching functions are equal, but often it is much easier to find the identfications by arguments involving the modular transformations of the characters and the automorphisms of $G_{k}$ and $H_{k'}$\cite{gepner89}, rather than by explicit calculation of the branching functions. 
%However in general there is some redundancy in this decomposition, which means that the coset model has a larger symmetry than the one divided out. This leads to extra degeneracies which manifest themselves through the fact that there are additional selection rules, i.e.~the fact that certain branching functions may be identical zero, while on the other hand it may be that branching functions for different  pairs of $(\Lambda ; \Lambda')$ are identical, meaning that we should think of equivalence classes of such pairs. 

An alternative way to find the branching rules and field identifications of coset theories is through the action of the \textit{identification group} $G_{id}$\cite{Schellekens1989}. For a $G_{k}/H_{k'}$ coset, this identification group is defined as the group of bosonic simple current primary fields in the tensor product theory $G_{k}\otimes \overline{H_{k'}}$. Here the bar indicates that we should use the conjugate representation of the usual mapping class group representation for the $H_{k'}$ theory. In particular, the conformal weight of a $G_{k}\otimes \overline{H_{k'}}$ primary field labeled by $(\Lambda ; \Lambda')$ is the difference $h_{\Lambda}-h_{\Lambda'}$ of the $G_{k}$ and $H_{k'}$ conformal weights and bosonic simple currents are those simple currents for which this difference is an integer. The group product on $G_{id}$ is given by the fusion of the simple currents. $G_{id}$ also acts on the labels of the branching functions by fusion. If the orbits of branching functions under the $G_{id}$ action all have the same number of elements, then one may describe the field identifications and branching rules of the coset in a very simple way: all fields in a single $G_{id}$ orbit are identified and the branching rules allow precisely those combinations $(\Lambda,\Lambda')$ such that the corresponding primary field of the $G_{k}\otimes \overline{H_{k'}}$ theory has trivial monodromy with the elements of $G_{id}$. There is obviously a strong similarity between this procedure for finding branching rules and field identifications in coset theories and the procedures we have described for quantum group symmetry breaking and confinement, particularly with the special case of our symmetry breaking scheme described at the beginning of section~\ref{sec:simplecurrents}, where the condensed fields are simple currents and the orbits under the action of these simple currents are all of maximal size. In such cases, the procedure for finding the spectrum, fusion and modular properties of coset fields reduces precisely to the procedure we have described for the condensation of the bosonic fields in the group $G_{id}$, in the TQFT corresponding to the $G_{k}\otimes \overline{H_{k'}}$ WZW theory. Field identifications appear at the symmetry breaking stage, as the $G_{k}\otimes \overline{H_{k'}}$ related by fusion with the condensed fields from $G_{id}$ turn out to have the same restriction, whereas the coset branching rules are due to confinement; only fields that have trivial monodromy with the fields in $G_{id}$ are not confined. 

As an illustration of this relation between breaking a quantum symmetry and the coset construction, we discuss the simplest example of the series (\ref{eq:suncosets}), the case $n=2$. In this case, the coset is the chiral Ising CFT, which plays a fundamental role in the construction of the Moore-Read fractional quantum Hall state, as well as in the hierarchy of non-Abelian Hall states based on it\cite{Bonderson07c}. We have to consider a boson condensate in $SU(2)_1 \otimes SU(2)_1 \otimes \overline{SU(2)}_2$. The properties of representations of the factors of this product are given below.
\begin{table}[hbt]
\begin{center}
$\begin{array}{cc}
 \begin{array}[t]{|c|}
\hline SU(2)_1 \\
\hline
\hline
\begin{array}{l|ll}
0&d_0=1&h_0=0\\
1&d_1=1&h_1=\frac{1}{4}\\
\end{array}\\
\hline
\hline
\begin{array}{l}
1\times 1=0 \\
\end{array}\\
\hline
\end{array}\;\;\;\;\;
&\;\;\;\;\;
\begin{array}[t]{|l|}
\hline SU(2)_2 \\
\hline
\hline
\begin{array}{l|ll}
0&d_0=1&h_0=0\\
1&d_1=\sqrt{2}&h_1=\frac{3}{16}\\
2&d_2=1&h_2=\frac{1}{2}\\
\end{array}\\
\hline
\hline
\begin{array}{ll}
1\times 1=0+2~& \\
1\times 2=1&2\times 2=0~ \\
\end{array} \\
\hline 
\end{array} 
\end{array}$
\end{center}
\caption{{\footnotesize Spins, quantum dimensions and nontrivial fusion rules for $SU(2)_1$ and $SU(2)_2$}}
\label{tab:su21and su22}
\end{table}

%\begin{table}[hbt]
% \begin{center}  
%\hline
%\end{array}$
%\end{center}
%\caption{\footnotesize Spins, quantum dimensions and nontrivial fusion rules for $SU(2)_2$ .}
%\label{tab:su22}
%\end{table}
%
\noindent All together there are $ 2 \times 2 \times 3 = 12$ fields which we denote by $(ij;k)$. There is one nontrivial boson $(11;2)$ which we assume to condense. It is a simple current because $(11;2)\otimes(11;2)= (00;0)$. We now have to identify the fields which form orbits under fusion with the condensed field:
\begin{eqnarray}
(00;0,1,2) \otimes (11;2) &=& (11;2,1,0) \\
(01;0,1,2) \otimes (11;2) &=& (10;2,1,0)
\end{eqnarray}
where we have used an obvious notation to save space. At this point we are left with $6$ fields which we will label as the ones on the left. Now we have to determine which of the remaining fields will be confined. Using the conformal weights given in the tables we see that for example 
that the identified fields (00;1) and (11;1) have conformal dimensions $h=-3/16$ and $h=5/16$ respectively which differ by 1/2. This as we explained before, means that this field has to be confined. Similarly one finds that $(01;0)$ and $(01;2)$ are confined. We are then left with three non-confined fields and as expected these correspond exactly to the coset model, which is the Ising model, as indicated in the following table.
\begin{table}[hbt]\begin{center} 
$\begin{array}{|l|}
\hline \mbox{Ising model} \\
\hline
\hline
\begin{array}{l|ll}
(00;0)\sim 1&d_1=1&h_1=0\\
(00;2)\sim\varepsilon&d_\varepsilon=1&h_\varepsilon=\frac{1}{2}\\
(01;1)\sim\sigma&d_\sigma=\sqrt{2}&h_\sigma=\frac{1}{16}\\
\end{array}\\
\hline
\hline
\begin{array}{ll}
\varepsilon\times \varepsilon= 1~& \\
\varepsilon \times\sigma=\sigma~&\sigma\times \sigma=1+\varepsilon~ \\
\end{array}\\
\hline
\end{array}$ \end{center}
\caption{\footnotesize Spins, quantum dimensions and nontrivial fusion rules for the Ising model .}
\label{tab:su22}
 \end{table} 
%% I added the above paragraphs on this coset model
%It is clear that our breaking procedure should be somehow  related to the procedure to construct coset models. The relation is not hard to establish if we keep in mind that the breaking of the quantum group basically works the opposite way and corresponds to \lq enlarging' the chiral symmetry. The identification group is basically generated by the condensate, while  the selection rule corresponds to the confinement condition we have applied, and finally  we see that the identifications follow in our picture from the consistency conditions imposed by the fusion rules of the residual symmetry algebra.

At this point it is natural to ask how the correspondence between the coset construction and quantum group symmetry breaking fits into the general picture of quantum group symmetry breaking as dual to conformal extension that we sketched before. It would appear that there is something of a mismatch. The quantum group symmetry breaking picture for the coset $G_{k}/H_{k'}$ starts from the $G_{k}\otimes\overline{H_{k'}}$ topological data and condenses the available bosonic simple currents (i.e.~the fields in $G_{id}$). Naively, the dual chiral algebra extension should start from the chiral algebra for a $G_{k}\otimes\overline{H_{k'}}$ WZW model and extend this by the currents in $G_{id}$. However, this chiral algebra is not the chiral algebra of the coset. In the construction of the coset theory, the $H_{k'}$ chiral algebra is embedded in the $G_{k}$ chiral algebra so there is a priori no tensor product of the two. In fact, the interpretation of the coset theory as a gauged WZW model and the identification of the coset primary field with branching functions strongly suggest that the chiral algebra of the coset theory should be the commutant of the $H_{k'}$ chiral algebra in (some extension of) the $G_{k}$ chiral algebra. Nevertheless the description of the coset based on the identification group strongly suggests that, while the coset theory and the $G_{k}\otimes\overline{H_{k'}}$ conformally extended by $G_{id}$ may be different as conformal field theories, they nevertheless have identical topological data. As a result, one should be able to describe the topological phase in $2+1$ dimensions whose $1+1$-dimensional boundary is described by the $G_{k}/H_{k'}$ coset model using the topological data obtained from breaking $G_{k}\otimes\overline{H_{k'}}$ by condensation of the bosons constituting $G_{id}$. This claim is also supported by the work of Moore and Seiberg\cite{mszoo}. They study the Chern-Simons theory based on the gauge group $(G \times H)/Z$, where $Z$ is the common center of $G$ and $H$, with Chern Simons terms at level $k$ for $G$ and at level $-k'$ for $H$, and they show that this theory has precisely the gauged WZW description of the $G_{k}/H_{k'}$ coset as its boundary theory.  

\subsection{Fixed points and maverick cosets}
\label{sec:mavericks}
  
So far we have only discussed the very simplest cosets, which have the property that the identification group orbits are all the same size. However in general, there will be orbits of different sizes. In this case one speaks of \lq field identification fixed points', since some of the elements of $G_{id}$ will now fix some of the pairs $(\Lambda,\Lambda')$ labeling the coset primaries. It turns out that in such cases it becomes necessary to introduce extra coset primary fields, and to view the branching functions corresponding to the identification fixed points as linear combinations of the characters for these fields. This is analogous to the situation we describe in the latter part of section~\ref{sec:simplecurrents}, where we show that fields that are fixed under fusion with a simple current condensate must split under restriction. In the context of coset CFTs, special techniques have been developed to deal with fixed points\cite{Schellekens1989,Fuchs1995,Frohlich2004}, but it appears quantum group symmetry breaking takes care of fixed points without any changes to the procedure we have described already (although of course the actual calculations involved in carrying out the procedure do become more complicated when fixed points appear). 

Fixed points are not the only complicating factors that may appear in the description of coset CFTs. There are in fact cosets for which there are more field identifications and more restrictive branching rules than one would expect from the action of the identification group (one may show this for example by explicit calculation of the branching functions). The first examples of such \emph{maverick cosets} were found in Refs.~\onlinecite{Dunbar1992,Dunbar1993} in 1992, and several more have been found since\cite{Pedrini1999}. In the quantum group symmetry breaking formalism, such maverick cosets can be explained by the condensation of a bosonic field which is not a simple current, again, without any change to the framework we have described. 

The simplest maverick coset is $SU(3)_2/SU(2)_8$. The central charge of this coset is $\frac{4}{5}$, which means the coset theory must be related to the unitary minimal model $\mathcal{M}(6,5)$. In fact it turns out that the coset primary fields are in one to one correspondence with the subset of the chiral primary fields of $\mathcal{M}(6,5)$ which appear in the description of the critical point of the three-state Potts model.  

We will work out the quantum group symmetry breaking point of view on this coset in some detail. We listed the conformal weights and quantum dimensions of the $SU(2)_8$ already in~\ref{su2_8_spintab} and we give those for the $SU(3)_2$ fields in table~\ref{su3_2_spintab}. We label the $SU(2)_8$ fields in the usual way (by their Dynkin labels) and we label the $SU(3)_2$ fields using a notation based on the dimensions of the corresponding $SU(3)$ representations. The correspondence between the Dynkin labels of the highest weights of these $SU(3)$ representations and the labels that we use is as follows: $1\equiv(0,0)$, $3\equiv(1,0)$, $\bar{3}\equiv(0,1)$, $6\equiv(2,0)$, $\bar{6}\equiv(0,2)$ and $8\equiv(1,1)$.
\begin{table}[hbt]
\begin{center}
$
\begin{array}[t]{|l|}
\hline SU(3)_{2}\\
\hline
\hline
\begin{array}{l|ll}
1&d_1=1&h_1=0\\
6&d_6=1&h_6=1\\
\bar{6}&d_{\bar{6}}=1 &h_{\bar{6}}=1\\
8&d_8= \frac{1+\sqrt{5}}{2}&h_8=\frac{3}{5}\\
3&d_3= \frac{1+\sqrt{5}}{2}&h_3=\frac{2}{3}\\
\bar{3}&d_{\bar{3}}=\frac{1+\sqrt{5}}{2}& h_{\bar{3}}=\frac{2}{3}\\
\end{array}\\
\hline
\end{array}
$
\end{center}
\caption{\footnotesize  Spins and quantum dimensions for $SU(3)_2$.}
\label{su3_2_spintab}
\end{table}

\noindent The fusion rules for $SU(2)_8$ can be read off from formula~(\ref{su2_k_fusion}). The fusion rules for $SU(3)_2$ are given in table~\ref{tab:SU(3)2_fusion}.
\begin{table}[hbt]
{\scriptsize 
\begin{center}
$
\begin{array}{|lllll|}
\hline
\rule{0pt}{0.25cm} 
3\times 3=\bar{3}+6~&&&& \\
3\times \bar{3}=1+8&\bar{3}\times \bar{3}=3+\bar{6}~&&& \\
3\times 8 = 3 +\bar{6}&\bar{3}\times 8=\bar{3}+6&8\times 8=1+8&&\\
3\times 6 = \bar{3}&\bar{3}\times 6 = 3&8\times 6=\bar{3}&6\times 6=\bar{6}& \\
3\times \bar{6} = 8& \bar{3}\times\bar{6}=8&8\times \bar{6}=3&6\times\bar{6}=1&
\bar{6}\times\bar{6}=6 \\ \hline
\end{array}
$
\end{center}}
\caption{\footnotesize Fusion rules for $SU(3)_2$.}
\label{tab:SU(3)2_fusion}
\end{table}
{}From table~\ref{su2_8_spintab}, we read off that there is a single nontrivial identification current in the $SU(3)_{2}\otimes\overline{SU(2)_{8}}$ theory, namely the field with labels $(1,8)$. in other words, $G_{id}=\{(1,0),(1,8)\}$. There is also a bosonic field which is not a simple current, the field labeled $(8,4)$. Let us first investigate what happens when we condense only the simple current field $(1,8)$ and not the field $(8,4)$. Since $(1,8)$ acts trivially on the $SU(3)$ part of the theory, we can just search for the restrictions of the pure $SU(2)_8$ fields, that is, the fields labeled $(1,\Lambda)$ for some $\Lambda$. The restrictions for more general fields will be similar. The problem of breaking $SU(2)_8$ by condensation of the $8$-sector was already considered in section~\ref{su28sec}, and the results of the breaking were presented in table~\ref{tab:SU(2)_8_broken}. After symmetry breaking and confinement, $SU(2)_8$ reduces to a $Fibonacci\otimes Fibonacci$ theory, with $4$ sectors, labeled in table~\ref{tab:SU(2)_8_broken} as $0$, $2$, $4_1$ and $4_2$. This means that condensation of the $(1,8)$ field in the $SU(3)_{2}\otimes\overline{SU(2)_{8}}$ theory will lead, after condensation and confinement, to a new theory with $24$ distinct sectors, each labeled by an $SU(3)_2$ representation and a label from the broken and confined remnant of the $SU(2)_8$ theory. This $24$ sector theory is clearly not the right description of the coset $SU(3)_2/SU(2)_8$. The full Virasoro minimal model $\mathcal{M}(5,6)$ at $c=\frac{4}{5}$ only has $10$ sectors, so $24$ sectors is clearly too many and also some of the conformal weights we find are not compatible with the conformal weights of $\mathcal{M}(5,6)$\footnote{Of course this mismatch does not necessarily mean that it would be impossible to find some CFT with the topological order of the $24$ sector model, it just means that this CFT would have to occur at some central charge which is different from $\frac{4}{5}$, necessarily by some nonzero multiple of $8$.}. 

To describe the coset at $c=\frac{4}{5}$, we must condense the non-simple current field $(8,4)$ in addition to the field $(1,8)$. This suggests that for the general description of coset models, we should condense all available bosons. Note however that in the case treated here, we will show that condensation of $(8,4)$ actually implies that $(1,8)$ condenses as well. We sketch the calculation of the branching rules. 

First of all, we note that there will be many additional splittings which do not occur when only $(1,8)$ is condensed. For example the fusions $(8,\Lambda)\times (8,\Lambda)$ for $2\le\Lambda\le 6$ all contain the field $(8,4)$, which means the restrictions of these fusions contain the vacuum at least twice and hence all the fields $(8,\Lambda)$ with $2\le \Lambda\le 6$ split. As a result the fields $(3,\Lambda)$ and $(\bar{3},\Lambda)$ with $\Lambda$ in the same range also split under restriction, since these fields can be obtained from the $(8,\Lambda)$ fields by fusion with the simple currents $(6,0)$ and $(\bar{6},0)$. 

There are also $18$ sectors which will certainly not split under restriction, because they have quantum dimensions less than $2$. These are all sectors with labels of the form $(x,0)$ or $(x,8)$ as well as the $6$ fields with labels $(1\vee6\vee \bar{6},1\vee 7)$. 

The sector labeled $(1,2)$ could in principle split into two sectors $(1,2)_1$ and $(1,2)_2$ of quantum dimensions $1$ and $\frac{1+\sqrt{5}}{2}$ respectively. Given such a splitting, we know that the restriction of $(1,2)\times (1,2)$ must contain the vacuum twice and we find   
{\small
\begin{eqnarray}
(1,2)\times (1,2) &\equiv&
1+1+\ldots \nonumber\\
(1,2)\times (1,2) &=& (1,0)+(1,2)+(1,4)\nonumber \\
~&=&(1,0)+(1,2)_1+(1,2)_2+(1,4)
\end{eqnarray}
}
and comparing the first and last lines, we notice that either $(1,4)$ or on of the components of $(1,2)$ must branch to the vacuum. However, this cannot be allowed, since $(1,4)$ and $(1,2)$ have nontrivial spin and hence cannot condense (alternatively we might say this would \lq confine the vacuum'). In conclusion, we find that the sector $(1,2)$  does not split under restriction. This is a crucial piece of information in what follows.
%Using that $(1,8)\times (1,2)=(1,6)$, we see that $(1,6)$ does not split either. 

Now let us consider the fusion of the condensed field $(8,4)$ with the $SU(2)_8$ type fields $(1,\Lambda)$. Whenever $\Lambda$ is even, we have $(8,4)\in (8,4)\times (1,\Lambda)$. Since $(8,4)$ has the trivial field in its restriction, we see that the restriction of $(8,4)$ must contain the restriction of the dual of some component of each of the sectors $(1,\Lambda)$ with $\Lambda$ even. Since the sectors $(1,\Lambda)$ are self-dual, we see in fact that the restriction of $(8,4)$ contains the restrictions of $(0,0)$, $(0,2)$, $(0,6)$ and $(0,8)$ and at least one component of the restriction of $(0,4)$. Similarly, we may consider the fusion of $(8,4)$ with fields labeled $(8,\Lambda)$ and derive that $(8,4)$ contains at least one component of the restrictions of $(8,0)$, $(8,2)$, $(8,4)$, $(8,6)$ and $(8,8)$.  

From this point simple arguments using quantum dimensions give us much information on splittings and field identifications. The quantum dimension of $(8,4)$ is $\frac{3+\sqrt{5}}{2}$. We know that the restriction of $(8,4)$ contains at least the vacuum $(1,0)$, which has quantum dimension $1$ and the full restriction of $(1,2)$, which has quantum dimension $\frac{3+\sqrt{5}}{2}$ (it must contain the full restriction, since $(1,2)$ does not split). Hence it follows that $(8,4)$ splits into $3$ components, the vacuum $(0,1)$ the restriction of $(0,2)$ which we just denote $(0,2$ as well and a third component $(8,4)_3$ of quantum dimension $\frac{1+\sqrt{5}}{2}$. But in the previous paragraph we noted that the restriction of $(8,4)$ contains the restriction of $(1,8)$ and since this has quantum dimension equal to $1$, it must be the same as the restriction of $(1,0)$, in other words, we have shown that $(1,8)$ is condensed, as we promised earlier. This immediately fixes the restrictions of the fields in the pure $SU(2)_8$ sector, as in our treatment in section~\ref{su28sec} and consequently also the restrictions of the fields with labels $(6,\Lambda)$ and $(\bar{6},\Lambda)$.

To find the restrictions of the remaining fields, consider the fusion $(8,0)\times(1,4)$. We have 
{\small
\begin{eqnarray}
(8,0)\times (1,4) &=& 
(8,4)\equiv (1,0)+(1,2)+(8,4)_3 \nonumber\\
(8,0)\times (1,4) &=& (8,0)\times(1,4)_1 +(8,0)\times (1,4)_2
\end{eqnarray}
}
where $(1,4)_1$ and $(1,4)_2$ are the two components of $(1,4)$ which result from the condensation  of $(1,8)$ (these correspond to $4_1$ and $4_2$ in table~\ref{tab:SU(2)_8_broken}). Comparing the two lines, we see that the restriction of $(8,0)$ must equal either $(1,4)_1$ or $(1,4)_2$. We can in fact make an arbitrary choice between these two options, because the fusion rules of the broken $SU(2)_8$ theory are invariant under the exchange of the sectors $(1,4)_1$ and $(1,4)_2$. Choosing $(1,8)\equiv (1,4)_1$, we can write
{\small
\begin{eqnarray}
(8,0)\times (1,4) &\equiv& (1,4)_1\times (1,4)_1 +(1,4)_1\times (1,4)_2 \nonumber \\
&=& (1,0)+(1,4)_1+(1,2),
\end{eqnarray}
}
using the fusion rules for the broken $SU(2)_8$ theory given in table~\ref{tab:SU(2)_8_broken}. Comparing this with the previous equation, we finally get the full branching of $(8,4)$, namely $(8,4)\equiv (1,0)+(1,2)+(1,4)_1$. From here, we can easily produce the branchings for $(8,1)$, $(8,2)$ and $(8,3)$. We have
{\small
\begin{eqnarray}
(8,1)&=&(8,0)\times (1,1) \equiv (1,4)_1\times (1,1)=(1,3) \nonumber\\ 
(8,2)&\equiv& (1,4)_1\times (1,2)=(1,2)+(1,4)_2 \nonumber\\
(8,3)&\equiv& (1,4)_1\times (1,3)=(1,1)+(1,3). 
\end{eqnarray}
}
We summarize the branchings for the fields of the forms $(1,\Lambda)$ and $(8,\Lambda)$ in table~\ref{tab:SU(3)2xSU(2)8broken}. They branch to a set of fields which is in one to one correspondence with the fields of the broken $SU(2)_8$ theory and which also has the same fusion rules (see table ~\ref{tab:SU(2)_8_broken}). The full set of fields for the broken $SU(3)_2\times\overline{SU(2)_8}$ theory has $18$ fields corresponding to the products of the $\ZZ_3$ group of simple currents $\{1,6,\bar{6}\}$ with the broken $SU(2)_8$ fields $\{0,1,2,3,4_1,4_2\}$. The branching for the full theory can easily be obtained from the branchings for the fields given in table~\ref{tab:SU(3)2xSU(2)8broken}. For example, for fields of the form $(3,\Lambda)$, we write $(3,\Lambda)=(\bar{6},1)\times(8,\Lambda)$ and conclude that if the branching for $(8,\Lambda)$ was given by $(8,\Lambda)\rightarrow \sum_{i}(1,x_i)$ then the branching for $(3,\Lambda)$ is given by $(3,\Lambda)\rightarrow \sum_{i}(\bar{6},x_i)$. A similar statement holds for all other branchings. The fusion rules of the full broken theory can also be described easily; they are just the product of the $\ZZ_3$ fusion rules for $\{1,6,\bar{6}\}$ and the broken $SU(2)_8$ fusion rules.

\begin{table}[hbt]
\begin{center}
$\begin{array}{|l|}
\hline SU(3)_2\times \overline{SU(2)_8} {\rm ~broken}\\
\hline
\hline
\begin{array}{ll}
(1,0),(1,8)\rightarrow (1,0)\equiv 1~& (8,0),(8,8)\rightarrow (1,4)_2\\
(1,1),(1,7)\rightarrow (1,1)~& (8,1),(8,7)\rightarrow (1,3)\\
(1,2),(1,6)\rightarrow (1,2)~& (8,2),(8,6)\rightarrow (1,2)+(1,4)_2\\
(1,3),(1,5)\rightarrow (1,3)~& (8,3),(8,5)\rightarrow (1,1)+(1,3)\\
(1,4)\rightarrow (1,4)_{1}+(1,4)_{2}~& (8,4)\rightarrow (1,0)+(1,4)_{1}+(1,2)
\end{array}\\
\hline
\end{array}$
\end{center}
\caption{\footnotesize Branching rules for $SU(3)_2\times \overline{SU(2)_8}$ after condensation in the $(8,4)$-sector. Branching rules for fields of the forms $(3,\Lambda)$, $(\bar{3},\Lambda)$, $(6,\Lambda)$ and $(\bar{6},\Lambda)$ can be easily produced from this table. Quantum dimensions and fusion of the sectors of the broken theory may be read off from the corresponding table for $SU(2)_8$ (table~\ref{tab:SU(2)_8_broken}).} 
\label{tab:SU(3)2xSU(2)8broken}
\end{table}

Using the branching rules, we may now check for confinement in the usual way. Note that some of the fields that were not confined in the broken $SU(2)_8$ theory are now confined, because they appear in more branching rules than before and no longer have well defined spin factors as a result. 

It turns out that there are $6$ non-confined fields, which have precisely the conformal weights and fusion rules of the chiral three state Potts model, or equivalently of the $6$ fields involved in the non-diagonal modular invariant for the $\mathcal{M}(6,5)$ minimal model. Hence, we have reproduced the topological data of this maverick coset, using precisely the same quantum group symmetry breaking formalism as for standard cosets.

\begin{table}[hbt]
\begin{center}
$
\begin{array}[t]{|l|}
\hline SU(3)_{2}/SU(2)_8\\
\hline
\hline
\begin{array}{l|ll}
~&d&h\\ \hline
(1,0)&1&0\\
(6,0)&1&\frac{2}{3}\\
(\bar{6},0)&1 &\frac{2}{3}\\
(1,4_2)&\frac{1+\sqrt{5}}{2}&\frac{2}{5}\\
(6,4_2)&\frac{1+\sqrt{5}}{2}&\frac{1}{15}\\
(\bar{6},4_2)&\frac{1+\sqrt{5}}{2}&\frac{1}{15}\\
\end{array}\\
\hline
\end{array}
$
\end{center}
\caption{\footnotesize  Spins and quantum dimensions for the coset $SU(3)_2/SU(2)_8$ as obtained from quantum group symmetry breaking. The result matches the spins and quantum dimensions for the chiral three state Potts model.}
\label{maverick_spintab}
\end{table}

%{\em Now also do/mention $SU(4)_1/SU(2)_{10}$, which should give the Ising model?}

\subsection{Conformal embeddings revisited}
\label{sec:embeddings2}

Since conformal embeddings $H_{k'}\subset G_{k}$ conserve the central charge, the cosets $G_{k}/H_{k'}$ coming from these embeddings have conformal central charge $c=0$ and must be trivial.
Still, confirming the triviality of these cosets using the identification group may not be so trivial, because the procedure can involve resolution of fixed points and even dealing with more complicated issues like those which occur for the (nontrivial) maverick cosets. From the quantum group symmetry breaking perspective, conformal embeddings are in fact just mavericks for which the coset happens to come out trivial. 

The quantum group symmetry breaking perspective on these cosets adds information to the usual treatment, because, while the effective theory for the non-confined excitations of the coset is of course trivial, the symmetry breaking approach also gives a description of the confined excitations, which we can view as boundary excitations between a phase with $G_{k}\otimes\overline{H_{k'}}$ topological order and a topologically trivial phase. This boundary theory will be nontrivial. In fact, one may expect that the boundary theory is the same as the boundary theory for a boundary between $G_{k}$ and $H_{k'}$ phases. The reason for this is that the $G_{k}\otimes\overline{H_{k'}}$ theory can be thought of as a two layer theory, where a piece of material with $H_{k'}$ topological order has been folded under a piece with $G_{k}$ topological order. This \lq folding' converts the boundary between the region with $G_{k}$ order and the region with $H_{k'}$ order into a boundary between a region with $G_{k}\otimes\overline{H_{k'}}$ topological order and a region with trivial topological order. However, since folding is just a geometric deformation of the medium it should not change the topological order on the boundary and so we expect the two boundary theories to be the same.  

We will now demonstrate triviality of the coset, as well as this correspondence of boundary theories for our favourite example, $SU(3)_1/SU(2)_4$. Forming the product $SU(3)_1\otimes \overline{SU(2)_4}$, we see that there are three nontrivial bosons, labeled $(1,4)$, $(3,2)$ and $(\bar{3},2)$ in our usual labeling conventions for $SU(3)_1$ and $SU(2)_4$. Only the field $(1,4)$ is a simple current and, as with the maverick $SU(3)_2/SU(2)_8$, we find that condensing only this simple current does not give the desired result, that is, the effective theory after breaking and confinement is still nontrivial. However, if we condense all bosonic fields, we find that we do obtain the correct (trivial) coset theory, as in the case of the maverick. We give the results of the symmetry breaking calculation in table~\ref{tab:su3_1xsu2_4broken}. After symmetry breaking, there are $4$ sectors left, the vacuum sector and the restrictions of $(3,0)$, $(\bar{3},0)$ and $(1,1)$. The fields $(1,4)$, $(3,2)$ and $(\bar{3},2)$ branch to the vacuum, so they are indeed condensed. Also, it is easy to check, using the weights of $SU(2)_4$ and of $SU(3)_1$ (see tables \ref{tab:su24} and \ref{tab:su3_1_and_so5_1}), that all nontrivial broken sectors are confined, leaving just the vacuum sector and hence confirming that the coset is trivial. Finally, the $4$ sectors of the boundary theory have quantum dimensions $1$, $1$, $1$ and $\sqrt{3}$, which fixes the fusion rules and indeed, we see that there is a one to one correspondence with the boundary theory between $SU(2)_4$ and $SU(3)_1$ given in table ~\ref{tab:su3_1xsu2_4broken}, as expected.
 
\begin{table}[hbt]
\begin{center}
$\begin{array}{|l|}
\hline SU(3)_1\times \overline{SU(2)_4} {\rm ~broken}\\
\hline
\hline
\begin{array}{ll}
(1,0),(1,4)\rightarrow (1,0)\equiv 1~&(1,2)\rightarrow (3,0)+(\bar{3},0)\\
(3,0),(3,4)\rightarrow (3,0)~&(3,2)\rightarrow (1,0)+(\bar{3},0)\\
(\bar{3},0),(\bar{3},4)\rightarrow (\bar{3},0)~&(\bar{3},2)\rightarrow (1,0)+(3,0)\\
\multicolumn{2}{l}{(1,1),(3,1),(\bar{3},1),(1,3),(3,3),(\bar{3},3)\rightarrow (1,1)}
\end{array}\\
\hline
\end{array}$
\end{center}
\caption{\footnotesize Branching rules for $SU(3)_1\times \overline{SU(2)_4}$ after condensation of all bosons.}
\label{tab:su3_1xsu2_4broken}
\end{table}

\section{Discrete gauge theory and Orbifolds}
\label{sec:orbifolds}
%{\em need to understand more general orbifolds (not from a trivial theory) better. Read Dijkgraaf's thesis again? Maybe do a study of a non-holomorphic orbifold, $SU(2)_1/ZZ_2$? (Dijkgraaf's thesis)}

In section~\ref{sec:bosons} we saw that many examples of anyon models with bosons can be obtained either from Kitaev's toric code construction\cite{Kitaev03} or from gauge theories by breaking the gauge group to a discrete subgroup\cite{dgt1,dgt2,dgt3,dgt4,dgt5}. These theories can also be realized as conformal field theories, namely as orbifolds of topologically trivial CFTs\cite{DVVV}. All the topological information in these models can be described using the representation theory of the quantum doubles of finite groups\cite{dpr}. Probably the simplest example which allows for non-Abelian braiding is the model based on the quantum double $D(D_3)$ of the smallest non-Abelian group $D_3$, the symmetry group of the regular triangle, or equivalently, the permutation group of three objects. This model has been shown to allow for universal quantum computation, if some measurements are allowed as operations in addition to braiding\cite{Mochon04}.  An implementation of this model using Josephson junctions has been proposed in Refs.~\onlinecite{Doucot04,Doucot05a,Doucot05c}. 
%The model may also be realized in CFT as a $D_3$ orbifold of the $E_8$ level $1$ WZW model (does E_3 have a D_3 subgroup?). 

There are $8$ topological sectors in the model, each labeled as described in section~\ref{sec:bosons}, by a conjugacy class of $D_3$ and a representation of the centralizer group of one of the elements in that conjugacy class (the elements all have isomorphic centralizer groups). $D_3$ has three conjugacy classes, the class of the trivial element $e$, a class we denote $r$, containing the nontrivial rotations of the triangle (three-cycles as permutations) and a class called $s$ which contains the reflections (exchanges).  The trivial class has all of $D_3$ as its centraliser, leading to three particle sectors $\Pi^{e}_{1}$,$\Pi^{e}_{J}$ and $\Pi^{e}_{\alpha}$ corresponding to the three irreducible representations $1$, $J$ and $\alpha$ of $D_3$. Here $1$ denotes the trivial representation, making $\Pi^{e}_{1}$ the vacuum sector, and $J$ and $\alpha$ denote the nontrivial one dimensional and two dimensional irreducible representations respectively. The centraliser of $r$ is the $Z_3$ generated by the rotations, giving sectors $\Pi^{r}_{0},\Pi^{r}_{1}$ and $\Pi^{r}_{2}$ and the centraliser of $s$ is a $\ZZ_2$ giving two sectors $\Pi^{s}_{1}$ and $\Pi^{s}_{\gamma}$. The spin factors and quantum dimensions of these sectors are given in table~\ref{D3_spintab}.
\label{D3sec}
\begin{table}[hbt]
\begin{center}
\[
\begin{array}{|c|c|c|c|c|c|c|} \hline
~&\Pi^{e}_{1}\equiv 1&\Pi^{e}_{J}&\Pi^{e}_{\alpha}&
\Pi^{r}_{l}&\Pi^{s}_{1}&\Pi^{s}_{\gamma}  \\ \hline
d^{A}_{\alpha}&1&1&2&2&3&3 \\
\theta^{A}_{\alpha}&1&1&1&e^{\frac{2\pi il}{3}}&1&-1 \\ \hline 
\end{array}
\]
\end{center}
\caption{\footnotesize dimensions and spin factors for the irreps of
$D(D_{3})$}
\label{D3_spintab}
\end{table}
The fusion rules of the irreps of $D(D_{2m+1})$ have been worked out in Ref.~\onlinecite{thesismark}. 
For $D(D_3)$, we have of course
$\Pi^{e}_{1}\times\Pi^{A}_{\alpha}=\Pi^{A}_{\alpha}$ (for all
$(A,\alpha)$) and furthermore
\begin{equation}
\begin{array}{ll}
\Pi^{e}_{J}\times\Pi^{e}_{J}=1&\\
\Pi^{e}_{J}\times\Pi^{e}_{\alpha}=\Pi^{e}_{\alpha}&~~
\Pi^{e}_{\alpha}\times\Pi^{e}_{\alpha}=1+\Pi^{e}_{J}+\Pi^{e}_{\alpha}\\
\Pi^{e}_{J}\times\Pi^{r}_{l}=\Pi^{r}_{l}&~~
\Pi^{e}_{\alpha}\times \Pi^{r}_{l}=\Pi^{r}_{m}+\Pi^{r}_{n}~~(l,m,n~{\rm distinct})\\
\Pi^{e}_{J}\times\Pi^{s}_{1}=\Pi^{s}_{\gamma}&~~
\Pi^{e}_{\alpha}\times \Pi^{s}_{1}=\Pi^{s}_{1}+\Pi^{s}_{\gamma} \\
\Pi^{e}_{J}\times\Pi^{s}_{\gamma}=\Pi^{s}_{1}&~~
\Pi^{e}_{\alpha}\times \Pi^{s}_{\gamma}=\Pi^{s}_{1}+\Pi^{s}_{\gamma}.
\end{array}
\end{equation}
For the fusion rules of the $\Pi^{r}_{l}$, we have 
\begin{equation}
\begin{array}{l}
\Pi^{r}_{l}\times\Pi^{r}_{l}=1+\Pi^{e}_{J}+\Pi^{r}_{l}\\
\Pi^{r}_{l}\times\Pi^{r}_{m}=\Pi^{e}_{\alpha}+\Pi^{r}_{n} ~~~(l,m,n~{\rm distinct})\\
\Pi^{r}_{l}\times\Pi^{s}_{1}=\Pi^{r}_{l}\times\Pi^{s}_{\gamma}=\Pi^{s}_{1}+ \Pi^{s}_{\gamma}.
\end{array}
\end{equation}
Finally, we have
\begin{eqnarray}
\Pi^{s}_{1}\times\Pi^{s}_{1}&=&
1+\Pi^{e}_{\alpha}+\Pi^{r}_{1}+\Pi^{r}_{2}+\Pi^{r}_{3}
\nonumber \\
\Pi^{s}_{\gamma}\times\Pi^{s}_{\gamma}&=&
1+\Pi^{e}_{\alpha}+\Pi^{r}_{1}+\Pi^{r}_{2}+\Pi^{r}_{3}
\nonumber \\
\Pi^{s}_{1}\times\Pi^{s}_{\gamma}&=&
\Pi^{e}_{J}+\Pi^{e}_{\alpha}+\Pi^{r}_{1}+\Pi^{r}_{2}+\Pi^{r}_{3}.
\end{eqnarray}
A look at the spin factors and fusion rules confirms immediately that there is a wealth of bosons in the theory. 
%As discussed in section~\ref{sec:bosons}, this is in fact a feature of all models based on the quantum doubles of groups. 
For the odd dihedral groups (including $D_3$), we have analyzed all possible choices of condensate using our earlier quantum group based approach\cite{BSS03} and we will not repeat that exercise here. However, it will be good to check in an example that we can actually reproduce the results obtained there. This will also serve to illustrate some of the more interesting things that may happen on condensation.  In particular, we will see an example 
%the choice of the condensed particle is not exactly the same as the choice of condensate. Also we will see that 
with non-Abelian fusion rules after symmetry breaking (but before confinement).  

We will investigate what happens when an excitation in the $\Pi^{e}_{\alpha}$ sector condenses. 
In our earlier treatment of quantum group symmetry breaking, there were two non-equivalent ways of condensing this sector, because we could choose different (non gauge equivalent) internal states of the $\Pi^{e}_{\alpha}$ particles to form the condensate; note that since all quantum dimensions are integers here, it makes sense to talk about internal Hilbert spaces for single particles. We will find that our current methods produce the same two unconfined theories. However at the level of the broken theory including confined excitations, we find one extra solution to the requirements set out in this paper, in addition to the two solutions produced by our previous methods. The extra solution is almost certainly spurious and due to the fact that the requirements we give here are not completely sufficient to determine the broken theory (before confinement) in this case. It is not surprising that this can sometimes happen, since we have restricted our attention to a relatively crude level of description of topological order in this paper, looking only at fusion rules and spin factors. For theories with only integer quantum dimensions our old methods allow an approach to the problem which makes full use of the underlying Hopf algebra theory. Also, theories with many integer quantum dimensions can be relatively complicated to handle with the methods of this paper because integers allow for so many different splittings into smaller integers.

Now let us sketch the calculations which lead to the above results. Given that $\Pi^{e}_{\alpha}$ condenses, we know that $\Pi^{e}_{\alpha}$ branches to the vacuum sector and some other one dimensional sector. Since $\Pi^{e}_{J}\times \Pi^{e}_{\alpha}=\Pi^{e}_{\alpha}$, it follows that in fact $\Pi^{e}_{\alpha}$ branches to $1$ plus the restriction of $\Pi^{e}_{J}$. Now we have two possibilities: either $\Pi^{e}_{J}$ restricts to the vacuum or it does not. Both possibilities  cases lead to a consistent theory for the confined and non-confined excitations. 

Let us first assume that $\Pi^{e}_{J}$ does not branch to the vacuum, but rather to some nontrivial sector which we will still call $\Pi^{e}_{J}$. Then $\Pi^{e}_{\alpha}\rightarrow 1+\Pi^{e}_{J}$. Comparing $\Pi^{e}_{\alpha}\times \Pi^{r}_{i}$ with $(1+\Pi^{e}_{J})\times \Pi^{r}_{i}$, we see immediately that all $\Pi^{r}_{i}$ must branch to the same new sector, which we will simply call $\Pi^{r}$. We also see that $\Pi^{e}_\alpha$ appears on the right hand side of the fusion rules for the fields $\Pi^{s}_{1}$ and $\Pi^{s}_{\gamma}$, so these must both split. After completing the calculation, we find the branching given in table~\ref{tab:DD3broken1}. The unconfined fields are the vacuum, $\Pi^{e}_{J}$, $\Pi^{s}_{11}$ and $\Pi^{s}_{\gamma 1}$. These have spins $1$, $1$, $1$ and $-1$ respectively and $\ZZ_2\times \ZZ_2$ fusion rules, which fixes the topological order to be that of the $\ZZ_2$ discrete gauge theory or toric code model. In the discrete gauge theory, we can interpret the transition as a Higgs effect which has broken the $D_3$ gauge group down to a $\ZZ_2$ subgroup. On the CFT side it looks like we have partially \lq unorbifolded' the original theory, leaving a CFT where only the $\ZZ_2$ subgroup of the $D_3$ is orbifolded. This could be interpreted as due to extension of the orbifold conformal algebra by some of the chiral primaries for the twisted sectors.

%\begin{widetext}
\begin{table}[hbt]
\begin{center}
{\scriptsize 
$
\begin{array}{|l|}
\hline D(D_3) {\rm ~broken}\\
\hline
\hline
\begin{array}{l|l}
\Pi^{e}_{J}\rightarrow \Pi^{e}_{J}&d^{e}_{J}=1\\
\Pi^{e}_{\alpha}\rightarrow 1+\Pi^{e}_{J}&d^{r}=2\\
\Pi^{r}_{l}\rightarrow \Pi^{r}&d^{s}_{11}=1\\
\Pi^{s}_{1}\rightarrow \Pi^{s}_{11}+\Pi^{s}_{12}&d^{s}_{12}=2\\
\Pi^{s}_{\gamma}\rightarrow \Pi^{s}_{\gamma 1}+\Pi^{s}_{12}&d^{s}_{\gamma 1}=1
\end{array}\\
\hline
\hline
\begin{array}{lll}
\Pi^{e}_{J}\times\Pi^{e}_{J}=1~& \\
\Pi^{e}_{J}\times\Pi^{r}=\Pi^{r}& \Pi^{r}\times\Pi^{r}=1+\Pi^{e}_{J}+\Pi^{r}\\
\Pi^{e}_{J}\times\Pi^{s}_{11}=\Pi^{s}_{\gamma 1}&\Pi^{r}\times\Pi^{s}_{11}=\Pi^{s}_{11}+\Pi^{s}_{\gamma 1}\\
\Pi^{e}_{J}\times\Pi^{s}_{\gamma 1}=\Pi^{s}_{11}&\Pi^{r}\times\Pi^{s}_{\gamma 1}=\Pi^{s}_{11}+\Pi^{s}_{\gamma 1}\\
\Pi^{e}_{J}\times\Pi^{s}_{12}=\Pi^{s}_{12}&\Pi^{r}\times\Pi^{s}_{12}=2 \Pi^{s}_{12}\\
&\\
\Pi^{s}_{11}\times\Pi^{s}_{11}=1& \\
\Pi^{s}_{11}\times\Pi^{s}_{\gamma 1}=\Pi^{e}_{J}&\Pi^{s}_{\gamma 1}\times\Pi^{s}_{\gamma 1}=1&\\
\Pi^{s}_{11}\times\Pi^{s}_{12}=\Pi^{r}&\Pi^{s}_{\gamma 1}\times\Pi^{s}_{12}=\Pi^{r} \;\;\;\;\;
\Pi^{s}_{12}\times\Pi^{s}_{12}=1+\Pi^{e}_{J}+\Pi^{r}\\
\end{array}\\
\hline
\end{array}
$
}
\end{center}
\caption{\footnotesize Branching rules, quantum dimensions and nontrivial fusion rules for $D(D_3)$ after condensation in the $\Pi^{e}_{\alpha}$-sector, with $\Pi^{e}_{J}$ not condensed. The four unconfined sectors have the fusion rules and spins of a $D(\ZZ_2)$ theory.}
\label{tab:DD3broken1}
\end{table}
%\end{widetext}

Now let us consider the possibility that not only $\Pi^{e}_{\alpha}$, but also $\Pi^{e}_{J}$ condenses. In this case the entire electric part of the spectrum becomes trivial, or in other words, the gauge symmetry is fully broken. As before, we note that the restrictions of the $\Pi^{r}_{i}$ must all be equal and now because $\Pi^{s}_{\gamma}=\Pi^{e}_{J}\times\Pi^{s}_{1}$, we see that the restrictions of $\Pi^{s}_{\gamma}$ and $\Pi^{s}_{1}$ also equal each other. In the fusion rules $\Pi^{r}_{i}\times\Pi^{r}_{i}$, we see that $\Pi^{e}_{J}\equiv 1$ appears on the right hand side, implying that the $\Pi^{r}_{i}$ split into two parts of quantum dimension $1$. Similarly, from $\Pi^{s}_{1}\times \Pi^{s}_{1}$, we see that $\Pi^{s}_{1}$ and $\Pi^{s}_{\gamma}$ split into three parts, also of quantum dimension $1$. We note that none of the $\Pi^{r}_{i}$ can branch to the vacuum because not all the $\Pi^{r}_{i}$ have trivial spin and they all have the same restriction. Similarly $\Pi^{s}_{\gamma}$ (and hence $\Pi^{s}_{1}$) cannot branch to the vacuum. Since all sectors after breaking are simple currents, they form a group under fusion. Moreover, one sees easily that the compomnents of the $\Pi^{r}_{i}$, together with the vacuum $1$, form a subgroup. If the components of the restriction of $\Pi^{r}_{i}$ were equal, this would have to be a $\ZZ_2$ subgroup, but this is inconsistent with the fusion rules of the $\Pi^{r}_{i}$, so there must be two different components in the restriction of $\Pi^{r}_{i}$ and with the vacuum, these form the group $\ZZ_3$ under fusion. From the fusion rules for $\Pi^{s}_{1}$ and $\Pi^{s}_{\gamma}$ we now read off that the three components of the restriction of these fields must also be distinct. This means in particular that at least one of them is a component that does not occur in the $\ZZ_3$ that we already uncovered. But since we are looking for a group with a $\ZZ_3$ subgroup that has at least $4$ and most $6$ elements, we now have only two possibilities left for the entire group of fields of the broken theory: it is isomorphic either to $\ZZ_6$ or to $D_3$ itself. In the first case, the $\ZZ_3$ would be the subgroup of even elements of $\ZZ_6$ and in the second case, it would be the $\ZZ_3$ subgroup of rotations (or three-cycles in the permutation representation of $D_3$). It turns out that both options are fully consistent with the fusion rules of the original theory. Also, both lead to the same unconfined theory, namely the trivial theory with one sector (in other words, all sectors of the broken theory save the vacuum sector are confined). However, from basic intuition about discrete gauge theory or toric code models, as well as from our formalism for theories with integer quantum dimensions based on Hopf algebra theory\cite{BSS03}, we know that the correct broken (but confined) theory should be the one whose sectors fuse according to the group multiplication of $D_3$. In other words, the $\ZZ_6$ is the spurious solution to the requirements posed in this paper that we already announced. The fact that we find a theory with fusion rules described by the group multiplication of $D_3$ is also interesting in itself, since it shows that our formalism can produce boundary theories which have non-Abelian fusion rules.   

Looking at the situation from a CFT perspective, it appears that we have fully \lq unorbifolded' the theory this time and this might lead one to conjecture that quantum group symmetry breaking accomplishes a sort of \lq inverse' of orbifolding. However, this seems premature and we expect that a more complicated picture emerges when one studies different condensates and orbifolds of topologically nontrivial CFTs.

\section{Other Constructions}
\label{sec:new_constructions}

One may envision many constructions of new topological field theories and corresponding conformal field theories based on the principle of quantum group symmetry breaking. Perhaps the simplest thing one may do is tensor a number of known TQFTs or CFTs together in such a way that the tensor product theory has some bosonic sectors and then condense some or all of these bosonic sectors. The coset construction is of course a special case of such a construction, but more generally we don't have to require that the tensor product is of the form $G_{k}\otimes \overline{H_{k'}}$ with $H_{k'}\subset G_{k}$. It is not difficult to come up with simple examples of such theories which are not cosets. 

An example of potential interest in the quantum-Hall context corresponds to the product $Ising \times \mathcal{M}(4,5)$, where $\mathcal{M}(4,5)$ is the unitary minimal model at $c=7/10$. After condensation of the single nontrivial simple current in this theory we obtain precisely the spins and fusion rules of the $SU(3)_2$ parafermions. This may be connected with the interface or transition between the spin-polarized Moore-Read state, which is based on the Ising model, and a non-abelian spin singlet state (NASS) proposed by Schoutens and Ardonne\cite{eddyenkjs}, based on the $SU(3)_2$ parafermionic CFT\cite{Grosfeld2008}. We will return to this in detail elsewhere\cite{BaisWIP}.

There are many other examples one can think of, e.g.~one may take $SU(2)_k\otimes SU(2)_{k+2}$ and condense the bosonic simple current $(k,k+2)$, or alternatively, one may take $SU(2)_{k+4}\otimes \overline{SU(2)_{k}}$ and condense the bosonic simple current $(k+4,k)$. It is often not at all obvious what the conformal field theories corresponding to these constructions should look like. For example the case $SU(2)_6\otimes \overline{SU(2)_2}$ yields topological central charge $c=\frac{3}{4}$, but if there is a corrresponding unitary CFT, then its conformal central charge cannot be equal to $\frac{3}{4}$, since there is no unitary minimal model with this central charge, and it must differ from $\frac{3}{4}$ by some multiple of $8$. On the other hand, we do expect, in analogy to the Chern-Simons description of coset theories\cite{mszoo}, that the topological data for such theories should be described by a Chern-Simons theory whose gauge group is a quotient of the product of the groups appearing in the construction, with Chern-Simons terms for these groups at the appropriate levels. The quotient would be by a finite group characterizing the simple currents which are condensed (for situations where the condensed sectors are not all simple currents, the situation may be more complicted). If this conjectured Chern Simons description is correct, then this suggests that a CFT with the same topological order can be obtained as a boundary theory of this Chern-Simons theory.

\subsection{Doubled Chern-Simons theories}
\label{sec:doubleC-S}

Many constructions of this type could start from products of the form $G_{k}\otimes \overline{G_{k}}$, which are just doubled Chern-Simons theories, or more generally from products $\MA\otimes\overline{\MA}$ where $\MA$ represents a TQFT which is not of the $G_{k}$ type. Such theories are important in the description of string net condensates\cite{Levin05a} and picture TQFTs\cite{freedman08}. As we discussed in section~\ref{sec:bosons}, these theories all have bosons, namely the \lq diagonal' sectors with labels of the form $(\Lambda, \Lambda)$. We should be able to produce many new theories by condensing some of these bosons. 

If we condense all diagonal fields, then we should expect that the broken phase is topologically trivial, while the boundary between the broken and unbroken phases will be described by $G_{k}$ (or more generally $\MA$) itself.  We give an intuitive argument for this first and then sketch a proof. 

Intuitively, condensing all bosons in the theory should implement the coset construction which in this case gives the completely trivial theory. The boundary between the completely trivial theory and the $\MA\otimes \overline{\MA}$ theory should be described by an $\MA$ or an $\overline{\MA}$ theory according to the \lq folding' principle for boundaries  that we introduced in section~\ref{sec:embeddings2}. That is, we can think of the $\MA\otimes\overline{\MA}$-theory as just a $\MA$-theory on a plane which has been folded over to give two layers. The boundary between this folded plane and empty space is not really a boundary on the plane before folding and so it should have the same excitations as the plane itself, in other words, it should be described by an $\MA$ theory or an $\overline{\MA}$ theory. The $\MA$ and $\overline{\MA}$ theories of course have the same fusion rules but opposite spins. The ambiguity in the spins explains the complete confinement of these boundary excitations. One might expect less severe confinement if the $\MA$ theory itself has bosons.

%Another good feature of the conjecture is that the total quantum dimension is reduced by a factor $\MD_{\MA}$ in symmetry breaking and again by a factor $\MD_{\MA}$ in the confinement step, in accordance with the general behaviour of $\MD$ we have observed.  

Now let us give an argument that comes closer to a proof. First of all consider the fusions $(\Lambda,0)\times(0,\Lambda)=(\Lambda,\Lambda)$. Since the right hand side has the vacuum in its restriction, we see that the restriction of $(0,\Lambda)$ must be identified with the restriction of $(\bar{\Lambda},0)$. Now more generally, we have that $(\Lambda_1,\Lambda_2)=(\Lambda_1,0)\times(0,\Lambda_2)\equiv (\Lambda_1\times\overline{\Lambda_2},0)$ and hence the restrictions of all sectors can be written in terms of the restrictions of the sectors $(\Lambda,0)$ alone, confirming that the broken theory before confinement is simply a single copy of the $\MA$ theory itself. Now let us consider confinement. In order for the restriction of $(\Lambda,0)$ to be unconfined, we must have $\theta_{\Lambda}=\theta_{(\Lambda,0)}=\theta_{(0,\bar{\Lambda})}=\bar{\theta}_{\Lambda}$. Hence unconfined particles have trivial spin. In fact, unconfined particles should satisfy much more stringent conditions. If $(\Lambda,0)$ is not confined, then for any $\Lambda_1,\Lambda_2$ such that $\Lambda\in \Lambda_1\times\overline{\Lambda_2}$, we must have $\theta_{\Lambda_1}\bar{\theta}_{\Lambda_2}=\theta_{\Lambda}=1$. However, from this, it follows also that for any $\Lambda_1$, $\Lambda_2$ such that $\Lambda_2\in \Lambda\times\Lambda_1$, we have $\theta_{\Lambda}\theta_{\Lambda_1}\bar{\theta}_{\Lambda_2}=1$ and using the ribbon equation, we see that $\Lambda$ must have totally trivial monodromy with all other fields in the $\MA$ theory. However, this can only happen if the tensor category describing $\MA$ is not modular (see Ref.~\onlinecite{Kitaev06a}, section E.5), so we have argued that there is indeed complete confinement if $\MA$ is modular. 

Perhaps the simplest example of condensation in a doubled theory which does not lead to complete confinement occurs in the doubled Ising model. We label the sectors of the Ising model in the usal way by $1$, $\sigma$ and $\epsilon$. Their spins and quantum dimensions are given in table~\ref{tab:su22}. The $Ising\times\overline{Ising}$ theory has three bosonic fields, the vacuum $(1,1)$ and the diagonal fields $(\sigma,\sigma)$ and $(\epsilon,\epsilon)$. It is possible to have condensation of $(\epsilon,\epsilon)$ without condensing $(\sigma,\sigma)$. This leads to the branching in table~\ref{tab:doubled_ising_broken}. The broken theory has $6$ sectors and its fusion is the same as that of an $Ising\otimes\ZZ_2$ tensor product theory. There are four simple currents,  $(1,1)$, $(\epsilon,1)$ and the two components of the restriction of $(\sigma,\sigma)$. These are also the unconfined fields and they form a $\ZZ_2\times\ZZ_2$ group under fusion. This is actually not completely straightforward to derive since the requirements for symmetry breaking we have stated in this paper would also be consistent with $\ZZ_4$ fusion rules. However, one may check that the values of the spins of these fields are consistent only with $\ZZ_2\times\ZZ_2$ fusion rules and in fact, from the spins and the fusion we see that the unconfined theory has precisely the topological order of a $\ZZ_2$ discrete gauge theory or toric code model. 

\begin{table}[hbt]
\begin{center}
$\begin{array}{|l|}
\hline Ising\times \overline{Ising} {\rm ~broken}\\
\hline
\hline
\begin{array}{ll}
(1,1),(\epsilon,\epsilon)\rightarrow (1,1)~& 
(\sigma,\epsilon),(\sigma,1)\rightarrow (\sigma,1)\\
(1,\epsilon),(\epsilon,1)\rightarrow (\epsilon,1)~& 
(1,\sigma),(\epsilon,\sigma)\rightarrow (1,\sigma)\\
(\sigma,\sigma)\rightarrow (\sigma,\sigma)_1+(\sigma,\sigma)_2~& ~
\end{array}\\
\hline
\end{array}$
\end{center}
\caption{\footnotesize Branching rules for $Ising\times \overline{Ising}$ after condensation in the $(\epsilon,\epsilon)$-sector. The fusion rules for the broken theory are of $Ising\times \ZZ_2$ type. The unconfined fields are $(1,1)$, $(\epsilon,1)$, $(\sigma,\sigma)_1$ and $(\sigma,\sigma)_1$. The unbroken theory has the same topological order as the $\ZZ_2$ discrete gauge theory or the $\ZZ_2$ toric code model.} 
\label{tab:doubled_ising_broken}
\end{table}

This suggests that in any local model which realizes the doubled Ising model one might expect a transition to an Abelian topological phase of $\ZZ_2\times\ZZ_2$ type. Loop gases with ground states reflecting topological order of doubled Ising type have been constructed\cite{Freedman05a,Freedman05b}, but it has been shown recently\cite{Troyer08} that these loop states cannot be ground states of a gapped local Hamiltonian \footnote{At least not with the conventional inner product on the Hilbert space of the loop models, otherwise there are new possibilities\cite{fendley08}.}. They are in fact associated with gapless critical points and it is possible to drive the models away from these critical points and into in a gapped Abelian topological phase with the same topological order as the $\ZZ_2$ toric code or $\ZZ_2$ discrete gauge theory. It would be interesting to study if this can be viewed as due to the condensation of bosonic excitations of $(\epsilon,\epsilon)$ type.

%{\em Something about new conformal embeddings and possibly even constructions of TQFTs that don't have a CFT corresponding to them (whatever that means). Examples are $SU(2)_{k+4m}\times \overline{SU(2)_{k}}$. Easy for $k$ odd, more interesting for $k$ even, for example have $c=\frac{3}{4}$ at $k=2$. Also maybe do $SU(2)_2\times SU(2)_2$ with condensation of $(2,2)$ (this goes)}

\section{Summary and Outlook}

We have given some simple principles and requirements relating the spectra and topological interactions of topological phases that may be obtained from one another by condensation of a bosonic quasiparticle, based on the the idea of quantum group symmetry breaking. These turn out to be surprisingly powerful and practical tools in determining the topological field theory describing the condensed phase from the TQFT that describes the phase without condensate and also in describing the boundaries between condensed and uncondensed topological phases. We have worked out a number of examples in detail and shown connections between our quantum group symmetry breaking scheme and various constructions in conformal field theory. 

Future developments should include 
\begin{itemize}
\item More detailed study of systems which are or may soon be accessible by experiment, notably the various proposed non-Abelian fractional quantum Hall states.  
\item Explicit realization of the predictions on phase transitions in topological models that we make here in local models that exhibit topological phases. One aspect of this would be the introduction of (necessarily non-local) order parameters which signal a non-zero condensate density\cite{baisromers2008}. String-net condensed phases would be a good laboratory for this. 
\item A more mathematically rigorous treatment of the material presented here. This would hopefully allow for the systematic construction of the full unitary braided tensor category describing the condensed theory. Also one would like to prove some of the observations on $c$ and $\MD$ in section~\ref{sec:observations}. 
\item More in-depth treatment of the relation between quantum group symmetry breaking and chiral algebra extension, i.e.~explicit investigation of the action of the chiral algebra on the Hilbert space of the CFT before and after condensation. Conformal embeddings would be the obvious place to start such a program, but it would be especially interesting if one could find a natural CFT counterpart for some of the TQFT constructions mentioned in section~\ref{sec:new_constructions}. 
\end{itemize}

\vspace*{3mm}
{\small
\setlength{\baselineskip}{0.9\baselineskip}
\setlength{\itemsep}{0.5\itemsep}
\vbox{
\noindent{\bf Acknowledgements}\vspace*{1mm}

\noindent
The authors would like to thank E.~Ardonne, A.~Kitaev, E.~Fradkin, N.~Read, P.~Bonderson and Z.~Wang for  illuminating discussions. F.A.B.~would like to acknowledge the hospitality of the Microsoft Project Q and KITP during the 2006 workshop \lq Topological Phases and Quantum Computation'.}}
%\end{acknowledgments}

%\bibliographystyle{apsrev}

%\bibliographystyle{unsrt}

%\bibliography{hallbreak}

\begin{thebibliography}{104}
\expandafter\ifx\csname natexlab\endcsname\relax\def\natexlab#1{#1}\fi
\expandafter\ifx\csname bibnamefont\endcsname\relax
  \def\bibnamefont#1{#1}\fi
\expandafter\ifx\csname bibfnamefont\endcsname\relax
  \def\bibfnamefont#1{#1}\fi
\expandafter\ifx\csname citenamefont\endcsname\relax
  \def\citenamefont#1{#1}\fi
\expandafter\ifx\csname url\endcsname\relax
  \def\url#1{\texttt{#1}}\fi
\expandafter\ifx\csname urlprefix\endcsname\relax\def\urlprefix{URL }\fi
\providecommand{\bibinfo}[2]{#2}
\providecommand{\eprint}[2][]{\url{#2}}

\bibitem[{\citenamefont{Kitaev}(2003)}]{Kitaev03}
\bibinfo{author}{\bibfnamefont{A.~Y.} \bibnamefont{Kitaev}},
  \bibinfo{journal}{Ann. Phys.} \textbf{\bibinfo{volume}{303}},
  \bibinfo{pages}{2} (\bibinfo{year}{2003}),
  \bibinfo{note}{\texttt{quant-ph/9707021}}.

\bibitem[{\citenamefont{Freedman}(1998)}]{Freedman98}
\bibinfo{author}{\bibfnamefont{M.~H.} \bibnamefont{Freedman}},
  \bibinfo{journal}{Proc. Natl. Acad. Sci. USA} \textbf{\bibinfo{volume}{95}},
  \bibinfo{pages}{98} (\bibinfo{year}{1998}).

\bibitem[{\citenamefont{Nayak et~al.}(2007)\citenamefont{Nayak, Simon, Stern,
  Freedman, and Sarma}}]{Nayak2007}
\bibinfo{author}{\bibfnamefont{C.}~\bibnamefont{Nayak}},
  \bibinfo{author}{\bibfnamefont{S.~H.} \bibnamefont{Simon}},
  \bibinfo{author}{\bibfnamefont{A.}~\bibnamefont{Stern}},
  \bibinfo{author}{\bibfnamefont{M.}~\bibnamefont{Freedman}}, \bibnamefont{and}
  \bibinfo{author}{\bibfnamefont{S.~D.} \bibnamefont{Sarma}},
  \emph{\bibinfo{title}{Non-abelian anyons and topological quantum
  computation}} (\bibinfo{year}{2007}),
  \bibinfo{note}{\texttt{arXiv:0707.1889}}.

\bibitem[{\citenamefont{Dolev et~al.}(2008)\citenamefont{Dolev, Heiblum,
  Umansky, Stern, and Mahalu}}]{Dolev2008}
\bibinfo{author}{\bibfnamefont{M.}~\bibnamefont{Dolev}},
  \bibinfo{author}{\bibfnamefont{M.}~\bibnamefont{Heiblum}},
  \bibinfo{author}{\bibfnamefont{V.}~\bibnamefont{Umansky}},
  \bibinfo{author}{\bibfnamefont{A.}~\bibnamefont{Stern}}, \bibnamefont{and}
  \bibinfo{author}{\bibfnamefont{D.}~\bibnamefont{Mahalu}},
  \emph{\bibinfo{title}{Towards identification of a non-abelian state:
  observation of a quarter of electron charge at $\nu=5/2$ quantum {H}all
  state}} (\bibinfo{year}{2008}), \bibinfo{note}{{\tt arXiv:0802.0930}}.

\bibitem[{\citenamefont{Radu et~al.}(2008)\citenamefont{Radu, Miller, Marcus,
  Kastner, Pfeiffer, and West}}]{Radu2008}
\bibinfo{author}{\bibfnamefont{I.~P.} \bibnamefont{Radu}},
  \bibinfo{author}{\bibfnamefont{J.~B.} \bibnamefont{Miller}},
  \bibinfo{author}{\bibfnamefont{C.~M.} \bibnamefont{Marcus}},
  \bibinfo{author}{\bibfnamefont{M.~A.} \bibnamefont{Kastner}},
  \bibinfo{author}{\bibfnamefont{L.~N.} \bibnamefont{Pfeiffer}},
  \bibnamefont{and} \bibinfo{author}{\bibfnamefont{K.~W.} \bibnamefont{West}},
  \emph{\bibinfo{title}{Quasiparticle tunneling in the fractional quantum
  {H}all state at $\nu = 5/2$}} (\bibinfo{year}{2008}), \bibinfo{note}{{\tt
  arXiv:0803.3530}}.

\bibitem[{\citenamefont{Willett et~al.}(2008)\citenamefont{Willett, Manfra,
  Pfeiffer, and West}}]{Willett2008}
\bibinfo{author}{\bibfnamefont{R.~L.} \bibnamefont{Willett}},
  \bibinfo{author}{\bibfnamefont{M.~J.} \bibnamefont{Manfra}},
  \bibinfo{author}{\bibfnamefont{L.~N.} \bibnamefont{Pfeiffer}},
  \bibnamefont{and} \bibinfo{author}{\bibfnamefont{K.~W.} \bibnamefont{West}},
  \emph{\bibinfo{title}{Interferometric measurement of filling factor 5/2
  quasiparticle charge}} (\bibinfo{year}{2008}),
  \bibinfo{note}{\texttt{arXiv.org:0807.0221}}.

\bibitem[{\citenamefont{Godfrey et~al.}(2007)\citenamefont{Godfrey, Jiang,
  Kang, Simon, Baldwin, Pfeiffer, and West}}]{Godfrey2007}
\bibinfo{author}{\bibfnamefont{M.~D.} \bibnamefont{Godfrey}},
  \bibinfo{author}{\bibfnamefont{P.}~\bibnamefont{Jiang}},
  \bibinfo{author}{\bibfnamefont{W.}~\bibnamefont{Kang}},
  \bibinfo{author}{\bibfnamefont{S.~H.} \bibnamefont{Simon}},
  \bibinfo{author}{\bibfnamefont{K.~W.} \bibnamefont{Baldwin}},
  \bibinfo{author}{\bibfnamefont{L.~N.} \bibnamefont{Pfeiffer}},
  \bibnamefont{and} \bibinfo{author}{\bibfnamefont{K.~W.} \bibnamefont{West}},
  \emph{\bibinfo{title}{{A}haronov-{B}ohm-like oscillations in quantum {H}all
  corrals}} (\bibinfo{year}{2007}),
  \bibinfo{note}{\texttt{arXiv.org:0708.2448}}.

\bibitem[{\citenamefont{Gladchenko et~al.}(2008)\citenamefont{Gladchenko,
  Olaya, Dupont-Ferrier, Doucot, Ioffe, and Gershenson}}]{Gladchenko2008}
\bibinfo{author}{\bibfnamefont{S.}~\bibnamefont{Gladchenko}},
  \bibinfo{author}{\bibfnamefont{D.}~\bibnamefont{Olaya}},
  \bibinfo{author}{\bibfnamefont{E.}~\bibnamefont{Dupont-Ferrier}},
  \bibinfo{author}{\bibfnamefont{B.}~\bibnamefont{Doucot}},
  \bibinfo{author}{\bibfnamefont{L.~B.} \bibnamefont{Ioffe}}, \bibnamefont{and}
  \bibinfo{author}{\bibfnamefont{M.~E.} \bibnamefont{Gershenson}},
  \emph{\bibinfo{title}{Superconducting nanocircuits for topologically
  protected qubits}} (\bibinfo{year}{2008}),
  \bibinfo{note}{\texttt{arXiv:0802.2295}}.

\bibitem[{\citenamefont{Einarsson}(1990)}]{Einarsson90}
\bibinfo{author}{\bibfnamefont{T.}~\bibnamefont{Einarsson}},
  \bibinfo{journal}{Phys. Rev. Lett.} \textbf{\bibinfo{volume}{64}},
  \bibinfo{pages}{1995} (\bibinfo{year}{1990}).

\bibitem[{\citenamefont{Wen and Niu}(1990)}]{Wen90}
\bibinfo{author}{\bibfnamefont{X.~G.} \bibnamefont{Wen}} \bibnamefont{and}
  \bibinfo{author}{\bibfnamefont{Q.}~\bibnamefont{Niu}},
  \bibinfo{journal}{Phys. Rev. B} \textbf{\bibinfo{volume}{41}},
  \bibinfo{pages}{9377} (\bibinfo{year}{1990}).

\bibitem[{\citenamefont{Levin and Wen}(2006)}]{Levin06}
\bibinfo{author}{\bibfnamefont{M.}~\bibnamefont{Levin}} \bibnamefont{and}
  \bibinfo{author}{\bibfnamefont{X.-G.} \bibnamefont{Wen}},
  \bibinfo{journal}{Phys.~Rev.~Lett.} \textbf{\bibinfo{volume}{96}},
  \bibinfo{pages}{110405} (\bibinfo{year}{2006}).

\bibitem[{\citenamefont{Kitaev and Preskill}(2006)}]{Kitaev06}
\bibinfo{author}{\bibfnamefont{A.}~\bibnamefont{Kitaev}} \bibnamefont{and}
  \bibinfo{author}{\bibfnamefont{J.}~\bibnamefont{Preskill}},
  \bibinfo{journal}{Phys.~Rev.~Lett.} \textbf{\bibinfo{volume}{96}},
  \bibinfo{pages}{110404} (\bibinfo{year}{2006}).

\bibitem[{\citenamefont{Elitzur}(1975)}]{Elitzur1975}
\bibinfo{author}{\bibfnamefont{S.}~\bibnamefont{Elitzur}},
  \bibinfo{journal}{Phys. Rev. D} \textbf{\bibinfo{volume}{12}},
  \bibinfo{pages}{3978} (\bibinfo{year}{1975}).

\bibitem[{\citenamefont{{`}t Hooft}(1978)}]{thooft:1977}
\bibinfo{author}{\bibfnamefont{G.}~\bibnamefont{{`}t Hooft}},
  \bibinfo{journal}{Nucl. Phys.} \textbf{\bibinfo{volume}{B138}},
  \bibinfo{pages}{1} (\bibinfo{year}{1978}).

\bibitem[{\citenamefont{Bais et~al.}(2002)\citenamefont{Bais, Schroers, and
  Slingerland}}]{BSS02}
\bibinfo{author}{\bibfnamefont{F.~A.} \bibnamefont{Bais}},
  \bibinfo{author}{\bibfnamefont{B.~J.} \bibnamefont{Schroers}},
  \bibnamefont{and} \bibinfo{author}{\bibfnamefont{J.~K.}
  \bibnamefont{Slingerland}}, \bibinfo{journal}{Phys.~Rev.~Lett.}
  \textbf{\bibinfo{volume}{89}}, \bibinfo{pages}{181601}
  (\bibinfo{year}{2002}).

\bibitem[{\citenamefont{Bais et~al.}(2003)\citenamefont{Bais, Schroers, and
  Slingerland}}]{BSS03}
\bibinfo{author}{\bibfnamefont{F.~A.} \bibnamefont{Bais}},
  \bibinfo{author}{\bibfnamefont{B.~J.} \bibnamefont{Schroers}},
  \bibnamefont{and} \bibinfo{author}{\bibfnamefont{J.~K.}
  \bibnamefont{Slingerland}}, \bibinfo{journal}{JHEP}
  \textbf{\bibinfo{volume}{0305}}, \bibinfo{pages}{068} (\bibinfo{year}{2003}).

\bibitem[{\citenamefont{Bais and Mathy}(2006)}]{baismathy06a}
\bibinfo{author}{\bibfnamefont{F.~A.} \bibnamefont{Bais}} \bibnamefont{and}
  \bibinfo{author}{\bibfnamefont{C.~J.~M.} \bibnamefont{Mathy}},
  \bibinfo{journal}{Phys.~Rev.~B} \textbf{\bibinfo{volume}{73}},
  \bibinfo{pages}{224120} (\bibinfo{year}{2006}), \bibinfo{note}{{\tt
  cond-mat/0602101}}.

\bibitem[{\citenamefont{Mathy and Bais}(2007)}]{baismathy06b}
\bibinfo{author}{\bibfnamefont{C.~J.~M.} \bibnamefont{Mathy}} \bibnamefont{and}
  \bibinfo{author}{\bibfnamefont{F.~A.} \bibnamefont{Bais}},
  \bibinfo{journal}{Ann. Phys.} \textbf{\bibinfo{volume}{322}},
  \bibinfo{pages}{709} (\bibinfo{year}{2007}), \bibinfo{note}{{\tt
  cond-mat/0602109}}.

\bibitem[{\citenamefont{Bais and Mathy}(2007)}]{baismathy06c}
\bibinfo{author}{\bibfnamefont{F.~A.} \bibnamefont{Bais}} \bibnamefont{and}
  \bibinfo{author}{\bibfnamefont{C.~J.~M.} \bibnamefont{Mathy}},
  \bibinfo{journal}{Ann. Phys.} \textbf{\bibinfo{volume}{322}},
  \bibinfo{pages}{552} (\bibinfo{year}{2007}), \bibinfo{note}{{\tt
  cond-mat/0602115}}.

\bibitem[{\citenamefont{Bombin and Martin-Delgado}(2007)}]{Bombin2007}
\bibinfo{author}{\bibfnamefont{H.}~\bibnamefont{Bombin}} \bibnamefont{and}
  \bibinfo{author}{\bibfnamefont{M.}~\bibnamefont{Martin-Delgado}},
  \emph{\bibinfo{title}{A family of non-{A}belian {K}itaev models on a lattice:
  Topological confinement and condensation}} (\bibinfo{year}{2007}),
  \bibinfo{note}{{\tt arXiv:0712.0190}}.

\bibitem[{\citenamefont{Bombin and Martin-Delgado}(2008)}]{Bombin2008}
\bibinfo{author}{\bibfnamefont{H.}~\bibnamefont{Bombin}} \bibnamefont{and}
  \bibinfo{author}{\bibfnamefont{M.}~\bibnamefont{Martin-Delgado}},
  \emph{\bibinfo{title}{Nested topological order}} (\bibinfo{year}{2008}),
  \bibinfo{note}{{\tt arXiv:0803.4299}}.

\bibitem[{\citenamefont{Kitaev}(2006)}]{Kitaev06a}
\bibinfo{author}{\bibfnamefont{A.}~\bibnamefont{Kitaev}},
  \bibinfo{journal}{Ann. Phys.} \textbf{\bibinfo{volume}{321}},
  \bibinfo{pages}{2} (\bibinfo{year}{2006}), \eprint{{\tt cond-mat/0506438}}.

\bibitem[{\citenamefont{Preskill}(2004)}]{Preskill-lectures}
\bibinfo{author}{\bibfnamefont{J.}~\bibnamefont{Preskill}}
  (\bibinfo{year}{2004}), \bibinfo{note}{lecture notes},
  \urlprefix\url{http://www.theory.caltech.edu/~preskill/ph219/topological.ps}.

\bibitem[{\citenamefont{Turaev}(1994)}]{Turaev94}
\bibinfo{author}{\bibfnamefont{V.~G.} \bibnamefont{Turaev}},
  \emph{\bibinfo{title}{Quantum Invariants of Knots and 3-Manifolds}}
  (\bibinfo{publisher}{Walter de Gruyter}, \bibinfo{address}{Berlin, New York},
  \bibinfo{year}{1994}).

\bibitem[{\citenamefont{Kassel}(1995)}]{Kassel95}
\bibinfo{author}{\bibfnamefont{C.}~\bibnamefont{Kassel}},
  \emph{\bibinfo{title}{Quantum Groups}} (\bibinfo{publisher}{Springer-Verlag},
  \bibinfo{address}{New York, Berlin, Heidelberg}, \bibinfo{year}{1995}).

\bibitem[{\citenamefont{Vafa}(1988)}]{Vafa1988}
\bibinfo{author}{\bibfnamefont{C.}~\bibnamefont{Vafa}}, \bibinfo{journal}{Phys.
  Lett.} \textbf{\bibinfo{volume}{B206}}, \bibinfo{pages}{421}
  (\bibinfo{year}{1988}).

\bibitem[{\citenamefont{Di~Francesco et~al.}(1997)\citenamefont{Di~Francesco,
  Mathieu, and S\'en\'echal}}]{dsm}
\bibinfo{author}{\bibfnamefont{F.}~\bibnamefont{Di~Francesco}},
  \bibinfo{author}{\bibfnamefont{P.}~\bibnamefont{Mathieu}}, \bibnamefont{and}
  \bibinfo{author}{\bibfnamefont{D.}~\bibnamefont{S\'en\'echal}},
  \emph{\bibinfo{title}{Conformal field theory}}
  (\bibinfo{publisher}{Springer}, \bibinfo{year}{1997}).

\bibitem[{\citenamefont{Fradkin et~al.}(1998)\citenamefont{Fradkin, Nayak,
  Tsvelik, and Wilczek}}]{Fradkin98}
\bibinfo{author}{\bibfnamefont{E.}~\bibnamefont{Fradkin}},
  \bibinfo{author}{\bibfnamefont{C.}~\bibnamefont{Nayak}},
  \bibinfo{author}{\bibfnamefont{A.}~\bibnamefont{Tsvelik}}, \bibnamefont{and}
  \bibinfo{author}{\bibfnamefont{F.}~\bibnamefont{Wilczek}},
  \bibinfo{journal}{Nucl. Phys. B} \textbf{\bibinfo{volume}{516}},
  \bibinfo{pages}{704} (\bibinfo{year}{1998}), \bibinfo{note}{{\tt
  cond-mat/9711087}}.

\bibitem[{\citenamefont{Das~Sarma et~al.}(2005)\citenamefont{Das~Sarma,
  Freedman, and Nayak}}]{DasSarma05}
\bibinfo{author}{\bibfnamefont{S.}~\bibnamefont{Das~Sarma}},
  \bibinfo{author}{\bibfnamefont{M.}~\bibnamefont{Freedman}}, \bibnamefont{and}
  \bibinfo{author}{\bibfnamefont{C.}~\bibnamefont{Nayak}},
  \bibinfo{journal}{Phys. Rev. Lett.} \textbf{\bibinfo{volume}{94}},
  \bibinfo{pages}{166802} (\bibinfo{year}{2005}), \bibinfo{note}{{\tt
  cond-mat/0412343}}.

\bibitem[{\citenamefont{Stern and Halperin}(2006)}]{Stern06a}
\bibinfo{author}{\bibfnamefont{A.}~\bibnamefont{Stern}} \bibnamefont{and}
  \bibinfo{author}{\bibfnamefont{B.~I.} \bibnamefont{Halperin}},
  \bibinfo{journal}{Phys. Rev. Lett.} \textbf{\bibinfo{volume}{96}},
  \bibinfo{pages}{016802} (\bibinfo{year}{2006}),
  \bibinfo{note}{\texttt{cond-mat/0508447}}.

\bibitem[{\citenamefont{Bonderson
  et~al.}(2006{\natexlab{a}})\citenamefont{Bonderson, Kitaev, and
  Shtengel}}]{Bonderson06a}
\bibinfo{author}{\bibfnamefont{P.}~\bibnamefont{Bonderson}},
  \bibinfo{author}{\bibfnamefont{A.}~\bibnamefont{Kitaev}}, \bibnamefont{and}
  \bibinfo{author}{\bibfnamefont{K.}~\bibnamefont{Shtengel}},
  \bibinfo{journal}{Phys. Rev. Lett.} \textbf{\bibinfo{volume}{96}},
  \bibinfo{pages}{016803} (\bibinfo{year}{2006}{\natexlab{a}}),
  \bibinfo{note}{\texttt{cond-mat/0508616}}.

\bibitem[{\citenamefont{Bonderson
  et~al.}(2006{\natexlab{b}})\citenamefont{Bonderson, Shtengel, and
  Slingerland}}]{Bonderson06b}
\bibinfo{author}{\bibfnamefont{P.}~\bibnamefont{Bonderson}},
  \bibinfo{author}{\bibfnamefont{K.}~\bibnamefont{Shtengel}}, \bibnamefont{and}
  \bibinfo{author}{\bibfnamefont{J.~K.} \bibnamefont{Slingerland}},
  \bibinfo{journal}{Phys. Rev. Lett.} \textbf{\bibinfo{volume}{97}},
  \bibinfo{pages}{016401} (\bibinfo{year}{2006}{\natexlab{b}}),
  \bibinfo{note}{\texttt{cond-mat/0601242}}.

\bibitem[{\citenamefont{Bonderson
  et~al.}(2006{\natexlab{c}})\citenamefont{Bonderson, Shtengel, and
  Slingerland}}]{Bonderson07a}
\bibinfo{author}{\bibfnamefont{P.}~\bibnamefont{Bonderson}},
  \bibinfo{author}{\bibfnamefont{K.}~\bibnamefont{Shtengel}}, \bibnamefont{and}
  \bibinfo{author}{\bibfnamefont{J.~K.} \bibnamefont{Slingerland}},
  \bibinfo{journal}{Phys. Rev. Lett.} \textbf{\bibinfo{volume}{98}},
  \bibinfo{pages}{070401} (\bibinfo{year}{2006}{\natexlab{c}}),
  \bibinfo{note}{\texttt{quant-ph/0608119}}.

\bibitem[{\citenamefont{Bonderson et~al.}(2007)\citenamefont{Bonderson,
  Shtengel, and Slingerland}}]{Bonderson07b}
\bibinfo{author}{\bibfnamefont{P.}~\bibnamefont{Bonderson}},
  \bibinfo{author}{\bibfnamefont{K.}~\bibnamefont{Shtengel}}, \bibnamefont{and}
  \bibinfo{author}{\bibfnamefont{J.~K.} \bibnamefont{Slingerland}},
  \emph{\bibinfo{title}{Interferometry of non-abelian anyons}}
  (\bibinfo{year}{2007}), \bibinfo{note}{\texttt{arXiv:0707.4206}}.

\bibitem[{\citenamefont{Chung and Stone}(2006)}]{Chung06}
\bibinfo{author}{\bibfnamefont{S.~B.} \bibnamefont{Chung}} \bibnamefont{and}
  \bibinfo{author}{\bibfnamefont{M.}~\bibnamefont{Stone}},
  \bibinfo{journal}{Phys. Rev. B} \textbf{\bibinfo{volume}{73}},
  \bibinfo{pages}{245311} (\bibinfo{year}{2006}), \bibinfo{note}{{\tt
  cond-mat/0601594}}.

\bibitem[{\citenamefont{Ilan et~al.}(2008)\citenamefont{Ilan, Grosfeld,
  Schoutens, and Stern}}]{Ilan2008}
\bibinfo{author}{\bibfnamefont{R.}~\bibnamefont{Ilan}},
  \bibinfo{author}{\bibfnamefont{E.}~\bibnamefont{Grosfeld}},
  \bibinfo{author}{\bibfnamefont{K.}~\bibnamefont{Schoutens}},
  \bibnamefont{and} \bibinfo{author}{\bibfnamefont{A.}~\bibnamefont{Stern}},
  \emph{\bibinfo{title}{Experimental signatures of non-abelian statistics in
  clustered quantum {H}all states}} (\bibinfo{year}{2008}),
  \bibinfo{note}{\texttt{arXiv:0803.1542}}.

\bibitem[{\citenamefont{de~C.~Chamon et~al.}(1997)\citenamefont{de~C.~Chamon,
  Freed, Kivelson, Sondhi, and Wen}}]{Chamon97}
\bibinfo{author}{\bibfnamefont{C.}~\bibnamefont{de~C.~Chamon}},
  \bibinfo{author}{\bibfnamefont{D.~E.} \bibnamefont{Freed}},
  \bibinfo{author}{\bibfnamefont{S.~A.} \bibnamefont{Kivelson}},
  \bibinfo{author}{\bibfnamefont{S.~L.} \bibnamefont{Sondhi}},
  \bibnamefont{and} \bibinfo{author}{\bibfnamefont{X.~G.} \bibnamefont{Wen}},
  \bibinfo{journal}{Phys. Rev. B} \textbf{\bibinfo{volume}{55}},
  \bibinfo{pages}{2331} (\bibinfo{year}{1997}), \bibinfo{note}{{\tt
  cond-mat/9607195}}.

\bibitem[{\citenamefont{Camino et~al.}(2005{\natexlab{a}})\citenamefont{Camino,
  Zhou, and Goldman}}]{Camino05a}
\bibinfo{author}{\bibfnamefont{F.~E.} \bibnamefont{Camino}},
  \bibinfo{author}{\bibfnamefont{W.}~\bibnamefont{Zhou}}, \bibnamefont{and}
  \bibinfo{author}{\bibfnamefont{V.~J.} \bibnamefont{Goldman}},
  \bibinfo{journal}{Phys. Rev. B} \textbf{\bibinfo{volume}{72}},
  \bibinfo{pages}{075342} (\bibinfo{year}{2005}{\natexlab{a}}),
  \eprint{cond-mat/0502406}.

\bibitem[{\citenamefont{Camino et~al.}(2005{\natexlab{b}})\citenamefont{Camino,
  Zhou, and Goldman}}]{Camino05b}
\bibinfo{author}{\bibfnamefont{F.~E.} \bibnamefont{Camino}},
  \bibinfo{author}{\bibfnamefont{W.}~\bibnamefont{Zhou}}, \bibnamefont{and}
  \bibinfo{author}{\bibfnamefont{V.~J.} \bibnamefont{Goldman}},
  \bibinfo{journal}{Phys. Rev. Lett.} \textbf{\bibinfo{volume}{95}},
  \bibinfo{pages}{246802} (\bibinfo{year}{2005}{\natexlab{b}}),
  \eprint{cond-mat/0504341}.

\bibitem[{\citenamefont{Kane and Fisher}(1992)}]{Kane1992}
\bibinfo{author}{\bibfnamefont{C.~L.} \bibnamefont{Kane}} \bibnamefont{and}
  \bibinfo{author}{\bibfnamefont{M.~P.~A.} \bibnamefont{Fisher}},
  \bibinfo{journal}{Phys. Rev. B} \textbf{\bibinfo{volume}{46}},
  \bibinfo{pages}{15233} (\bibinfo{year}{1992}).

\bibitem[{\citenamefont{Fendley
  et~al.}(1995{\natexlab{a}})\citenamefont{Fendley, Ludwig, and
  Saleur}}]{Fendley1994}
\bibinfo{author}{\bibfnamefont{P.}~\bibnamefont{Fendley}},
  \bibinfo{author}{\bibfnamefont{A.~W.~W.} \bibnamefont{Ludwig}},
  \bibnamefont{and} \bibinfo{author}{\bibfnamefont{H.}~\bibnamefont{Saleur}},
  \bibinfo{journal}{Phys.~Rev.~Lett.} \textbf{\bibinfo{volume}{74}},
  \bibinfo{pages}{3005} (\bibinfo{year}{1995}{\natexlab{a}}),
  \bibinfo{note}{\texttt{cond-mat/9408068}}.

\bibitem[{\citenamefont{Fendley
  et~al.}(1995{\natexlab{b}})\citenamefont{Fendley, Ludwig, and
  Saleur}}]{Fendley1995}
\bibinfo{author}{\bibfnamefont{P.}~\bibnamefont{Fendley}},
  \bibinfo{author}{\bibfnamefont{A.~W.~W.} \bibnamefont{Ludwig}},
  \bibnamefont{and} \bibinfo{author}{\bibfnamefont{H.}~\bibnamefont{Saleur}},
  \bibinfo{journal}{Phys.~Rev.~B} \textbf{\bibinfo{volume}{52}},
  \bibinfo{pages}{8934} (\bibinfo{year}{1995}{\natexlab{b}}),
  \bibinfo{note}{\texttt{cond-mat/9503172}}.

\bibitem[{\citenamefont{Bais and Slingerland}()}]{BaisWIP}
\bibinfo{author}{\bibfnamefont{F.~A.} \bibnamefont{Bais}} \bibnamefont{and}
  \bibinfo{author}{\bibfnamefont{J.~K.} \bibnamefont{Slingerland}},
  \bibinfo{note}{in preparation}.

\bibitem[{\citenamefont{Etingof et~al.}(2005)\citenamefont{Etingof, Nikshych,
  and Ostrik}}]{Etingof2005}
\bibinfo{author}{\bibfnamefont{P.}~\bibnamefont{Etingof}},
  \bibinfo{author}{\bibfnamefont{D.}~\bibnamefont{Nikshych}}, \bibnamefont{and}
  \bibinfo{author}{\bibfnamefont{V.}~\bibnamefont{Ostrik}},
  \bibinfo{journal}{Ann. Math.} \textbf{\bibinfo{volume}{162}},
  \bibinfo{pages}{581} (\bibinfo{year}{2005}).

\bibitem[{\citenamefont{Pauli}(1940)}]{Pauli40}
\bibinfo{author}{\bibfnamefont{W.}~\bibnamefont{Pauli}},
  \bibinfo{journal}{Phys. Rev.} \textbf{\bibinfo{volume}{58}},
  \bibinfo{pages}{716} (\bibinfo{year}{1940}).

\bibitem[{\citenamefont{Streater and Wightman}(2000)}]{StreaterandWightman}
\bibinfo{author}{\bibfnamefont{R.~F.} \bibnamefont{Streater}} \bibnamefont{and}
  \bibinfo{author}{\bibfnamefont{A.~S.} \bibnamefont{Wightman}},
  \emph{\bibinfo{title}{P{CT}, spin and statistics, and all that}}, Princeton
  Landmarks in Physics (\bibinfo{publisher}{Princeton University Press},
  \bibinfo{address}{Princeton, NJ}, \bibinfo{year}{2000}), ISBN
  \bibinfo{isbn}{0-691-07062-8}, \bibinfo{note}{corrected third printing of the
  1978 edition}.

\bibitem[{\citenamefont{Moore and Read}(1991)}]{mooreread}
\bibinfo{author}{\bibfnamefont{G.}~\bibnamefont{Moore}} \bibnamefont{and}
  \bibinfo{author}{\bibfnamefont{N.}~\bibnamefont{Read}},
  \bibinfo{journal}{Nucl. Phys.} \textbf{\bibinfo{volume}{B360}},
  \bibinfo{pages}{362} (\bibinfo{year}{1991}).

\bibitem[{\citenamefont{Read and Rezayi}(1999)}]{readrez}
\bibinfo{author}{\bibfnamefont{N.}~\bibnamefont{Read}} \bibnamefont{and}
  \bibinfo{author}{\bibfnamefont{E.}~\bibnamefont{Rezayi}},
  \bibinfo{journal}{Phys.Rev.} \textbf{\bibinfo{volume}{B59}},
  \bibinfo{pages}{8084} (\bibinfo{year}{1999}),
  \bibinfo{note}{\texttt{cond-mat/9809384}}.

\bibitem[{\citenamefont{Levin and Wen}(2005)}]{Levin05a}
\bibinfo{author}{\bibfnamefont{M.~A.} \bibnamefont{Levin}} \bibnamefont{and}
  \bibinfo{author}{\bibfnamefont{X.-G.} \bibnamefont{Wen}},
  \bibinfo{journal}{Phys. Rev. B} \textbf{\bibinfo{volume}{71}},
  \bibinfo{eid}{045110} (pages~\bibinfo{numpages}{21}) (\bibinfo{year}{2005}),
  \bibinfo{note}{\texttt{cond-mat/0404617}}.

\bibitem[{\citenamefont{Freedman et~al.}(2004)\citenamefont{Freedman, Nayak,
  Shtengel, Walker, and Wang}}]{Freedman04a}
\bibinfo{author}{\bibfnamefont{M.}~\bibnamefont{Freedman}},
  \bibinfo{author}{\bibfnamefont{C.}~\bibnamefont{Nayak}},
  \bibinfo{author}{\bibfnamefont{K.}~\bibnamefont{Shtengel}},
  \bibinfo{author}{\bibfnamefont{K.}~\bibnamefont{Walker}}, \bibnamefont{and}
  \bibinfo{author}{\bibfnamefont{Z.}~\bibnamefont{Wang}},
  \bibinfo{journal}{Ann. Phys.} \textbf{\bibinfo{volume}{310}},
  \bibinfo{pages}{428} (\bibinfo{year}{2004}), \eprint{{\tt cond-mat/0307511}}.

\bibitem[{\citenamefont{Freedman
  et~al.}(2005{\natexlab{a}})\citenamefont{Freedman, Nayak, and
  Shtengel}}]{Freedman05a}
\bibinfo{author}{\bibfnamefont{M.}~\bibnamefont{Freedman}},
  \bibinfo{author}{\bibfnamefont{C.}~\bibnamefont{Nayak}}, \bibnamefont{and}
  \bibinfo{author}{\bibfnamefont{K.}~\bibnamefont{Shtengel}},
  \bibinfo{journal}{Phys. Rev. Lett.} \textbf{\bibinfo{volume}{94}},
  \bibinfo{pages}{066401} (\bibinfo{year}{2005}{\natexlab{a}}), \eprint{{\tt
  cond-mat/0312273}}.

\bibitem[{\citenamefont{Freedman
  et~al.}(2005{\natexlab{b}})\citenamefont{Freedman, Nayak, and
  Shtengel}}]{Freedman05b}
\bibinfo{author}{\bibfnamefont{M.}~\bibnamefont{Freedman}},
  \bibinfo{author}{\bibfnamefont{C.}~\bibnamefont{Nayak}}, \bibnamefont{and}
  \bibinfo{author}{\bibfnamefont{K.}~\bibnamefont{Shtengel}},
  \bibinfo{journal}{Phys. Rev. Lett.} \textbf{\bibinfo{volume}{94}},
  \bibinfo{pages}{147205} (\bibinfo{year}{2005}{\natexlab{b}}), \eprint{{\tt
  cond-mat/0408257}}.

\bibitem[{\citenamefont{Fendley}(2008)}]{fendley08}
\bibinfo{author}{\bibfnamefont{P.}~\bibnamefont{Fendley}},
  \emph{\bibinfo{title}{Topological order from quantum loops and nets}}
  (\bibinfo{year}{2008}), \bibinfo{note}{\texttt{arXiv.org:0804.0625}}.

\bibitem[{\citenamefont{Bais et~al.}(1992)\citenamefont{Bais, van Driel, and
  de~Wild~Propitius}}]{dgt1}
\bibinfo{author}{\bibfnamefont{F.~A.} \bibnamefont{Bais}},
  \bibinfo{author}{\bibfnamefont{P.}~\bibnamefont{van Driel}},
  \bibnamefont{and}
  \bibinfo{author}{\bibfnamefont{M.}~\bibnamefont{de~Wild~Propitius}},
  \bibinfo{journal}{Phys. Lett.} \textbf{\bibinfo{volume}{B280}},
  \bibinfo{pages}{63} (\bibinfo{year}{1992}),
  \bibinfo{note}{\texttt{hep-th/9203046}}.

\bibitem[{\citenamefont{Bais et~al.}(1993{\natexlab{a}})\citenamefont{Bais, van
  Driel, and de~Wild~Propitius}}]{dgt2}
\bibinfo{author}{\bibfnamefont{F.~A.} \bibnamefont{Bais}},
  \bibinfo{author}{\bibfnamefont{P.}~\bibnamefont{van Driel}},
  \bibnamefont{and}
  \bibinfo{author}{\bibfnamefont{M.}~\bibnamefont{de~Wild~Propitius}},
  \bibinfo{journal}{Nucl. Phys.} \textbf{\bibinfo{volume}{B393}},
  \bibinfo{pages}{547} (\bibinfo{year}{1993}{\natexlab{a}}),
  \bibinfo{note}{\texttt{hep-th/9203047}}.

\bibitem[{\citenamefont{Bais et~al.}(1993{\natexlab{b}})\citenamefont{Bais,
  Morozov, and de~Wild~Propitius}}]{dgt3}
\bibinfo{author}{\bibfnamefont{F.~A.} \bibnamefont{Bais}},
  \bibinfo{author}{\bibfnamefont{A.}~\bibnamefont{Morozov}}, \bibnamefont{and}
  \bibinfo{author}{\bibfnamefont{M.}~\bibnamefont{de~Wild~Propitius}},
  \bibinfo{journal}{Phys. Rev. Lett.} \textbf{\bibinfo{volume}{71}},
  \bibinfo{pages}{2383} (\bibinfo{year}{1993}{\natexlab{b}}),
  \bibinfo{note}{\texttt{hep-th/9303150}}.

\bibitem[{\citenamefont{Bais and de~Wild~Propitius}(1994)}]{dgt4}
\bibinfo{author}{\bibfnamefont{F.~A.} \bibnamefont{Bais}} \bibnamefont{and}
  \bibinfo{author}{\bibfnamefont{M.}~\bibnamefont{de~Wild~Propitius}},
  \bibinfo{journal}{Theor. Math. Phys.} \textbf{\bibinfo{volume}{98}},
  \bibinfo{pages}{357} (\bibinfo{year}{1994}),
  \bibinfo{note}{\texttt{hep-th/9311162}}.

\bibitem[{\citenamefont{de~Wild~Propitius and Bais}(1998)}]{dgt5}
\bibinfo{author}{\bibfnamefont{M.}~\bibnamefont{de~Wild~Propitius}}
  \bibnamefont{and} \bibinfo{author}{\bibfnamefont{F.~A.} \bibnamefont{Bais}},
  in \emph{\bibinfo{booktitle}{Particles and {F}ields}}, edited by
  \bibinfo{editor}{\bibfnamefont{G.}~\bibnamefont{Semenoff}} \bibnamefont{and}
  \bibinfo{editor}{\bibfnamefont{L.}~\bibnamefont{Vinet}}
  (\bibinfo{publisher}{Springer Verlag}, \bibinfo{address}{New York},
  \bibinfo{year}{1998}), {C}{R}{M} Series in {M}athematical {P}hysics, pp.
  \bibinfo{pages}{353--439}, \bibinfo{note}{\texttt{hep-th/9511201}}.

\bibitem[{\citenamefont{Moore and Seiberg}(1988)}]{MS}
\bibinfo{author}{\bibfnamefont{G.}~\bibnamefont{Moore}} \bibnamefont{and}
  \bibinfo{author}{\bibfnamefont{N.}~\bibnamefont{Seiberg}},
  \bibinfo{journal}{Nucl. Phys.} \textbf{\bibinfo{volume}{300B}},
  \bibinfo{pages}{451} (\bibinfo{year}{1988}).

\bibitem[{\citenamefont{Bonderson and Slingerland}()}]{BondersonWIP}
\bibinfo{author}{\bibfnamefont{P.}~\bibnamefont{Bonderson}} \bibnamefont{and}
  \bibinfo{author}{\bibfnamefont{J.~K.} \bibnamefont{Slingerland}},
  \bibinfo{note}{in preparation}.

\bibitem[{\citenamefont{Rowell et~al.}(2007)\citenamefont{Rowell, Stong, and
  Wang}}]{Rowell2007}
\bibinfo{author}{\bibfnamefont{E.}~\bibnamefont{Rowell}},
  \bibinfo{author}{\bibfnamefont{R.}~\bibnamefont{Stong}}, \bibnamefont{and}
  \bibinfo{author}{\bibfnamefont{Z.}~\bibnamefont{Wang}},
  \emph{\bibinfo{title}{On classification of modular tensor categories}}
  (\bibinfo{year}{2007}), \bibinfo{note}{\texttt{arXiv.org:0712.1377}}.

\bibitem[{\citenamefont{Moore and Seiberg}(1989)}]{mszoo}
\bibinfo{author}{\bibfnamefont{G.~W.} \bibnamefont{Moore}} \bibnamefont{and}
  \bibinfo{author}{\bibfnamefont{N.}~\bibnamefont{Seiberg}},
  \bibinfo{journal}{Phys. Lett.} \textbf{\bibinfo{volume}{B220}},
  \bibinfo{pages}{422} (\bibinfo{year}{1989}).

\bibitem[{\citenamefont{Goddard et~al.}(1985)\citenamefont{Goddard, Kent, and
  Olive}}]{Goddard1984}
\bibinfo{author}{\bibfnamefont{P.}~\bibnamefont{Goddard}},
  \bibinfo{author}{\bibfnamefont{A.}~\bibnamefont{Kent}}, \bibnamefont{and}
  \bibinfo{author}{\bibfnamefont{D.~I.} \bibnamefont{Olive}},
  \bibinfo{journal}{Phys. Lett.} \textbf{\bibinfo{volume}{B152}},
  \bibinfo{pages}{88} (\bibinfo{year}{1985}).

\bibitem[{\citenamefont{Witten}(1984)}]{Witten1984}
\bibinfo{author}{\bibfnamefont{E.}~\bibnamefont{Witten}},
  \bibinfo{journal}{Commun. Math. Phys.} \textbf{\bibinfo{volume}{92}},
  \bibinfo{pages}{455} (\bibinfo{year}{1984}).

\bibitem[{\citenamefont{Gepner and Witten}(1986)}]{Gepner1986}
\bibinfo{author}{\bibfnamefont{D.}~\bibnamefont{Gepner}} \bibnamefont{and}
  \bibinfo{author}{\bibfnamefont{E.}~\bibnamefont{Witten}},
  \bibinfo{journal}{Nucl. Phys.} \textbf{\bibinfo{volume}{B278}},
  \bibinfo{pages}{493} (\bibinfo{year}{1986}).

\bibitem[{\citenamefont{Tsuchiya and Kanie}(1987)}]{tsuka87}
\bibinfo{author}{\bibfnamefont{A.}~\bibnamefont{Tsuchiya}} \bibnamefont{and}
  \bibinfo{author}{\bibfnamefont{Y.}~\bibnamefont{Kanie}},
  \bibinfo{journal}{Lett. Math. Phys.} \textbf{\bibinfo{volume}{13}},
  \bibinfo{pages}{303} (\bibinfo{year}{1987}).

\bibitem[{\citenamefont{Tsuchiya and Kanie}(1988)}]{tsuka88}
\bibinfo{author}{\bibfnamefont{A.}~\bibnamefont{Tsuchiya}} \bibnamefont{and}
  \bibinfo{author}{\bibfnamefont{Y.}~\bibnamefont{Kanie}}, in
  \emph{\bibinfo{booktitle}{Conformal field theory and solvable lattice
  models}}, edited by \bibinfo{editor}{\bibfnamefont{M.}~\bibnamefont{Jimbo}},
  \bibinfo{editor}{\bibfnamefont{T.}~\bibnamefont{Miwa}}, \bibnamefont{and}
  \bibinfo{editor}{\bibfnamefont{A.}~\bibnamefont{Tsuchiya}}
  (\bibinfo{publisher}{Academic {P}ress}, \bibinfo{year}{1988}),
  vol.~\bibinfo{volume}{16} of \emph{\bibinfo{series}{Advanced studies in pure
  mathematics}}, pp. \bibinfo{pages}{297--372}, \bibinfo{note}{erratum in {\em
  "Integrable Systems in quantum field theory and statistical mechanics,"}
  Advanced studies in pure mathematics 19 (1989), 675-682}.

\bibitem[{\citenamefont{Bouwknegt et~al.}(1991)\citenamefont{Bouwknegt,
  McCarthy, and Pilch}}]{bmccp91}
\bibinfo{author}{\bibfnamefont{P.}~\bibnamefont{Bouwknegt}},
  \bibinfo{author}{\bibfnamefont{J.}~\bibnamefont{McCarthy}}, \bibnamefont{and}
  \bibinfo{author}{\bibfnamefont{K.}~\bibnamefont{Pilch}},
  \bibinfo{journal}{Commun. Math. Phys.} \textbf{\bibinfo{volume}{131}},
  \bibinfo{pages}{125} (\bibinfo{year}{1991}).

\bibitem[{\citenamefont{Gomez and Sierra}(1990)}]{gomsier90}
\bibinfo{author}{\bibfnamefont{C.}~\bibnamefont{Gomez}} \bibnamefont{and}
  \bibinfo{author}{\bibfnamefont{G.}~\bibnamefont{Sierra}},
  \bibinfo{journal}{Phys. Lett.} \textbf{\bibinfo{volume}{240B}},
  \bibinfo{pages}{149} (\bibinfo{year}{1990}).

\bibitem[{\citenamefont{Gomez and Sierra}(1991)}]{gomsier91}
\bibinfo{author}{\bibfnamefont{C.}~\bibnamefont{Gomez}} \bibnamefont{and}
  \bibinfo{author}{\bibfnamefont{G.}~\bibnamefont{Sierra}},
  \bibinfo{journal}{Nucl. Phys} \textbf{\bibinfo{volume}{B352}},
  \bibinfo{pages}{791} (\bibinfo{year}{1991}).

\bibitem[{\citenamefont{Slingerland and Bais}(2001)}]{sliba}
\bibinfo{author}{\bibfnamefont{J.~K.} \bibnamefont{Slingerland}}
  \bibnamefont{and} \bibinfo{author}{\bibfnamefont{F.~A.} \bibnamefont{Bais}},
  \bibinfo{journal}{Nucl. Phys.} \textbf{\bibinfo{volume}{B612}},
  \bibinfo{pages}{229} (\bibinfo{year}{2001}), \bibinfo{note}{{\tt
  cond-mat/0104035}}.

\bibitem[{\citenamefont{Bais and Bouwknegt}(1987)}]{Bais1987}
\bibinfo{author}{\bibfnamefont{F.~A.} \bibnamefont{Bais}} \bibnamefont{and}
  \bibinfo{author}{\bibfnamefont{P.~G.} \bibnamefont{Bouwknegt}},
  \bibinfo{journal}{Nucl. Phys.} \textbf{\bibinfo{volume}{B279}},
  \bibinfo{pages}{561} (\bibinfo{year}{1987}).

\bibitem[{\citenamefont{Schellekens and Warner}(1986)}]{Schellekens1986}
\bibinfo{author}{\bibfnamefont{A.~N.} \bibnamefont{Schellekens}}
  \bibnamefont{and} \bibinfo{author}{\bibfnamefont{N.~P.}
  \bibnamefont{Warner}}, \bibinfo{journal}{Phys. Rev.}
  \textbf{\bibinfo{volume}{D34}}, \bibinfo{pages}{3092} (\bibinfo{year}{1986}).

\bibitem[{\citenamefont{Dixon et~al.}(1985)\citenamefont{Dixon, Harvey, Vafa,
  and Witten}}]{Dixon1985}
\bibinfo{author}{\bibfnamefont{L.~J.} \bibnamefont{Dixon}},
  \bibinfo{author}{\bibfnamefont{J.~A.} \bibnamefont{Harvey}},
  \bibinfo{author}{\bibfnamefont{C.}~\bibnamefont{Vafa}}, \bibnamefont{and}
  \bibinfo{author}{\bibfnamefont{E.}~\bibnamefont{Witten}},
  \bibinfo{journal}{Nucl. Phys.} \textbf{\bibinfo{volume}{B261}},
  \bibinfo{pages}{678} (\bibinfo{year}{1985}).

\bibitem[{\citenamefont{Dixon et~al.}(1986)\citenamefont{Dixon, Harvey, Vafa,
  and Witten}}]{Dixon1986a}
\bibinfo{author}{\bibfnamefont{L.~J.} \bibnamefont{Dixon}},
  \bibinfo{author}{\bibfnamefont{J.~A.} \bibnamefont{Harvey}},
  \bibinfo{author}{\bibfnamefont{C.}~\bibnamefont{Vafa}}, \bibnamefont{and}
  \bibinfo{author}{\bibfnamefont{E.}~\bibnamefont{Witten}},
  \bibinfo{journal}{Nucl. Phys.} \textbf{\bibinfo{volume}{B274}},
  \bibinfo{pages}{285} (\bibinfo{year}{1986}).

\bibitem[{\citenamefont{Dixon et~al.}(1987)\citenamefont{Dixon, Friedan,
  Martinec, and Shenker}}]{Dixon1986b}
\bibinfo{author}{\bibfnamefont{L.~J.} \bibnamefont{Dixon}},
  \bibinfo{author}{\bibfnamefont{D.}~\bibnamefont{Friedan}},
  \bibinfo{author}{\bibfnamefont{E.~J.} \bibnamefont{Martinec}},
  \bibnamefont{and} \bibinfo{author}{\bibfnamefont{S.~H.}
  \bibnamefont{Shenker}}, \bibinfo{journal}{Nucl. Phys.}
  \textbf{\bibinfo{volume}{B282}}, \bibinfo{pages}{13} (\bibinfo{year}{1987}).

\bibitem[{\citenamefont{Bais and Taormina}(1986)}]{Bais1986}
\bibinfo{author}{\bibfnamefont{F.~A.} \bibnamefont{Bais}} \bibnamefont{and}
  \bibinfo{author}{\bibfnamefont{A.}~\bibnamefont{Taormina}},
  \bibinfo{journal}{Phys. Lett.} \textbf{\bibinfo{volume}{B181}},
  \bibinfo{pages}{87} (\bibinfo{year}{1986}).

\bibitem[{\citenamefont{Bouwknegt and Nahm}(1987)}]{Bouwknegt1986}
\bibinfo{author}{\bibfnamefont{P.}~\bibnamefont{Bouwknegt}} \bibnamefont{and}
  \bibinfo{author}{\bibfnamefont{W.}~\bibnamefont{Nahm}},
  \bibinfo{journal}{Phys. Lett.} \textbf{\bibinfo{volume}{B184}},
  \bibinfo{pages}{359} (\bibinfo{year}{1987}).

\bibitem[{\citenamefont{Cappelli
  et~al.}(1987{\natexlab{a}})\citenamefont{Cappelli, Itzykson, and
  Zuber}}]{Cappelli1986}
\bibinfo{author}{\bibfnamefont{A.}~\bibnamefont{Cappelli}},
  \bibinfo{author}{\bibfnamefont{C.}~\bibnamefont{Itzykson}}, \bibnamefont{and}
  \bibinfo{author}{\bibfnamefont{J.~B.} \bibnamefont{Zuber}},
  \bibinfo{journal}{Nucl. Phys.} \textbf{\bibinfo{volume}{B280}},
  \bibinfo{pages}{445} (\bibinfo{year}{1987}{\natexlab{a}}).

\bibitem[{\citenamefont{Cappelli
  et~al.}(1987{\natexlab{b}})\citenamefont{Cappelli, Itzykson, and
  Zuber}}]{Cappelli1987}
\bibinfo{author}{\bibfnamefont{A.}~\bibnamefont{Cappelli}},
  \bibinfo{author}{\bibfnamefont{C.}~\bibnamefont{Itzykson}}, \bibnamefont{and}
  \bibinfo{author}{\bibfnamefont{J.~B.} \bibnamefont{Zuber}},
  \bibinfo{journal}{Commun. Math. Phys.} \textbf{\bibinfo{volume}{113}},
  \bibinfo{pages}{1} (\bibinfo{year}{1987}{\natexlab{b}}).

\bibitem[{\citenamefont{Gepner and Kapustin}(1995)}]{Gepner94}
\bibinfo{author}{\bibfnamefont{D.}~\bibnamefont{Gepner}} \bibnamefont{and}
  \bibinfo{author}{\bibfnamefont{A.}~\bibnamefont{Kapustin}},
  \bibinfo{journal}{Phys. Lett.} \textbf{\bibinfo{volume}{B349}},
  \bibinfo{pages}{71} (\bibinfo{year}{1995}),
  \bibinfo{note}{\texttt{hep-th/9410089}}.

\bibitem[{\citenamefont{Gawedzki and Kupiainen}(1989)}]{Gawedzki88}
\bibinfo{author}{\bibfnamefont{K.}~\bibnamefont{Gawedzki}} \bibnamefont{and}
  \bibinfo{author}{\bibfnamefont{A.}~\bibnamefont{Kupiainen}},
  \bibinfo{journal}{Nucl. Phys.} \textbf{\bibinfo{volume}{B320}},
  \bibinfo{pages}{625} (\bibinfo{year}{1989}).

\bibitem[{\citenamefont{Karabali et~al.}(1989)\citenamefont{Karabali, Park,
  Schnitzer, and Yang}}]{Karabali88}
\bibinfo{author}{\bibfnamefont{D.}~\bibnamefont{Karabali}},
  \bibinfo{author}{\bibfnamefont{Q.-H.} \bibnamefont{Park}},
  \bibinfo{author}{\bibfnamefont{H.~J.} \bibnamefont{Schnitzer}},
  \bibnamefont{and} \bibinfo{author}{\bibfnamefont{Z.}~\bibnamefont{Yang}},
  \bibinfo{journal}{Phys. Lett.} \textbf{\bibinfo{volume}{B216}},
  \bibinfo{pages}{307} (\bibinfo{year}{1989}).

\bibitem[{\citenamefont{Bardakci et~al.}(1988)\citenamefont{Bardakci,
  Rabinovici, and S{\"a}ring}}]{Bardakci88}
\bibinfo{author}{\bibfnamefont{K.}~\bibnamefont{Bardakci}},
  \bibinfo{author}{\bibfnamefont{E.}~\bibnamefont{Rabinovici}},
  \bibnamefont{and}
  \bibinfo{author}{\bibfnamefont{B.}~\bibnamefont{S{\"a}ring}},
  \bibinfo{journal}{Nucl. Phys. B} \textbf{\bibinfo{volume}{299}},
  \bibinfo{pages}{151} (\bibinfo{year}{1988}), ISSN \bibinfo{issn}{0550-3213}.

\bibitem[{\citenamefont{Gepner}(1989)}]{gepner89}
\bibinfo{author}{\bibfnamefont{D.}~\bibnamefont{Gepner}},
  \bibinfo{journal}{Phys. Lett. B} \textbf{\bibinfo{volume}{222}},
  \bibinfo{pages}{207} (\bibinfo{year}{1989}).

\bibitem[{\citenamefont{Schellekens and Yankielowicz}(1990)}]{Schellekens1989}
\bibinfo{author}{\bibfnamefont{A.~N.} \bibnamefont{Schellekens}}
  \bibnamefont{and}
  \bibinfo{author}{\bibfnamefont{S.}~\bibnamefont{Yankielowicz}},
  \bibinfo{journal}{Nucl. Phys.} \textbf{\bibinfo{volume}{B334}},
  \bibinfo{pages}{67} (\bibinfo{year}{1990}).

\bibitem[{\citenamefont{Bonderson and Slingerland}(2007)}]{Bonderson07c}
\bibinfo{author}{\bibfnamefont{P.}~\bibnamefont{Bonderson}} \bibnamefont{and}
  \bibinfo{author}{\bibfnamefont{J.~K.} \bibnamefont{Slingerland}},
  \emph{\bibinfo{title}{Fractional quantum {H}all hierarchy and the second
  {L}andau level}} (\bibinfo{year}{2007}),
  \bibinfo{note}{\texttt{arXiv.org:0711.3204}}.

\bibitem[{\citenamefont{Fuchs et~al.}(1996)\citenamefont{Fuchs, Schellekens,
  and Schweigert}}]{Fuchs1995}
\bibinfo{author}{\bibfnamefont{J.}~\bibnamefont{Fuchs}},
  \bibinfo{author}{\bibfnamefont{B.}~\bibnamefont{Schellekens}},
  \bibnamefont{and}
  \bibinfo{author}{\bibfnamefont{C.}~\bibnamefont{Schweigert}},
  \bibinfo{journal}{Nucl. Phys.} \textbf{\bibinfo{volume}{B461}},
  \bibinfo{pages}{371} (\bibinfo{year}{1996}),
  \bibinfo{note}{\texttt{hep-th/9509105}}.

\bibitem[{\citenamefont{Fr\"ohlich et~al.}(2004)\citenamefont{Fr\"ohlich,
  Fuchs, Runkel, and Schweigert}}]{Frohlich2004}
\bibinfo{author}{\bibfnamefont{J.}~\bibnamefont{Fr\"ohlich}},
  \bibinfo{author}{\bibfnamefont{J.}~\bibnamefont{Fuchs}},
  \bibinfo{author}{\bibfnamefont{I.}~\bibnamefont{Runkel}}, \bibnamefont{and}
  \bibinfo{author}{\bibfnamefont{C.}~\bibnamefont{Schweigert}},
  \bibinfo{journal}{Fortschritte der Physik} \textbf{\bibinfo{volume}{52}},
  \bibinfo{pages}{672} (\bibinfo{year}{2004}),
  \bibinfo{note}{\texttt{hep-th/0309269}}.

\bibitem[{\citenamefont{Dunbar and Joshi}(1993{\natexlab{a}})}]{Dunbar1992}
\bibinfo{author}{\bibfnamefont{D.~C.} \bibnamefont{Dunbar}} \bibnamefont{and}
  \bibinfo{author}{\bibfnamefont{K.~G.} \bibnamefont{Joshi}},
  \bibinfo{journal}{Int. J. Mod. Phys.} \textbf{\bibinfo{volume}{A8}},
  \bibinfo{pages}{4103} (\bibinfo{year}{1993}{\natexlab{a}}),
  \bibinfo{note}{\texttt{hep-th/9210122}}.

\bibitem[{\citenamefont{Dunbar and Joshi}(1993{\natexlab{b}})}]{Dunbar1993}
\bibinfo{author}{\bibfnamefont{D.~C.} \bibnamefont{Dunbar}} \bibnamefont{and}
  \bibinfo{author}{\bibfnamefont{K.~G.} \bibnamefont{Joshi}},
  \bibinfo{journal}{Mod. Phys. Lett.} \textbf{\bibinfo{volume}{A8}},
  \bibinfo{pages}{2803} (\bibinfo{year}{1993}{\natexlab{b}}),
  \bibinfo{note}{\texttt{hep-th/9309093}}.

\bibitem[{\citenamefont{Pedrini et~al.}(1999)\citenamefont{Pedrini, Schweigert,
  and Walcher}}]{Pedrini1999}
\bibinfo{author}{\bibfnamefont{B.}~\bibnamefont{Pedrini}},
  \bibinfo{author}{\bibfnamefont{C.}~\bibnamefont{Schweigert}},
  \bibnamefont{and} \bibinfo{author}{\bibfnamefont{J.}~\bibnamefont{Walcher}},
  \bibinfo{journal}{Phys. Lett.} \textbf{\bibinfo{volume}{B466}},
  \bibinfo{pages}{206} (\bibinfo{year}{1999}),
  \bibinfo{note}{\texttt{hep-th/9908185}}.

\bibitem[{\citenamefont{Dijkgraaf et~al.}(1989)\citenamefont{Dijkgraaf, Vafa,
  Verlinde, and Verlinde}}]{DVVV}
\bibinfo{author}{\bibfnamefont{R.}~\bibnamefont{Dijkgraaf}},
  \bibinfo{author}{\bibfnamefont{C.}~\bibnamefont{Vafa}},
  \bibinfo{author}{\bibfnamefont{E.}~\bibnamefont{Verlinde}}, \bibnamefont{and}
  \bibinfo{author}{\bibfnamefont{H.}~\bibnamefont{Verlinde}},
  \bibinfo{journal}{Commun. Math. Phys.} \textbf{\bibinfo{volume}{123}},
  \bibinfo{pages}{485} (\bibinfo{year}{1989}).

\bibitem[{\citenamefont{Dijkgraaf et~al.}(1990)\citenamefont{Dijkgraaf,
  Pasquier, and Roche}}]{dpr}
\bibinfo{author}{\bibfnamefont{R.}~\bibnamefont{Dijkgraaf}},
  \bibinfo{author}{\bibfnamefont{V.}~\bibnamefont{Pasquier}}, \bibnamefont{and}
  \bibinfo{author}{\bibfnamefont{P.}~\bibnamefont{Roche}},
  \bibinfo{journal}{Nucl. Phys. B (Proc. Suppl.)}
  \textbf{\bibinfo{volume}{18B}}, \bibinfo{pages}{60} (\bibinfo{year}{1990}).

\bibitem[{\citenamefont{Mochon}(2004)}]{Mochon04}
\bibinfo{author}{\bibfnamefont{C.}~\bibnamefont{Mochon}},
  \bibinfo{journal}{Phys. Rev. A} \textbf{\bibinfo{volume}{69}},
  \bibinfo{pages}{032306} (\bibinfo{year}{2004}),
  \bibinfo{note}{\texttt{quant-ph/0306063}}.

\bibitem[{\citenamefont{Dou\c{c}ot et~al.}(2004)\citenamefont{Dou\c{c}ot,
  Ioffe, and Vidal}}]{Doucot04}
\bibinfo{author}{\bibfnamefont{B.}~\bibnamefont{Dou\c{c}ot}},
  \bibinfo{author}{\bibfnamefont{L.~B.} \bibnamefont{Ioffe}}, \bibnamefont{and}
  \bibinfo{author}{\bibfnamefont{J.}~\bibnamefont{Vidal}},
  \bibinfo{journal}{Phys. Rev. B} \textbf{\bibinfo{volume}{69}},
  \bibinfo{pages}{214501} (\bibinfo{year}{2004}),
  \bibinfo{note}{\texttt{cond-mat/0302104}}.

\bibitem[{\citenamefont{Dou\c{c}ot et~al.}(2005)\citenamefont{Dou\c{c}ot,
  Feigel'man, Ioffe, and Ioselevich}}]{Doucot05a}
\bibinfo{author}{\bibfnamefont{B.}~\bibnamefont{Dou\c{c}ot}},
  \bibinfo{author}{\bibfnamefont{M.~V.} \bibnamefont{Feigel'man}},
  \bibinfo{author}{\bibfnamefont{L.~B.} \bibnamefont{Ioffe}}, \bibnamefont{and}
  \bibinfo{author}{\bibfnamefont{A.~S.} \bibnamefont{Ioselevich}},
  \bibinfo{journal}{Phys. Rev. B} \textbf{\bibinfo{volume}{71}},
  \bibinfo{pages}{024505} (\bibinfo{year}{2005}),
  \bibinfo{note}{\texttt{cond-mat/0403712}}.

\bibitem[{\citenamefont{Doucot and Ioffe}(2005)}]{Doucot05c}
\bibinfo{author}{\bibfnamefont{B.}~\bibnamefont{Doucot}} \bibnamefont{and}
  \bibinfo{author}{\bibfnamefont{L.~B.} \bibnamefont{Ioffe}},
  \bibinfo{journal}{New Journal of Physics} \textbf{\bibinfo{volume}{7}},
  \bibinfo{pages}{187} (\bibinfo{year}{2005}), \bibinfo{note}{{\tt
  cond-mat/0510612}}.

\bibitem[{\citenamefont{de~Wild~Propitius}(1995)}]{thesismark}
\bibinfo{author}{\bibfnamefont{M.}~\bibnamefont{de~Wild~Propitius}}, Ph.D.
  thesis, \bibinfo{school}{University of Amsterdam} (\bibinfo{year}{1995}).

\bibitem[{\citenamefont{Ardonne and Schoutens}(1999)}]{eddyenkjs}
\bibinfo{author}{\bibfnamefont{E.}~\bibnamefont{Ardonne}} \bibnamefont{and}
  \bibinfo{author}{\bibfnamefont{K.}~\bibnamefont{Schoutens}},
  \bibinfo{journal}{Phys. Rev. Lett.} \textbf{\bibinfo{volume}{82}},
  \bibinfo{pages}{5096} (\bibinfo{year}{1999}),
  \bibinfo{note}{\texttt{cond-mat/9811352}}.

\bibitem[{\citenamefont{Grosfeld and Schoutens}()}]{Grosfeld2008}
\bibinfo{author}{\bibfnamefont{E.}~\bibnamefont{Grosfeld}} \bibnamefont{and}
  \bibinfo{author}{\bibfnamefont{K.}~\bibnamefont{Schoutens}},
  \bibinfo{note}{in preparation}.

\bibitem[{\citenamefont{Freedman et~al.}()\citenamefont{Freedman, Nayak,
  Walker, and Wang}}]{freedman08}
\bibinfo{author}{\bibfnamefont{M.}~\bibnamefont{Freedman}},
  \bibinfo{author}{\bibfnamefont{C.}~\bibnamefont{Nayak}},
  \bibinfo{author}{\bibfnamefont{K.}~\bibnamefont{Walker}}, \bibnamefont{and}
  \bibinfo{author}{\bibfnamefont{Z.}~\bibnamefont{Wang}},
  \emph{\bibinfo{title}{On picture (2+1)-{TQFT}s}}, \bibinfo{note}{{\tt
  arXiv:0806.1926}}.

\bibitem[{\citenamefont{Troyer et~al.}(2008)\citenamefont{Troyer, Trebst,
  Shtengel, and Nayak}}]{Troyer08}
\bibinfo{author}{\bibfnamefont{M.}~\bibnamefont{Troyer}},
  \bibinfo{author}{\bibfnamefont{S.}~\bibnamefont{Trebst}},
  \bibinfo{author}{\bibfnamefont{K.}~\bibnamefont{Shtengel}}, \bibnamefont{and}
  \bibinfo{author}{\bibfnamefont{C.}~\bibnamefont{Nayak}},
  \emph{\bibinfo{title}{Local interactions and non-abelian quantum loop gases}}
  (\bibinfo{year}{2008}), \bibinfo{note}{{\tt arXiv.org:0805.2177}}.

\bibitem[{\citenamefont{Bais and Romers}()}]{baisromers2008}
\bibinfo{author}{\bibfnamefont{F.~A.} \bibnamefont{Bais}} \bibnamefont{and}
  \bibinfo{author}{\bibfnamefont{J.~C.} \bibnamefont{Romers}},
  \bibinfo{note}{in preparation}.

\end{thebibliography}

\end{document}